\documentclass[12pt]{iopart}
\pdfoutput=1
\usepackage{graphicx}
\usepackage{iopams}
\begin{document}

\title{Colloidal Gels: \\
Equilibrium and
Non-Equilibrium Routes}

\author{Emanuela
Zaccarelli}
\address{ Dipartimento di Fisica and CNR-INFM-SOFT, \\
Universit\`{a} di
Roma La Sapienza, P.le A. Moro 2, I-00185 Rome, Italy\\
E-mail: emanuela.zaccarelli@phys.uniroma1.it
}

\begin{abstract}
We attempt a classification of different colloidal gels based on
colloid-colloid interactions. We discriminate primarily between
non-equilibrium and equilibrium routes to gelation, the former case
being slaved to thermodynamic phase separation while the latter is
individuated in the framework of competing interactions and of patchy
colloids. Emphasis is put on recent numerical simulations of colloidal
gelation and their connection to experiments. Finally we underline
typical signatures of different gel types, to be looked at, in more
detail, in experiments.  
\tableofcontents
\end{abstract}

\section{Introduction}
\label{sec:1}

In recent years, dynamical arrest in colloidal, and more generally in
soft matter systems, has gained increasing scientific attention
\cite{network}.  Colloidal suspensions have unambiguous advantages with
respect to their atomic counterparts.  Characteristic space and time
scales are much larger, allowing for experimental studies in the light
scattering regime and for a better time resolution. The large
dimension of the particles allows for direct observation with confocal
microscopy techniques, down to the level of single-particle
resolution\cite{Weeks07}.  In addition, the `tunability' of the system
and of particle-particle interactions is almost arbitrary, as
opposed to standard atomic interactions fixed by elementary chemistry.
Therefore, it opens up the extraordinary possibility to `engineer'
colloidal model systems, for example by synthesizing ad-hoc particles
with specific properties\cite{Manoh_03} or simply changing the
solution composition by appropriate additives and salt ions. This is
accompanied by a great control and a fine tuning of the interparticle
potential parameters\cite{Blaad_03}. The explored field of research is
rapidly growing\cite{vanblaad} to include all kinds of spherical
interactions, as well as to address the role of anisotropy, either due
to the shape of the particles or to the presence of different chemical
subunits in the colloidal particles, the so-called `patches', with
different properties with respect to the rest of the particles.
  
Colloidal suspensions, despite being very complex in nature and number
of components, can be well described theoretically via simple
effective potentials\cite{Lik01b}. Indeed, the solvent and additives
degrees of freedom are generally much faster than those of the
colloidal particles, so that they can be
effectively `integrated out'. This provides the possibility of describing the
complexity of the solutions via simple effective one-component models
for the colloids only, the most famous of which are the DLVO
potential\cite{dlvo} or the Asakura-Oosawa model\cite{Asa58a}.  In
this respect, from a fundamental point of view, colloidal systems and
soft matter can be considered as `ideal' model systems with `desired
interactions' to be tested with rapidly advancing experimental
techniques (for a recent review of this topic, see \cite{Cip05a}), and
often closely compared with theory and simulations.

Much effort has been devoted so far to clarify the dynamical behaviour
at large packing fractions, where dynamical arrest, commonly identified as a
glass transition, takes place. In this respect, already other
reviewers have described the state of the art \cite{Daw02a,advances}.
Here, we aim to give a perception of what happens when the system
slows down and arrests at much smaller densities.  An experimental
review of this topic, focusing on elasticity concepts, has appeared
recently\cite{Tra04a}.  Dynamic arrest at low densities, in terms of
dominating mechanisms and various interplay, is still very poorly
understood.  A review of the low-density behaviour in attractive
colloids was reported about a decade ago by Poon\cite{Poon98a}.  This
work focused on the view of colloids as
`super-atoms', for which a thermodynamic description can still be
applied, and mainly reported about the relation between phase
separation and gelation, in particular to address the often-invoked
point that a similarity, in equilibrium phase diagrams and arrest
transitions, should hold between colloids and globular proteins, of
deep importance because of protein crystallization
issues\cite{Pia00a,Sear06}.

The problems in understanding deeply the low-density region of the
colloids phase diagram are multiple. Experimentally, there is a zoo of
results, often in contradiction with each other.  Sometimes the
studied systems are highly complicated to be used as prototypes of the
gel transition (see for example Laponite) or to make general claims
about the nature of the arrest transition and phase diagram.  In other
cases, the system is not enough well characterized, to be sure of the
responsible interactions determining some type of aggregation instead
of phase separation and so on.  For example, only recently the
important role of residual charges on colloidal particles
\cite{Gro03a} has been elucidated in PMMA spheres
gelation\cite{Din02aJPCM,Bartlett04}.  Theoretically the situation is
not better, as, in most cases, there is not yet a unifying theoretical
framework capable to roughly locate and describe the colloidal gel
transition, as it was for example the Flory theory for chemical
gelation\cite{Florybook} or the ideal Mode Coupling Theory
(MCT)\cite{goetze} for colloidal glasses.  MCT is applicable for
low-density arrested solids only to a certain extent, as for example
to describe Wigner glasses\cite{Sci04a}.
Finally, the role of numerical simulations
is quite important at present, since a number of models are being
studied to incorporate the minimal, necessary ingredients to
discriminate between gelation, phase separation, cluster or glass
formation.

In our opinion, the principal question to ask is the very basic
definition of what a colloidal gel is and of its, possibly existing, 
universal features. Moreover, it is not clear if a gel can be described 
in an unifying framework including glasses and non-ergodic states
in general. Sometimes the terminology gel/glass is interchanged. In
this review, we will try to assess under which conditions each should
be used. Moreover, we will propose a classification scheme between
different gelation mechanisms.
In this respect, the role of interparticle potential will be important
in characterizing the different gel states. We will put particular
emphasis on the difference between non-equilibrium and
equilibrium approach to gelation.

In a thinking framework, the creation of an ideal model for
equilibrium gels, as canonical as the hard sphere model for glasses
would be important for future studies.  Very recently, some efforts
towards the individuation of the basic ingredients that are necessary
to design such model are being carried out.  Strong evidence, from
experiments\cite{Man05a} and simulations\cite{zaccapri,Fof05a}, has
proven that for hard-core plus spherically-symmetric pair-wise
attractive potentials, arrest at low density occurs only through an
interrupted phase separation. In the limit of very large attraction
strength and very small density, this scenario crosses continuously to
Diffusion-Limited-Cluster-Aggregation (DLCA)\cite{Vicsekbook}.
Modification of simple attraction is invoked to produce gelation in
equilibrium. This turns out to be the case when long-range repulsion,
induced by excessive surface charges in solution, complements the
short-range depletion attraction\cite{Sciobartlett}, as well in the
new family of patchy\cite{Bianchi_06} or limited-valency
potentials\cite{Zac05aPRL}.  The present review will try to describe
some of the models and their predictions for gelation, focusing mainly
on recent advances in modeling and simulations. Finally we will try to
characterize, within the limits of the present knowledge, the basic
features of the different encountered gels in connection to
experiments. Our aim is to provide a reference framework for future
understanding of this complicated state of matter, that is ubiquitous
in applications, and frequent in everyday life from the kitchen table
to our own body.

\section{Definitions and scope}
\label{sec:intro} 

To present a coherent picture of the state of the art in the field of
colloidal gelation, we introduce and classify in this Section
different phenomena that have similarities, interplay, or are at the
essence of colloidal gelation. In particular, we start by discussing
chemical gelation and percolation theory.  Then we describe physical
gels and we illustrate the gel-formation process with respect to
percolation and phase separation. We also briefly mention DLCA
gels. We will emphasize the role of the `bond lifetime' as key concept
to identify a gelation mechanism. We illustrate equilibrium and
non-equilibrium routes to physical gelation, introducing the concept
of `ideal gels' and drawing typical phase diagrams as a reference for
the different types of systems.  Two brief paragraphs will conclude
this section, with the specific goals to (i) clarify the role of
percolation towards gelation and other types of arrested low-density
solids and (ii) highlight the repulsive and attractive glass
transition at high densities. Both these topics are very relevant to
the following discussion, especially to understand their relation, in
properties and location, with respect to the phase diagram and
(eventually) gel formation.

In the next Section \ref{sec:bond}, we focus on the role of the bond
lifetime as the parameter connecting chemical to physical gelation,
reporting results from numerical models which have focused on this
aspect. In Section \ref{sec:routes}, we will discuss three
different routes to gelation: (i)
non-equilibrium gelation as arrested phase separation ; (ii)
equilibrium gelation resulting from a competition between short-range
attraction and long-range repulsion; (iii) equilibrium (and ideal)
gels made of particles with patchy (or directional) interactions. In
Section \ref{sec:experiments} we try to individuate common and
different signatures of the three types of gels in connection to
experimental observations (past or future).  Finally, we draw our
conclusions and perspectives of future studies.

\subsection{Basic definition of a gel}

Let us start with the basic definition of a gel from Britannica
encyclopedia: {\it coherent mass consisting of a liquid in which
particles are either dispersed or arranged in a fine network
throughout the mass. A gel may be notably elastic and jellylike (as
gelatin or fruit jelly), or quite solid and rigid (as silica
gel)}\cite{britannica}.  From this general definition it follows that
a low density disordered arrested state which does not flow but
possess solid-like properties as a yield stress, is commonly named a
gel. Similarly to glasses, the gel structure, does not show any
significant order and, in this respect, it is similar to that of a
liquid.  However, for dilute systems, a gel often displays large
length scale signal associated to the fractal properties of its
structure.
The terminology of sol-gel transition refers to a liquid mixture where
solute (sol) particles (ranging from monomers to biological
macromolecules) are suspended in a solvent. Initially the sol
particles are separated, but, under appropriate conditions, they aggregate
until a percolating network is formed.  In the following the
conditions under which such percolating network can be defined as a
gel will be discussed. Colloidal gels are often formed by particles
dispersed in a liquid solvent.  However, in polymers and silica-gels
the solvent is not a liquid or it is missing.

\subsection{Chemical Gelation and Percolation}

Chemical gelation studies were initiated in the framework of
cross-linking polymers, whose gelation transition was 
associated to the formation of an infinite network with finite shear
modulus and infinite zero-shear viscosity. At the gelation point, the
system stops flowing. One possible example of polymer
gel-forming systems is provided by epoxy resins\cite{Florybook}. In
these systems, polymer chains grow step-wise by reactions mediated by
end-groups or cross-linkers (step polymerization). As the
 reaction
takes place, chemical (hence irreversible) bonds between different
chains are formed. If the (average) functionality of the monomers is
greater than two, to allow the establishment of a branched structure
with junction points, a fully connected network, spanning the whole
space, is built\cite{corezzi05JPCM} and a gel is obtained. 
Another example is rubber
whose gelation
process is usually called vulcanization, where entangled polymers are
not bonded at first, and, with time of reaction, covalent bonds are
chemically induced.

The irreversible (chemical) gelation process is well described in
terms of percolation theory, since --- due the infinite lifetime of
the bonds ---the gel point coincides strictly with the appearance of
an infinite spanning network.  The mean-field theory of percolation
was developed by Flory\cite{flory} and Stockmayer\cite{stock1,stock2},
under the following two assumptions: independent bonds and absence of
bonding loops.  Each possible bond is formed with a probability $p$
and the percolation threshold is identified in terms of a critical
bond probability $p_c$, analytically calculated on the Bethe
lattice\cite{flory,Sta92book}.  Close to $p_c$, the cluster size
distribution $n(s)$ is found to scale as a power law of the cluster
size $s$: $n(s)\sim s^{-\tau} f[s^{\sigma}(p-p_c)]$, while the mean
cluster size $S \equiv \sum s^2 n(s)/[\sum s n(s)]$ is found to
diverge at percolation as $S \sim (p-p_c)^{-\gamma}$. The probability
to belong to the spanning cluster $P_{\infty}$ is found to grow from
the transition as $P_{\infty}\sim (p-p_c)^{\beta}$. Finally, the
cluster radius of gyration $R_g$ is found to scale with the cluster
size as $R_g \sim s^{1/d_f}$, where $d_f$ is the cluster fractal
dimension. Here, $\tau$, $\gamma$, $\beta$ and $\sigma$ are universal
exponents satisfying appropriate scaling relations, as
$\gamma=(3-\tau)/\sigma$ and $\beta=(\tau-2)/\sigma$, while $f(z)$ is
a system-dependent scaling function\cite{Sta92book}. In 3d, the
exponents have been calculated numerically for many systems, resulting
in $\tau=2.18, \sigma=0.45$ and $d_f=2.53$, which are the exponents of
the random percolation universality class.

Percolation is defined in term of bonds, i.e. it is based on the
connective properties of the system.  It does not require information
on the physical properties of the bond, on the temperature dependence
of the bond probability or, even more importantly, on the lifetime of
the bonds as well as of the spanning cluster. In this respect, its extension
to non-covalent (non-permanent) bonds requires caution.

\begin{figure}[h]
\begin{center}
\includegraphics[width=8cm,angle=270.,clip]{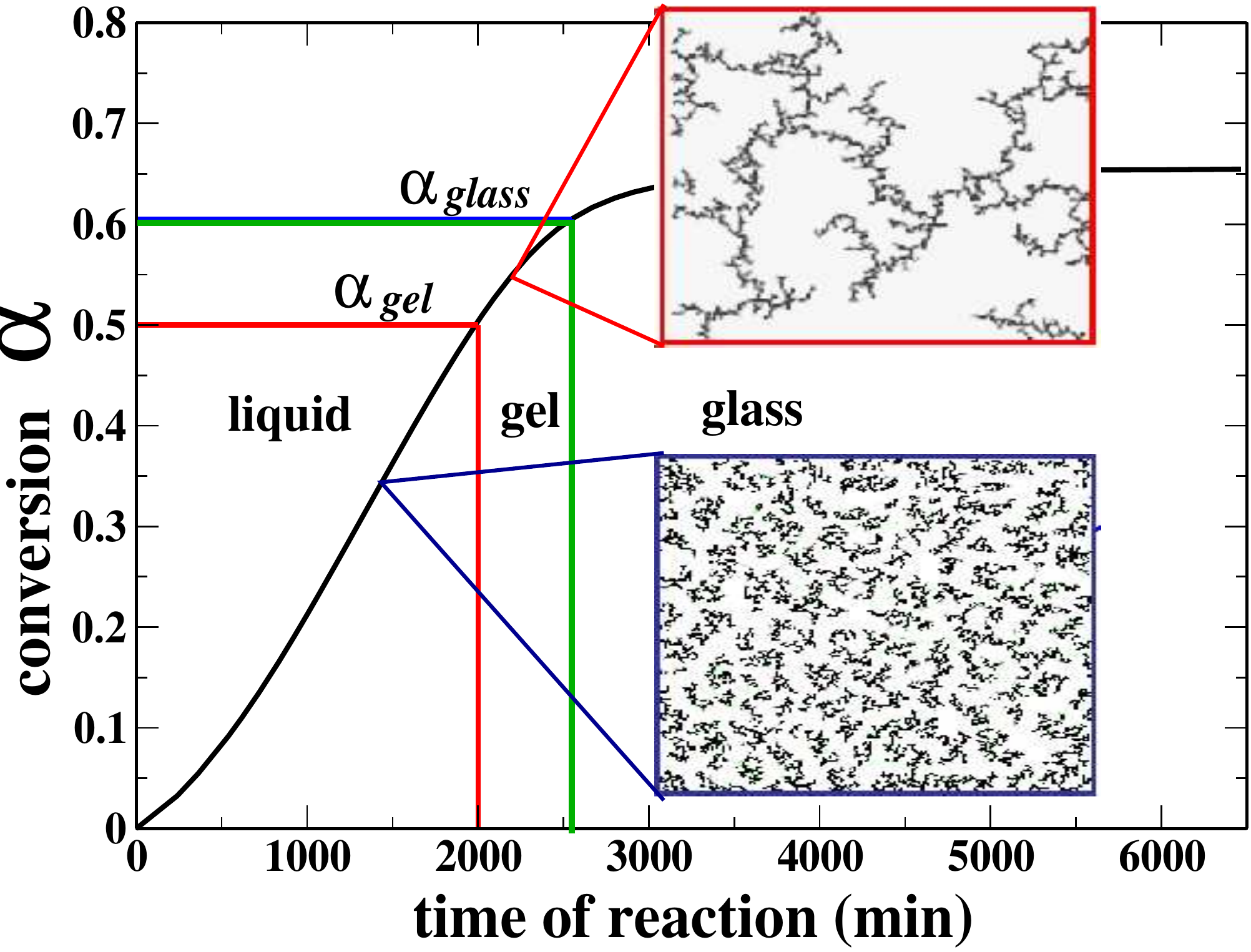}
\end{center}
\caption{Chemical conversion $\alpha$, indicating the fraction of
formed bonds during a chemical reaction, versus time of reaction. It
commonly saturates at a finite value well below $1$. Before reaching a
plateau value, the system encounters the gel transition at
$\alpha_{gel}$ and the glass one at $\alpha_{glass}$. The curve refers
to a mixture of epoxy resins with different functionalities. The
images show a representation of the liquid phase and of the gel
network. Note that different scales are used for resolution of the two
images: the particle volume fraction does not change along the
reaction. Courtesy of S. Corezzi.  }
\label{fig:alpha}
\end{figure}

In the case of chemical bonds, a clear distinction can be formulated
between chemical gelation and chemical vitrification. As shown in
Fig.~\ref{fig:alpha}, with the proceeding of a polymerization process,
an increasing fraction of bonds $\alpha$, commonly named chemical
conversion, is formed.  Gelation is found at the time of reaction
where the systems stops flowing. At this point the system percolates
and only the fraction $\alpha_{gel}$ of possible bonds is formed,
which can be well predicted by Flory theory\cite{Florybook}. With
further proceeding of the reaction, other bonds are formed until a
point where $\alpha$ saturates to a plateau value, well below the
fully connected state ($\alpha=1$). This indicates that the system
becomes trapped into a metastable minimum of the free energy and
undergoes a glass transition at the typical conversion
$\alpha_{glass}$.  In this case, the system becomes non-ergodic, the
density auto-correlation function displays a plateau in time and the
structural relaxation time becomes larger than the experimental time
window\cite{corezzi03,corezzi06}, as found in standard glasses.  A
length-scale dependent analysis of the chemical gel and glassy states
should be able to discriminate between the two cases.  Indeed, while
the glass is non-ergodic at all relevant length-scales, the gel only
has a correlation, dictated by the infinite network, strictly at
$q\rightarrow 0$, while all other length-scales retain a
quasi-ergodicity.

Experimental and simulation works on chemical gelation have reported
\cite{Martin88a,Martin88b,Martin91,Ikk99,Del03aEL}: (i) a slow
relaxation approaching the gel transition, that can be well fitted by
a stretched exponential decay; (ii) a power-law decay of the density
and stress auto-correlation functions close to percolation.  An
experimental study of the dynamical behaviour well within the gel
region is also performed in Ref.\cite{Martin91}, where the power-law
decay is also found in the gel phase for $q$-values well in the
diffusive regime. Given the limited investigated range in $q$ and in
gel states, no extensive characterization of the wave-vector
dependence of the gel and percolation transition was performed, also
in relation to the evolution of the non-ergodic properties approaching
the glass transition.

\subsection{Physical gelation} 

Physical gels are gels in which bonds originate from physical
interactions of the order of $k_BT$, so that bonds can reversibly
break and form many times during the course of an experiment.  This
provides a fundamental difference in the nature of chemical with
respect to physical gels. The latter are usually formed by colloidal
and soft particles as well as associative polymers, and bonds are
induced via depletion interactions, hydrogen bonds, hydrophobic
effects to name a few.  This difference allows us to classify
generally as chemical gels those characterized by irreversible bond
formation, and as physical gels those in which the bonds are
transient, i.e. are characterized by a finite (although large)
lifetime.

Non-exhaustive examples of transient gel-forming systems are:
colloid-polymer
mixtures\cite{Ile95aPRE,Poo97aPhysA,Ver97aPhysA,Seg01aPRL}, in which
polymers act as depletants, and hence polymer concentration $c_p$
controls the attraction strength; colloidal silica spheres that are
sterically stabilized by grafting polymer chains onto their surface
\cite{Grant93,Ver94a,Rueb97,Sol01aPRE,Nara2006PRE}, 
where temperature, changing the solvent quality of the polymer chains, acts
as the control parameter for an effective adhesive attractions between
the colloidal spheres; telechelic micelles with functionalized
end-groups\cite{Vlass99,Lafleche,federica} or a ternary mixture of oil-in-water
microemulsion in suspension with telechelic
polymers\cite{Michel00Lang}, where bridging of micelles is provided by
hydrophobic end-caps; among gel-forming protein systems, the case of
sickle cell hemoglobin\cite{Manno04,vekilov}, where attraction should
be as in typical globular proteins short-range, probably patchy, and
arising from a combination of hydrophobic effects and van der Waals
attraction.

In the framework of thermoreversible gelation for associative
polymers, a long-standing debate involves the association of the
percolative (network-forming) transition to a thermodynamic
transition.  This question arises naturally from the different
assumptions implied respectively in the Flory and in the Stockmayer
approach in the post-gel regime.  A recent review focused on this
question \cite{Rub99a} and suggested, based on several studies of
different associating systems, that the gel transition is not
thermodynamic, but rather {\it connective} in nature.  In this
review, we provide evidence that no signature of a thermodynamic
transition is found in colloidal gelation, a result consistent with
the finite lifetime of the bonds.  Moreover, we point out that, in
general, when the bond lifetime is much shorter than the experimental
time-scale, the establishment of a network, i.e. percolation, is not
even associated to a dynamic transition.

In standard percolation studies, the bond lifetime, and hence the lifetime of
the spanning cluster, is not taken into account. For chemical gels,
the bond lifetime is infinite and thus percolation theory has been the
theoretical framework for describing the gel transition. In the
case of chemical bonds, where bond formation and bond duration are
coupled, the percolation concept is connected to the dynamics and
thus, it can describe the chemical gelation transition.  For colloidal
gels, bonds are transient. Clusters break and reform
continuously. Percolation theory can thus be applied only to describe
static connectivity properties. Neglecting dynamic information, it is
still possible to locate the line in the phase diagram where a
spanning transient cluster first appears, which plays the role of
percolation transition locus.  Analysis of the cluster size
distribution and of all other percolation observables
($S,P_{\infty},R_g$) close to such a line are consistent with the
universality class of random percolation\cite{Sta92book,Tor02book}.

\begin{figure}
\begin{center}
\includegraphics[width=8cm,angle=270.,clip]{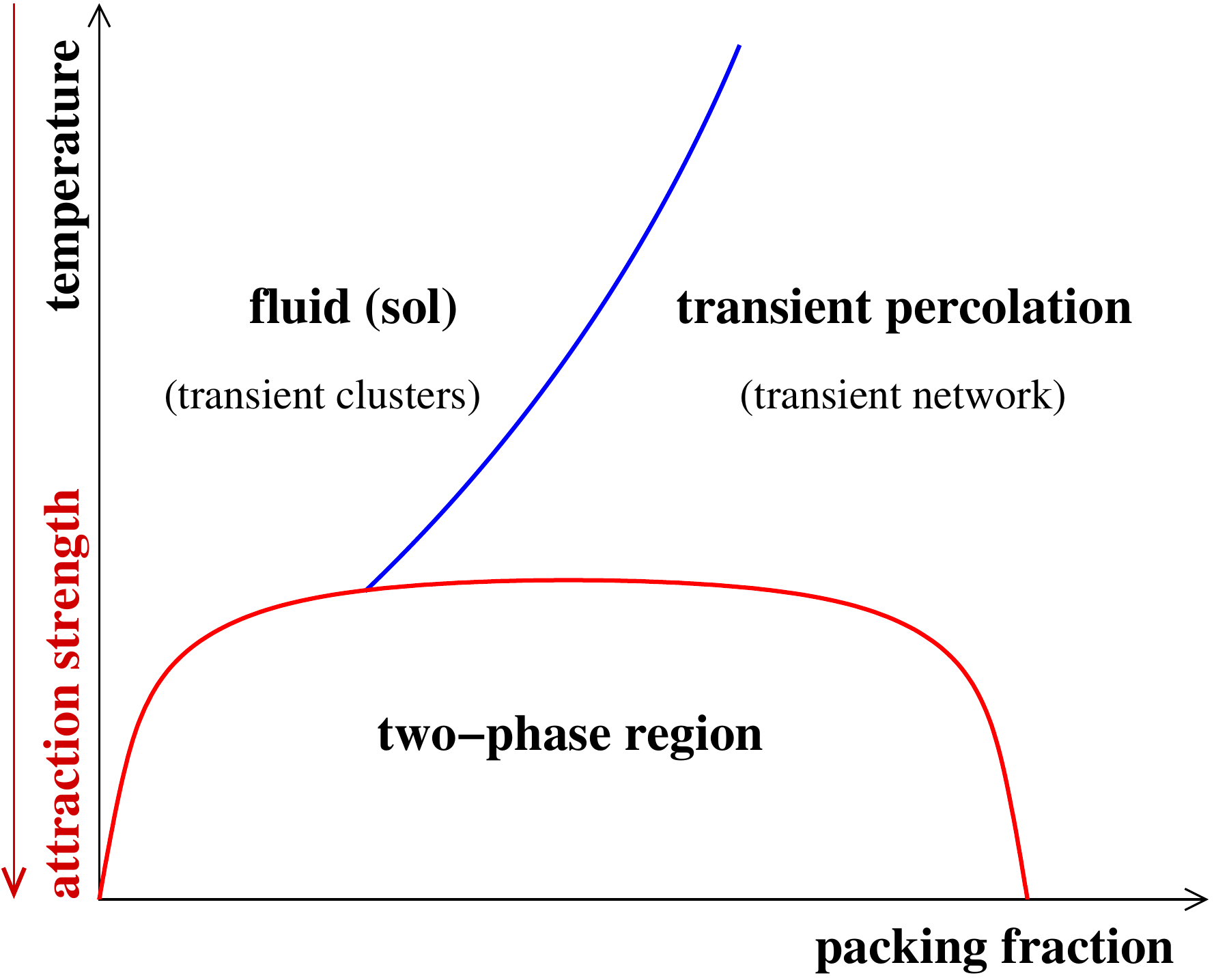}
\end{center}
\caption{Schematic picture of the percolation transition in physical
gels, where the formation of transient network does not have
implication for gelation.}
\label{fig:phase1}
\end{figure}

A schematic plot of the phase diagram for a simple attractive
potential, including beside the phase separation locus also the
percolation line, is shown in Fig.~\ref{fig:phase1}.  No dynamical
ingredients are taken into account within this picture, and hence no
information on the location of the arrested states is provided. Only
if the lifetime of the bonds close to the percolation locus is longer
than the experimental observation time it would be possible to
conclude that the system becomes non ergodic at the percolation
line. Among the studies pointing out the irrelevance of the
percolation transition for reversible gelation was a theoretical
description of thermoreversible gelation for associating polymers by
Rubinstein and Semenov\cite{Rub99b}, soon followed by a lattice model
simulation by Kumar and Douglas\cite{Kum01a}.

The colloidal gel-forming systems are often based on spherically
symmetric attractive potentials. One famous example is the
Asakura-Oosawa (AO) \cite{Asa58a} effective potential for
colloid-colloid attraction entropically induced by the polymers.
Bonds can here be defined between any pair of particles with a
relative distance smaller than the attraction range.  When attraction
strength is increased, the system prefers to adapt locally dense
configurations, so that energy can be properly minimized.  Under these
conditions, a liquid condensation (a colloidal liquid) is favored, as
discussed in more details below.  The presence of a phase-separation
region in the phase diagram is thus often intimately connected to the
presence of a percolation
locus\cite{coniglio-klein,coniglio-klein-stanley-prl}.

\subsection{Interplay between  Phase Separation and Physical Gelation}
\label{sec:gamma}
Percolation in physical gel-forming systems does not correspond to
gelation due to finite bond lifetime.  Long-living bonds necessarily
require large attraction strength.  In systems in which the hard-core
interaction is complemented by spherically symmetric attraction, very
large attraction strengths not only increase the bond lifetime but
also inevitably lead to the onset of liquid-gas (colloid rich-colloid
poor) phase separation.  We can rationalize the tendency to phase
separate through Hill's work on liquid condensation in term of
physical clusters\cite{hill}. Indeed, the free energy $F_N$ of a
cluster of $N$ particles can be written as contribution of a bulk and
a surface term, respectively proportional to $N$ and to $N^{2/3}$.
Thus $F_N/N =f_{bulk}+\gamma N^{-1/3}$, where $\gamma$ is proportional
to the surface tension and $f_{bulk}$ is the free energy per particle
in bulk conditions.  If $\gamma > 0$, then $F_N/N$ is minimized for
$N\rightarrow\infty$ and hence a condensed liquid phase is expected.
At sufficiently low $T$, where entropic terms can be neglected,
$\gamma \propto (e_{surface}- e_{bulk})$, where $e_{surface}$ and
$e_{bulk}$ are the energy of a particle on the surface and in the
interior of a cluster respectively.  For spherically symmetric
attractive potentials $ e_{bulk} < e_{surface}$ and hence $\gamma>0$
(see for example the calculation for cluster ground state energy for
various widths of attraction from Lennard-Jones to narrow
wells\cite{wales97,Mos04a}), so that lowering the temperature will
always induce phase separation.  If $\gamma\leq 0$\cite{genova} a bulk
liquid-gas separation will be disfavored.  We will analyze the
separate cases $\gamma < 0$ and $\gamma \simeq 0$ later on.

\begin{figure}[h]
\begin{center}
\includegraphics[width=7cm,angle=270.,clip]{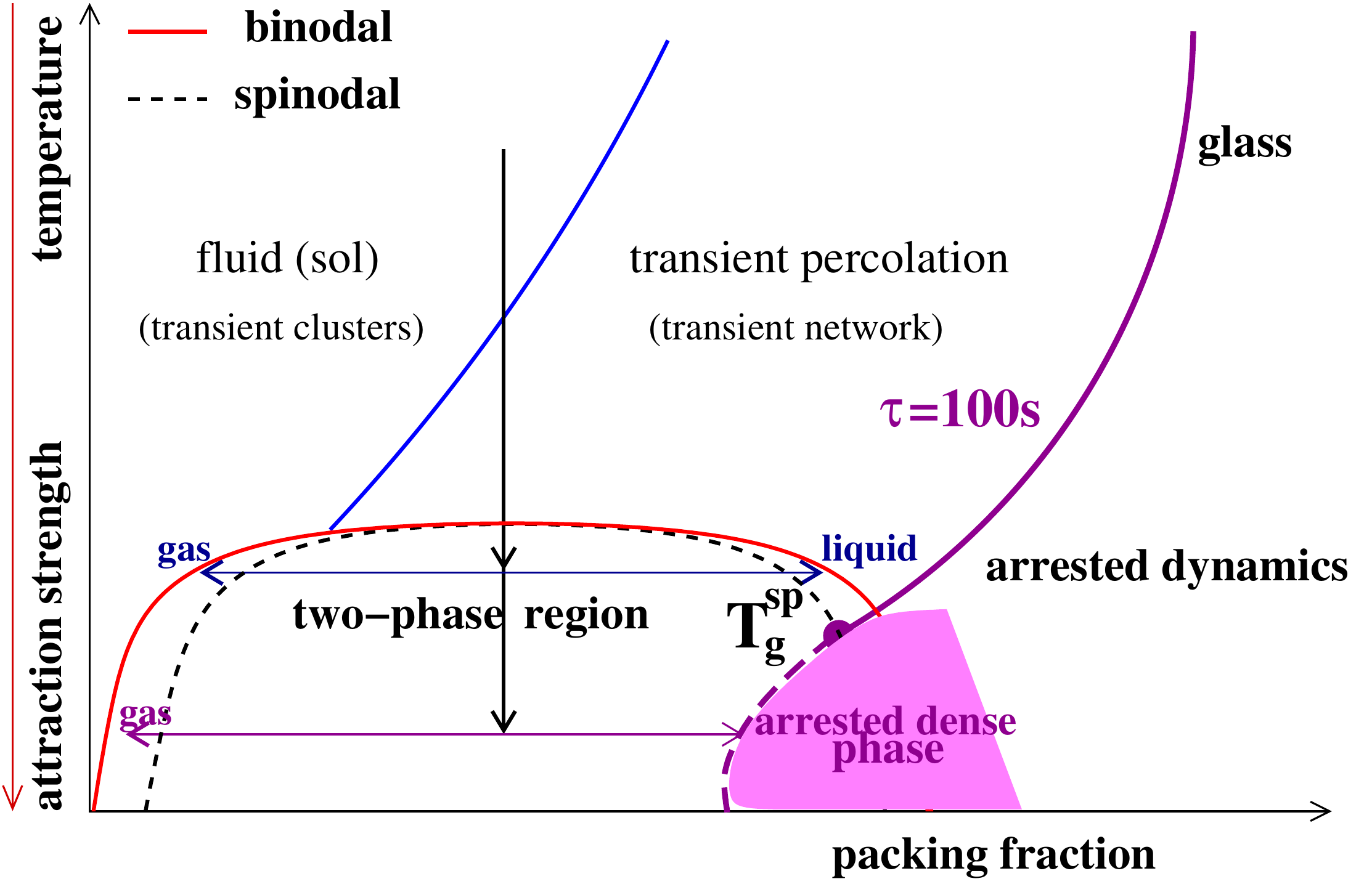}
\end{center}
\caption{Schematic picture of the interrupted phase separation or
arrested spinodal scenario. A quench into the two-phase region may
lead to an arrest of the denser phase. It is not yet clear how the
glass line continues within the spinodal region. The figure shows the
case where the density fluctuations freeze before they reach the final
spinodal value, a scenario that is supported by a study on
lysozyme\protect\cite{cardinauxpreprint}. Alternatively, the glass
line might merge with the spinodal on the high density branch.  }
\label{fig:phase2}
\end{figure}

On the basis of these considerations we can suggest a first
crucial distinction between different types of arrest at low density by
discriminating whether the system undergoes gelation with or without
the intervening of phase separation.  

If the phase separation boundary is crossed before dynamical arrest
takes place (for example through a quench inside the spinodal
decomposition region) the system will experience liquid
condensation. The coarsening process will induce the formation of
dense regions which might arrest due to the crossing of a glass
transition boundary.  In this case we talk of `arrested (or
interrupted) phase separation' or `arrested spinodal
decomposition'\cite{zaccapri,Cat04aJPCM}. This route to gelation is a
non-equilibrium route, as it is made possible through an irreversible
process, i.e. spinodal decomposition, and it is pictorially represented
in Fig.~\ref{fig:phase2}, and discussed in details for short-ranged
attractive colloids, in particular colloid-polymer mixtures, in
subsection \ref{sec:low}.

\subsection{DLCA gels}
A remarkable case of arrested spinodal mechanism is that of 
Diffusion-Limited-Cluster-Aggregation (DLCA)\cite{Vicsekbook},
that is realized when a very low density colloidal system is quenched
to a state point with large attraction strength, combining in this
limit aspects of chemical and physical gelation. Indeed, in this
limit, attraction is so large that bonds are effectively irreversible.
The aggregation process is mediated by diffusion of the growing
clusters, which irreversibly stick when touching, forming a well
characterized fractal structure (with $d_f\simeq 1.75$).  Arrest is
achieved by inhomogeneous filling of all available space with clusters
of progressively smaller density.  The percolation transition is here
mediated by clusters, rather than particles as in chemical gelation.

Several experimental studies have focused on gelation in the DLCA
limit\cite{WeitzDLCA84,Car92a,Kra98aPRL}. In these strongly
aggregating colloids, the bond energy is much larger than
$k_BT$. These types of gels are found to exhibit fractal properties
and aging dynamics\cite{Cip00aPRL,Ram05aPRL}.  Interestingly, several
types of fundamental questions on the internal dynamics, restructuring
and limits of stability of such low-density gels can be tackled by
these kind of
studies\cite{Man04aPRL,Man05aPRL,Man05bPRL,CipellettiJSTAT}. In these
types of gels, phase separation is kinetically interrupted by the
freezing of the bonds, hence we can also consider these gels to belong
to the category of `out-of-equilibrium' gels.

Also, many numerical studies have addressed DLCA, at first onto a
lattice with particular interest on understanding the cluster
properties and the fractal dimension
\cite{Kolb83PRL,Meak83PRL,Vic84PRL,Vicsekbook,Gim99a}. Later on,
studies have addressed the full gelation process, to also examine the
fractal properties and structure of the gel\cite{Gim95a,Gim99a}. To do
so, off-lattice realizations of DLCA were
employed\cite{hasmy93,Rot04a,Rot04b}, to allow for a more realistic
characterization of the structure of the clusters as well as of the
percolating network.

\subsection{Equilibrium approaches to Gelation}

If phase separation is not intervening (for example via the
realization of the condition $\gamma \leq 0$ in Hill's formalism) the
system is able to form a stable particle network, through a series of
equilibrium states. We call this scenario `equilibrium gelation',
since the gel state is reached continuously from an ergodic phase,
always allowing an equilibration time, much longer than the bond
lifetime, for the system to rearrange itself.
It is important to point out that the experimental determination of a
gel transition requires an arbitrary definition of time-scale, in
analogy with the glass case. The glass transition is commonly
signaled with the point where the viscosity of a glass-forming system
becomes larger than typically $10^{13}$ poise, or equivalently, when
the non-ergodic behaviour persists for an observation time-scale of
$10^2 s$.

Also in the case of gels, the dynamical arrest process will be
strictly connected to the observation time window.  Indeed, being the
bond-lifetime finite, there always exists a longer time-scale over
which the system will eventually relax.  Therefore, it is useful to
adopt an `operative' definition of gelation transition.  We could
define, similarly to glasses, an equilibrium gel as a low-density
state when the percolating network lifetime is larger than $10^2
s$. Of course, if one waits long enough time, i.e. more than this
established minimal lifetime of a percolating network, the system will
possibly still restructure itself, due to bond rearrangements. Hence,
strictly speaking, a true ideal gel transition should only take place
at infinite network lifetime.  When the bond lifetime is governed by
an Arrhenius dependence on the attraction strength, the ideal gel
state would arise at infinite attraction strength (vanishing $T$ for
temperature-activated bonds). In the following we will refer to
equilibrium `gel' states as those approached continuously from the 
fluid phase and exhibiting a long (even if not infinite) lifetime,
retaining the `ideal gel' concept only to those extrapolated states
where lifetime becomes infinite. In these respects, percolation is a
necessary pre-requisite (since the infinite spanning network is
present only after percolation is reached) but it is not sufficient for
defining a gel state.

\begin{figure}[h]
\begin{center}
\includegraphics[width=7cm,angle=270.,clip]{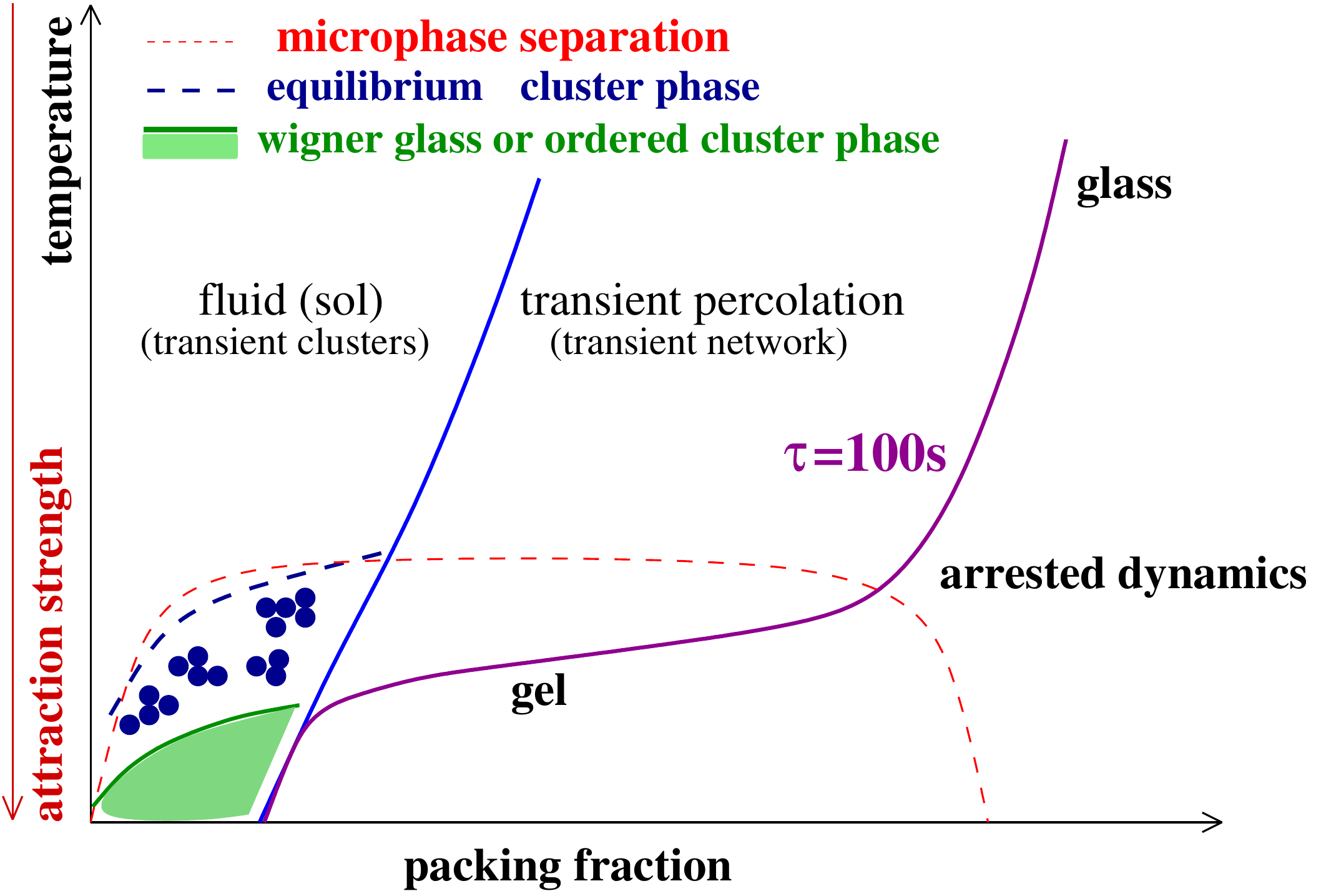}
\end{center}
\caption{Schematic picture of the stabilization of an equilibrium
cluster phase and gel, through the inhibition of the phase separation
region by an enhanced bond lifetime, when additional long-range
repulsion are taken into account. Equilibrium clusters are formed due
to the microphase separation. At low $T$ and low $\phi$ (filled area),
such clusters form either a disordered (Wigner glass) or an
increasingly ordered phase (cluster crystal, columnar phase) depending
on residual cluster-cluster interactions. At low $T$ and larger
$\phi$, gelation results as percolation of the long-lived clusters.}
\label{fig:phase3}
\end{figure}

We can distinguish again two different topological phase diagrams for
equilibrium gelation.
Firstly, in one case the phase separation is pushed towards
higher attraction strength \cite{charbonneau} and can be replaced by
microphase separation. This can be
achieved through an enhancement of the bond lifetime, as for example
by considering various sorts of stabilizing barrier in the potential
with\cite{hurtado} or
without\cite{Pue02a,Pue03aPRE,Zac03a,Voi04a,Zac04b} a clear
microscopic interpretation. A similar effect can be obtained when
considering the effects of residual charges onto colloidal particles
(or proteins) in suspension, which give rise to an additional
long-range repulsion in the effective interaction potential.  In this
case, the condition $\gamma <0$ in Hill's terms\cite{hill} can be
realized through the addition of a sufficiently long-ranged repulsion.
Hence, a finite optimal size $N^*$ of clusters exists which minimizes
the free energy (microphase separation), generating a so-called
equilibrium cluster phase\cite{Gro03a,Sci04a,Strad04}. This behaviour
will be discussed in details in
subsection \ref{sec:repulsive}. For the present description, such a
modification of the potential opens up a window of stability for the
equilibrium gel by pushing at larger attraction strengths the phase
separation. In the micro-phase separating region, at low density,
equilibrium clusters are found, merging somehow into a percolating
network at larger densities. A qualitative picture is proposed in
Fig.~\ref{fig:phase3}, where the $\tau=100s$-line signals the slow
dynamics, connecting the gel and the (attractive) glass line at higher
densities. The only case where a similar phase diagram has been
discussed for a wide range of densities, encompassing both gel and
glass states, is found in the works of Puertas {\it et
al}\cite{Pue03aPRE,Pue04aJPCB}.  Although the authors claim down the
role of the repulsive barrier which is just employed ad-hoc to prevent
phase separation, they find evidence of a gel phase at an intermediate
packing fraction $\approx 0.40$ which, by MCT analysis, is compatible
with attractive glass features\cite{Henrich07,puertas2007}.  Finally, we note
that, if $\xi$ is sufficiently long, the phase separation can be
completely absent (as in the limit of unscreened Coulomb repulsion),
so that at very low $\phi$, below the percolation threshold, and very
low $T$, a Wigner glass of clusters is expected\cite{Sci04a}.

\begin{figure}[h]
\begin{center}
\includegraphics[width=7cm,angle=270.,clip]{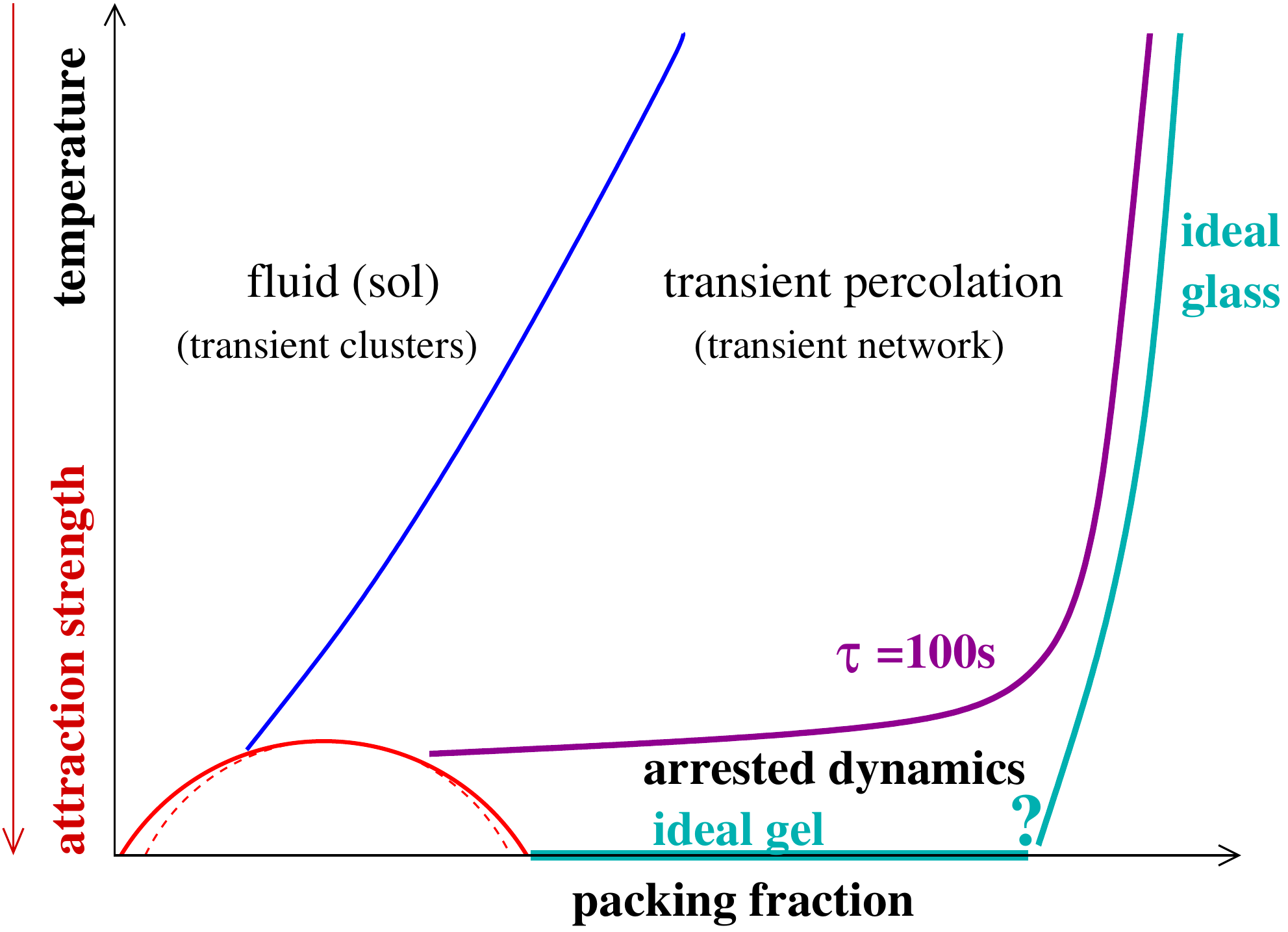}
\end{center}
\caption{Schematic picture of the shift to lower packing fractions of
the phase separation region and emergence of an equilibrium gel phase,
as well as of the ideal gel phase at $T=0$. The question mark refers
to the unknown details of the crossover from gel-to-glass dynamics.}
\label{fig:phase4}
\end{figure}

The other case that can lead to equilibrium gelation is realized when
a mechanism for inhibition of phase separation not only to lower
temperatures, but most importantly to lower packing fractions, is at
hand. This is achieved by inducing directional interactions between
colloidal particles, preferably of low-coordination. We will see that
lowering the (average) coordination number is the essential condition
to push the critical point to lower and lower packing fraction. In
this case, we can consider that $\gamma\rightarrow 0$ in Hill's
formalism, as at low $T$ the driving force for compact aggregation
becomes very small, since the energy is the same in the interior and
on the surface of a cluster, thus enhancing saturated network
structures.  In this case, a completely new topology of the phase
diagram is found. A wide region of stability of an equilibrium
network, to become a gel at lower and lower $T$, opens up at
low/intermediate densities.  Through a careful equilibration procedure
down to very low $T$, almost-ideal gel states may become
accessible. This new topology of the phase diagram and arrest
transitions is sketched in Fig.~\ref{fig:phase4}, where the line of
arrest, again drawn as a $\tau=100s$ line, joins the slow gel states
with the glassy states at large $\phi$, but in truth these two lines
are distinct and the ideal gel and glass lines are reported, with a
question mark about the nature of the crossover/meeting between the
two lines.  We will elucidate this scenario in the framework of patchy
models in subsection \ref{sec:patchy}.

\subsection{Question: is percolation strictly necessary to form a gel?}
We have seen so far that percolation is not a sufficient condition for
physical gelation. However, it should be at least a necessary
condition, if one follows the idea that a gel arises from a stable
percolating network.  Within this picture, attraction should be a
necessary ingredient for gel formation.  However, some systems may
form arrested states at extremely low densities, and their properties
be not at all related to percolation. This happens primarily in
systems with sufficiently long-range repulsion, that in the end acts
as the stabilizing mechanism for arrest.  Essentially two classes of
systems that we are aware are found to belong to this
category. Firstly, soft or ultrasoft systems, like star polymers,
micelles and other aggregates where effective interactions between
different objects can be slightly repulsive close to contact,
essentially for entropic reasons. When two of these objects become
close enough that the end-monomers feel the effects of self-avoidance,
these systems become solid.  Secondly, highly charged systems at low
screening conditions that, independently from the presence of a
short-range attraction, feel at longer distances (comparable to the
average distance dictated by number density) a strong repulsion.  Both
these classes of systems can form a low-density non-ergodic disordered
solid, that is governed by repulsive interactions. The prototype
model for such a low-density arrest transition is the Yukawa
potential, which describes both star-polymer like systems and charged
colloids in suspensions.  For charged systems, the arrested state is
usually called a Wigner glass and can be formed by particles (in
purely Yukawa systems)\cite{Lai97aPRE,Bos98aPRL} or by clusters (in
the presence of an additional short-ranged attraction)\cite{Sci04a},
or perhaps by both in different regions of the phase diagram as
recently speculated in Laponite suspensions at low ionic strength
\cite{Bon99aEPL,Ruz04APRL}.  In star-polymer and soft micellar
systems, the arrest transition is described in the literature as a gel
or jamming or glass
transition\cite{Kap00PRL,Sti02aPRL,Nicolai04micelles,Laurati05PRL,Zacca2005}
and it can be theoretically interpreted both in an effective
hard-sphere picture\cite{Fof03aPRL} and in a purely Yukawa
treatment\cite{unpub}.  The question that naturally arises is: should
these states be considered gels or glasses in general terms?  It is
certainly, once again, a matter of definition how to interpret the
arrest, so that the resulting arrested state is often named gel
without discrimination whether its origin is purely network formation
or not. This happens primarily because it is sometimes hard to call
glass a solid observed at, for example, a packing fraction of few
percent, where particles are very far from each other. We may propose
that a gel should necessarily have attraction as the leading mechanism
for gelation, while a glass can be driven either by repulsion
(hard-sphere or Wigner glass), or by attraction just in the high
density region (attractive glass).

Hence, while in theory and simulations, the knowledge of the governing
interactions would render easy to discriminate a gel from a glass at
low density, in experiments, if the interactions at hand are not clear
as for example in the case of laponite, this can be a hard task.  An
interesting test that could be performed experimentally to provide an
answer to this question could be a sort of `dilution test'.  The
low-density solid could be smoothly diluted (without being altered in
nature) and if persisting, at least for some dilution range,
attraction should be identified as relevant mechanism, thus invoking
for a gel state, while if breaking apart repulsion could be the
responsible mechanism for a Wigner glass state. Of course, care should
be taken that, for example in charged systems, the counterion
concentration is not dramatically affected by dilution in order to
avoid a drastic change in the Debye screening length $\xi$, which governs the
repulsive glass state.

\subsection{Attractive and Repulsive Glass Transition and Mode Coupling Theory}
\label{sec:high}
To correctly locate and interpret the different gel lines, we
need to clarify the high density behaviour for short-ranged attractive
colloids and in particular to address the two glass transitions
arising in these systems: repulsive and attractive glasses.  This
issue has been recently reviewed by other authors
\cite{Daw02a,advances,Cip05a} and, to avoid redundancy, we report here
only a brief summary of the main findings.

The canonical model for glass transition in colloids is the hard
sphere (HS) model, realized experimentally with PMMA particles in an
appropriately index-matched organic solvent
(toluene+cisdecaline)\cite{Pus87a,vanM93,Wil01a}.  Its study allowed
the first direct comparison between MCT\cite{goetze} of the ideal
glass transition and experiments.  MCT provides equations of motion for the
dynamical evolution of the (normalized) density autocorrelation functions,
\begin{equation}
F_q(t)=\frac{\langle\rho^*_{q}(0)\rho_q(t)\rangle}{N S(q)}
\end{equation}
where $N$ is the number of particles, $\rho_q(t)=\sum_{j=1}^N
\exp{(i{\bf q}\cdot {\bf r}_j(t))}$ is the Fourier transform of the local
density variable and $S(q)=\langle |\rho_q|^2\rangle/N$ is the static
structure factor.  Despite uncontrolled approximations in its
derivation\cite{goetze,Zac01a}, the theory is capable to predict the
full dynamical behaviour of the system, starting only from the
knowledge of equilibrium properties, such as $S(q)$ and the number
density $\rho=N/V$. For simple pair interaction potentials, the use of
integral equation closures can be used to obtain a good estimate of
$S(q)$. Alternatively, the `exact' $S(q)$ can directly be evaluated
from numerical simulations.  We remind the reader to previous
reviews\cite{goetze,advances} for details of the
equations and predictions of the theory.

Light scattering measurements at different angles directly provide the
same observable $F_q(t)$ to be compared with MCT.  For HS, a
quantitative comparison was carried out by van Megen {\it et
al}\cite{vanM93} for different values of the packing fraction
$\phi=\pi \rho \sigma^3/6$, with $\sigma$ being the diameter of the
particles, and of the scattering vector $q$. Taking into account a
shift of the glass transition point --- roughly $\approx 0.58$ in the
experiments, while it is underestimated by $10\%$ within MCT --- they
found a strikingly similar behaviour between theory and experiments
and were able to verify the main predictions of MCT.  Avoiding
crystallization thanks to the intrinsic polydispersity of colloidal
particles, the HS glass transition is approached upon
super-compressing the system, being the packing fraction $\phi$ the
only control parameter. Hence, a typical two-step relaxation in
$F_q(t)$ develops with increasing $\phi$. An initial microscopic
relaxation, corresponding to the vibrations of particles around its
initial configuration, is followed by a plateau which becomes longer
and longer upon increasing $\phi$. The presence of a plateau indicates
that particles are trapped in cages formed by their nearest
neighbours. The height of the plateau, coinciding with the long-time
limit of $F_q(t)$, is defined as the non-ergodicity parameter
$f_q$. When the particle is capable of breaking such a cage and escape
from its initial configuration, ergodicity is restored and a final
relaxation is observed, named $\alpha$-relaxation. Otherwise,
the system remains trapped in a non-ergodic state, i.e. a glass
(at least on the time-scale of experiments, as said above typically of
$10^2 s$). 
A similar picture emerges from examining the mean squared displacement
(MSD) $\langle r^2(t) \rangle$, which also displays an intermediate
plateau between short-time Brownian diffusion (or ballistic motion for
Newtonian dynamics) and long-time diffusion. The plateau in the MSD
allows to obtain a direct measurement of the cage in which particles
are confined, and for HS glass it is of the order of $10-15\%$ of the
particle diameter.

\begin{figure}
\begin{center}
\includegraphics[width=10cm,angle=270.,clip]{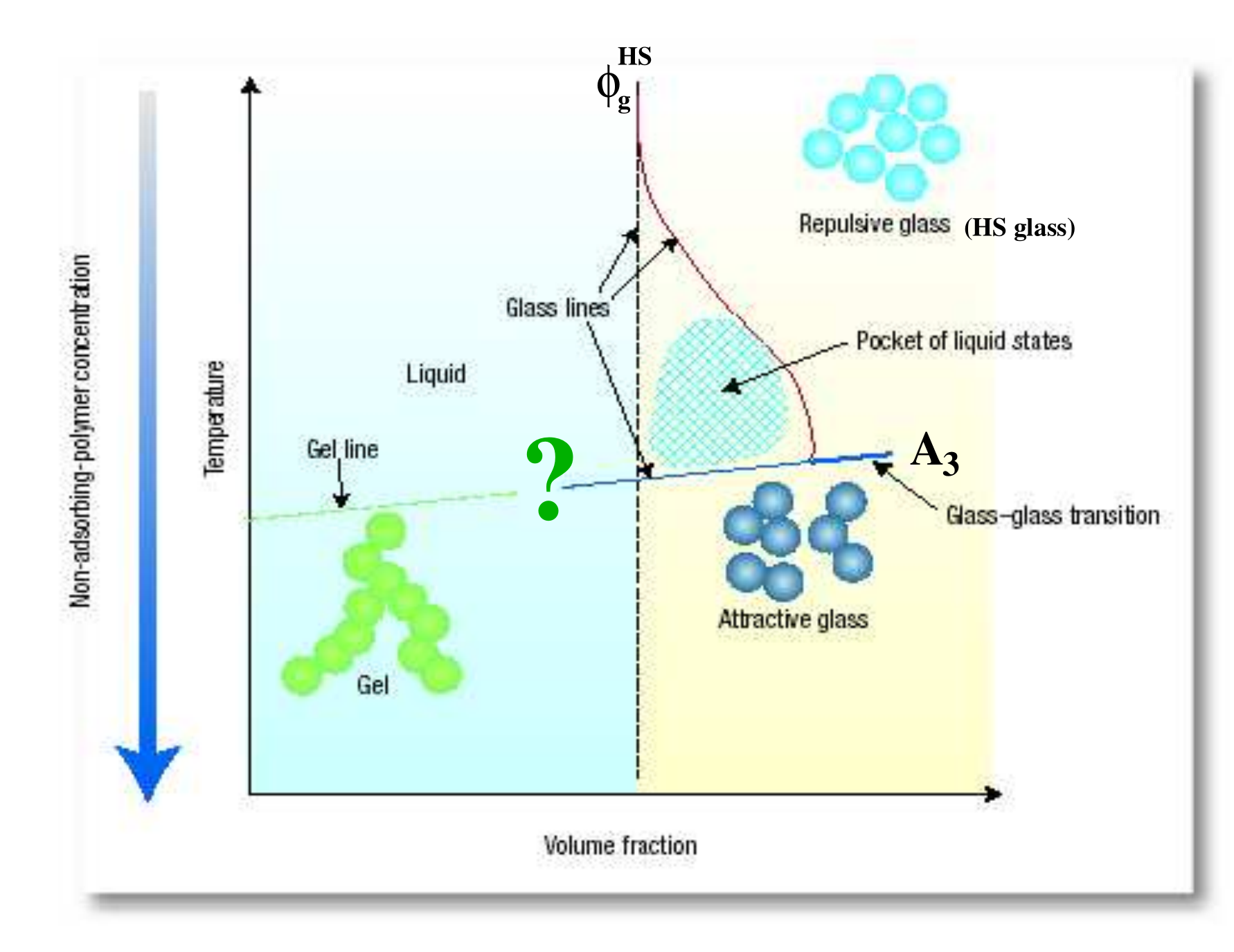}
\end{center}
\caption{Cartoon of the reentrant repulsive and attractive glass
transitions at high density for short-ranged attractive colloids. Adapted with permission from Macmillan Publishers Ltd:  \protect\cite{Sciortino2002b}, copyright 2002. }
\label{fig:natmat}
\end{figure}

These experiments opened up the way for a systematic application of
MCT in colloidal systems.  The next step was to consider the
effect of a short-range attraction complementing the hard-core
repulsion. This type of modification of the interactions can be easily
produced in hard-sphere colloidal suspensions simply by adding
non-adsorbing polymers, thereby inducing an effective attractive force
between the colloids via depletion interactions. This was known since
the pioneering works of Asakura-Oosawa \cite{Asa58a} and Vrij
\cite{Vrij}. It turns out that the width of the attraction $\Delta$ can be
simply controlled by changing the size of the polymers and its
magnitude simply by changing the polymers concentration.
New unexpected features emerged from the study of short-ranged
attractive colloids within MCT\cite{Fab99a,Ber99a,Daw00a}.
These results were found to be independent 
both on the detailed shape of the short-range attractive potential
(SW, hard-core attractive Yukawa, AO etc.), 
as well as of the approximation used to calculate
$S(q)$. They can be summarized as follows and pictorially represented in 
Fig.~\ref{fig:natmat}, redrawn from
\protect\cite{Sciortino2002b}.

At high densities, two distinct glassy phases are identified. Along a
fixed isochore with $\phi > \phi_g^{HS}$, where $\phi_g^{HS}$ is the
HS glass transition threshold, the HS glass is found at high
temperatures, named also repulsive glass.  At low temperatures, a new
glass, named attractive glass, appears. This is induced by the
attractive bonds between the particles. In between these two glasses,
at intermediate temperatures, there is a reentrant pocket of liquid
states, which exists at higher $\phi$ with respect to the HS
glass. The phenomenon at hand is achieved when the range of attraction
is sufficiently smaller than the typical localization length of a HS
glass. In this situation, decreasing the temperature, some particles
will tend to get closer within the attractive range, thus opening up
free volume in the system. In this way, dynamics is speeded up by an
increase of attraction strength. A further decrease of temperature
localizes most particles within the bonds, until they are trapped
within the bond distance. Here, a second glassification process arises
driven by energy, as opposed to the repulsive glass which is driven by
entropy. It is therefore the competition between these two glasses
that determines the reentrance in the glassy phase diagram as well as
an anomalous dynamical behaviour for these
systems\cite{Sciortino2002b,Fre02a}.

Confirmations of the reentrant liquid regime was provided by several
experiments on different
systems\cite{Pha02a,Eck02a,Mal00aPRL,Che03aPRE,Pha04a,Pon03a,Gra04a,Pham06EPL}
and by numerical simulations\cite{Pue02a,Fof02a,Zac02a,Pue03aPRE},
thereby making very robust the MCT predictions for this class of
potentials.  
The two glasses can be differentiated by their respective
non-ergodicity factors and localization
lengths\cite{Pha02a,Zac02a}. The attractive glass is confined by the
short-ranged attractive bonds, implying that $f_q$ is consistently
higher than the HS one at all wave-vectors, and that the MSD plateau
is of order $\Delta^2 << (0.1\sigma)^2$. Moreover, the two glasses are
characterized by utterly different rheological
properties\cite{Zaccamechanical,Pue05a,Nara2006PRL,Nara2006PRE}.
In Fig.~\ref{fig:natmat}, the attractive glass line is
virtually extended to low densities to indicate a possible merging to
the gel line. We will address this point in the routes to gelation
section.

When the two glass lines meet, a discontinuous glass-glass transition
is predicted. It is to be noticed that this is a purely kinetic
transition, given the fact that $S(q)$ are virtually identical at the
transition\cite{Zac04b,Nara2006PRL}. The glass-glass transition line
terminates into a higher order singularity point\cite{goetze} ($A_3$),
beyond which the two glasses become indistinguishable and the
transition is continuous.  There exists a particular state point
$(\phi^*, T^*, \Delta^*)$ for which the higher order singularity point
coincides with the crossing point of the two glass lines. In this
case, the glass-glass line becomes just a single point, and the higher
order singularity is approached from the liquid side, and not buried
within the glassy regime. Associated to such higher order singularity,
MCT predicts a new type of dynamics for the intermediate scattering
function and the MSD\cite{gotzesperl,sperl} that was confirmed in
numerical simulations\cite{Sci03a}. Instead of observing a two-step
relaxation with an intermediate plateau, the relaxation is governed by
a logarithmic behaviour, arising from the competition of the two
glassy states.  Thus, the MSD displays a subdiffusive regime $\propto
t^\alpha$, with $\alpha <1$ being state-point dependent, and $F_q(t)$
can be fitted in terms of a polynomial in $log(t)$.  The influence of
the $A_3$ higher order singularity on the dynamics is also found in
the reentrant liquid region, thereby numerous confirmations of
logarithmic behaviours have been provided in experiments and
simulations\cite{Mal00aPRL,Bar92a,Pue02a}. Finally, when the range of
attraction increases, the two glasses tend to become
identical\cite{Daw00a} as there is no distinction between the bond
(energetic) cage and the geometrical (free-volume) cage. For very
large $\Delta$, attraction tends to stabilize the glass to lower
densities and the slope of the glass line in the $(\phi,T)$ plane for
large $T$ is opposite to that reported in Fig.~\ref{fig:natmat}.  A
detailed review of the glassy phase diagram and associated dynamics
was already reported in \cite{Cip05a,advances}.

\section{Connecting Chemical to Physical Gelation: 
the Bond Lifetime as a Tunable Parameter}
\label{sec:bond}

To describe physical gelation, models were developed at first by
building on existing knowledge about DLCA and  chemical gelation.  The
reversibility concept was initially introduced  to study
thermoreversible polymer gels \cite{Liu96JCP} or to address the
properties of a reversible DLCA-like process in 2d \cite{Jin96}, where
a different structure of the clusters, e.g. a different fractal
dimension with respect to irreversible formation, was found.

To our knowledge, the first study where the concept of a finite bond
lifetime was introduced, to mimic colloidal gel formation, is due to
Liu and Pandey \cite{Liu97b}. On a simple cubic lattice, the dynamics
of aggregation of functionalized sites was followed under two
different conditions: irreversible aggregation, and reversible
aggregation, where reversibility was modulated by a finite bond
breaking probability $p_b$. The results of such study were limited to
a shift of the gel transition with varying $p_b$, associated to
different scaling properties and exponents.  Building on DLCA-like
models, Gimel {\it et al}\cite{Gim01,Gim02} studied the interplay
between gel formation and phase separation for a 3d lattice model with
MonteCarlo dynamics, where a bond probability $p_b$ is assigned to
neighbouring sites.

More recently, a lattice model was extensively studied by Del Gado and
coworkers \cite{Del03aEL,Del04aPRE} to connect chemical and colloidal
gels by means of a tunable bond lifetime parameter.  They studied
tetrafunctional monomers with a fraction of randomly quenched bonds,
mimicking the irradiation process of a polymer solution that induces
chemical bonds. The bonds are formed with probability $p_b$ and are
permanent in the case of chemical gelation, while they can be broken
with a finite probability in the case of colloidal gelation. Fixing
the bond lifetime to $\tau_B$, bonds are broken with a frequency
$1/\tau_B$ so that a constant number of bonds is always present, in
order to compare dynamics for permanent and transient bonds. In the
analysis of the decay of the density correlation functions, the
authors observe a power-law decay close to percolation for
irreversible bonds, as found in experiments for chemical
gels. However, when $\tau_B$ is finite, a crossover to a standard
glassy dynamics is found, with a typical two-step decay well described
by the MCT Von Schweidler law\cite{goetze}.  A plot of the
$\alpha$-relaxation time for different values of bond lifetimes at
various $\phi$ (see Fig.~2 in \cite{Del03aEL}) reveals quite
strikingly this crossover, which takes place at larger $\phi$
with increasing $\tau_B$.  Very recently, the same authors also
proposed to use this framework to explain the viscosity behaviour with
density of rheological measurements for L64 block copolymer
micelles\cite{Mal05aCM}.

A revisiting of the model by Del Gado {\it et al} in terms of a simple
off-lattice model was proposed by Saika-Voivod {\it et
al}\cite{Voi04a}.  This model consists of a modification of a simple
SW model, adapted to a binary mixture to suppress crystallization at
high densities\cite{Zac02a}, but with the addition of an
infinitesimally thin barrier of arbitrary height $u_h$. Such a model
was first introduced\cite{Zac03a,Zac04b} in the case of infinitely
high barrier, to mimic the irreversible bond formation and study the
effect of hopping in attractive glasses.  An unambiguous advantage of
the model is that thermodynamic and static properties of the system
are strictly the same, either in presence or in absence of the barrier, because
of its zero-measure in phase space. However, the height of the barrier
does have an effect on the dynamics, by setting the timescale of
barrier crossing via the ratio $k_BT/u_h$. Being the equilibrium
states the same with and without the barrier, the system can be
readily equilibrated without the barrier, and then dynamics followed
with barrier, averaging over several initial
configurations\cite{Voi04a}.  MD simulations of this system confirmed
the results of Del Gado {\it et al} \cite{Del03aEL}, but also allowed
for a careful study of the wave-vector dependence of the density
correlators. Saika-Voivod {\it et al} showed that, in the case of
infinite barrier height, the percolation transition generates a
breaking of ergodicity for the system only at $q\rightarrow 0$,
supporting the view that gelation in attractive systems corresponds to
the formation of a network of infinite connectivity
length\cite{Tra04a}.  Indeed, the cluster spanning the system at the
transition is still very tenuous and almost massless (strictly so in
the thermodynamic limit), so that it provides a non-ergodic
confinement only at infinite length scale. Beyond the percolation
transition, since the percolating cluster size $P_{\infty}$ grows
rapidly (as $(p-p_c)^{\beta}$), also the non-ergodic behaviour extends
up to much larger $q$, until all particles are in the largest cluster
and the system becomes highly non-ergodic.
\begin{figure}[h]
\begin{center}
\includegraphics[width=7cm,angle=0.,clip]{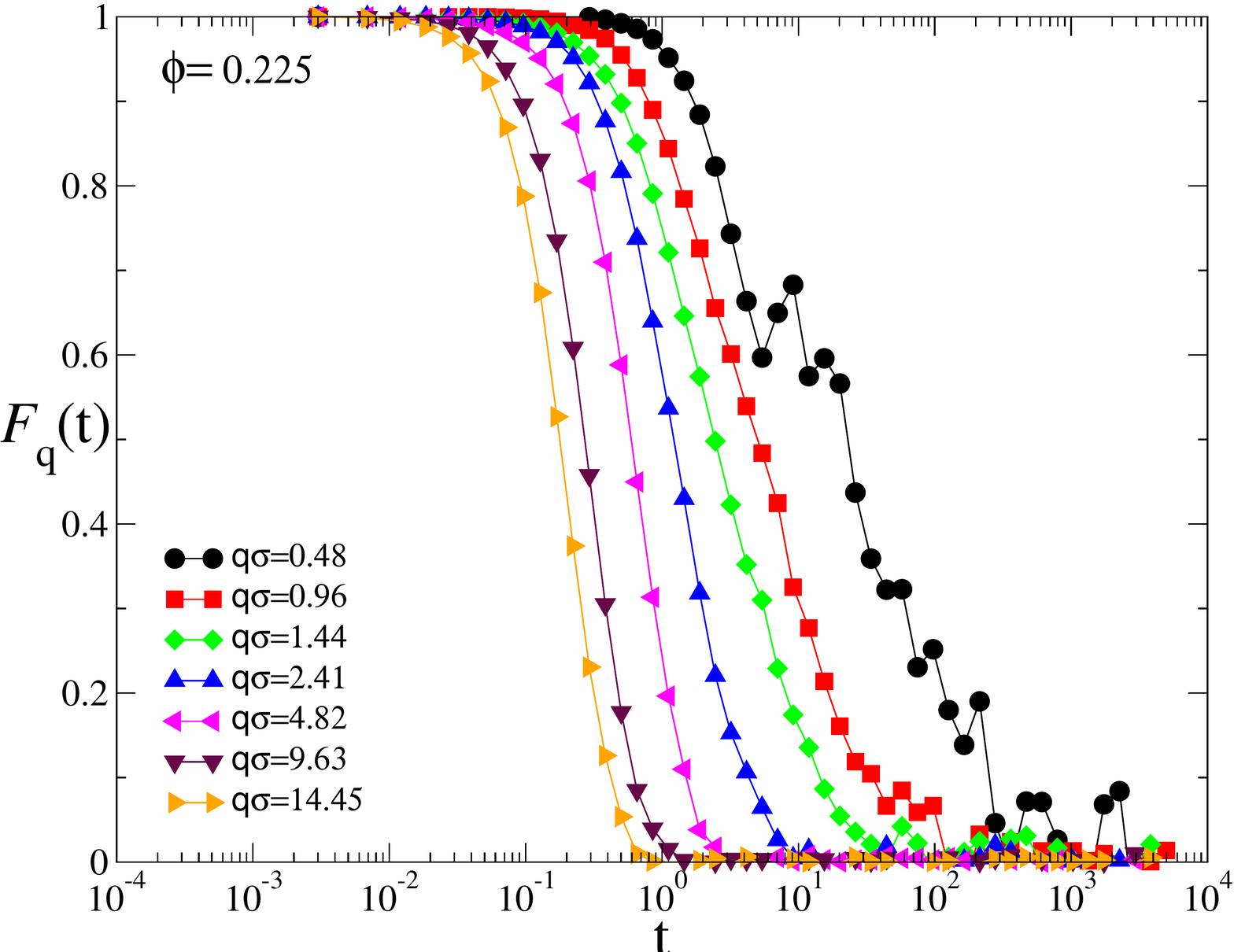}
\includegraphics[width=7cm,angle=0.,clip]{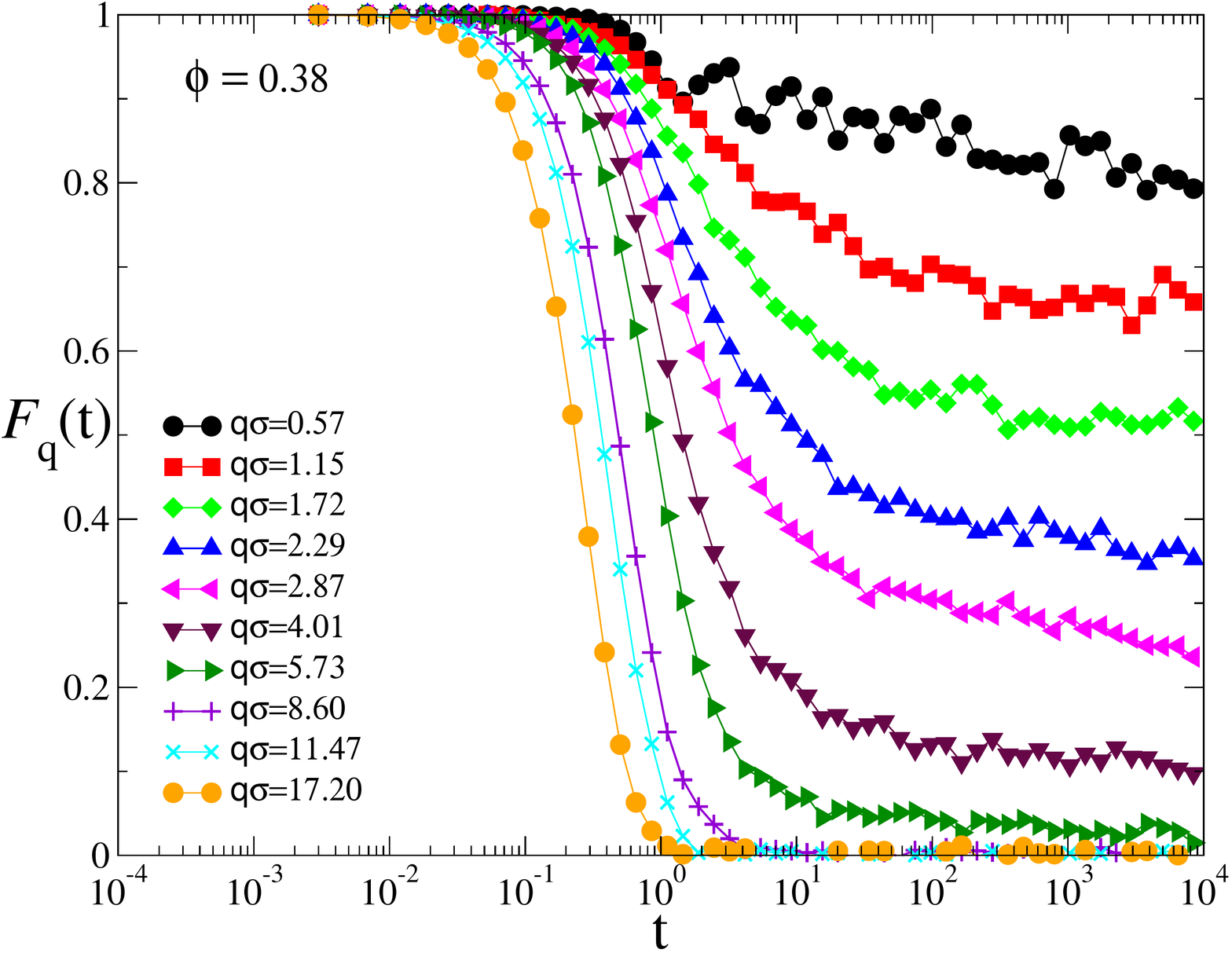}
\end{center}
\caption{Wave-vector dependence of density correlation functions
$F_q(t)$ for chemical gelation at two fixed values of $\phi$: just
below percolation (left) and well within percolation
(right).  $\phi_p=0.23$ for this model. Data taken from
\protect\cite{Voi04a}.}
\label{fig:sqt-barrier}
\end{figure}

\begin{figure}[h]
\begin{center}
\includegraphics[width=7cm,angle=0.,clip]{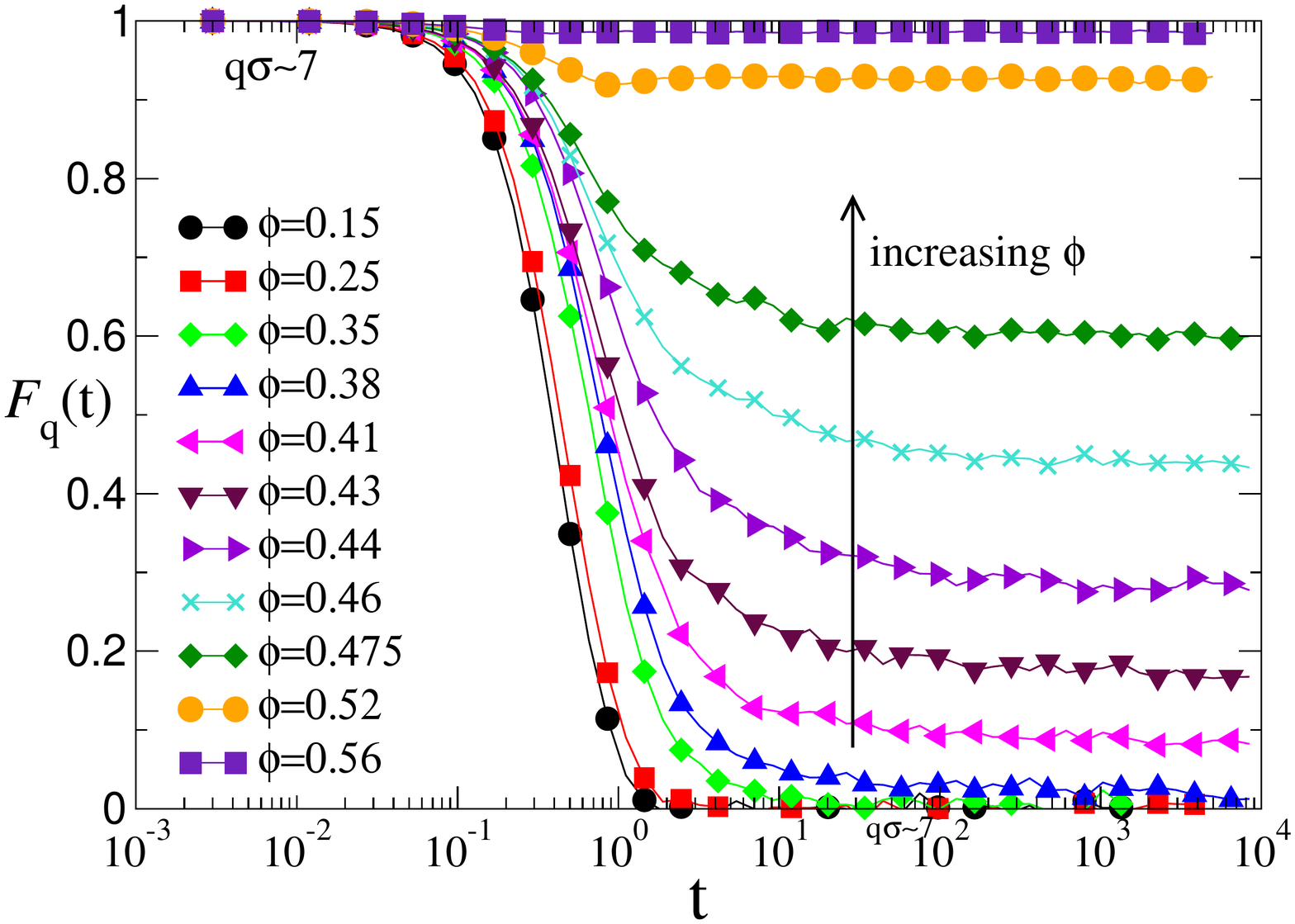}
\includegraphics[width=7cm,angle=0.,clip]{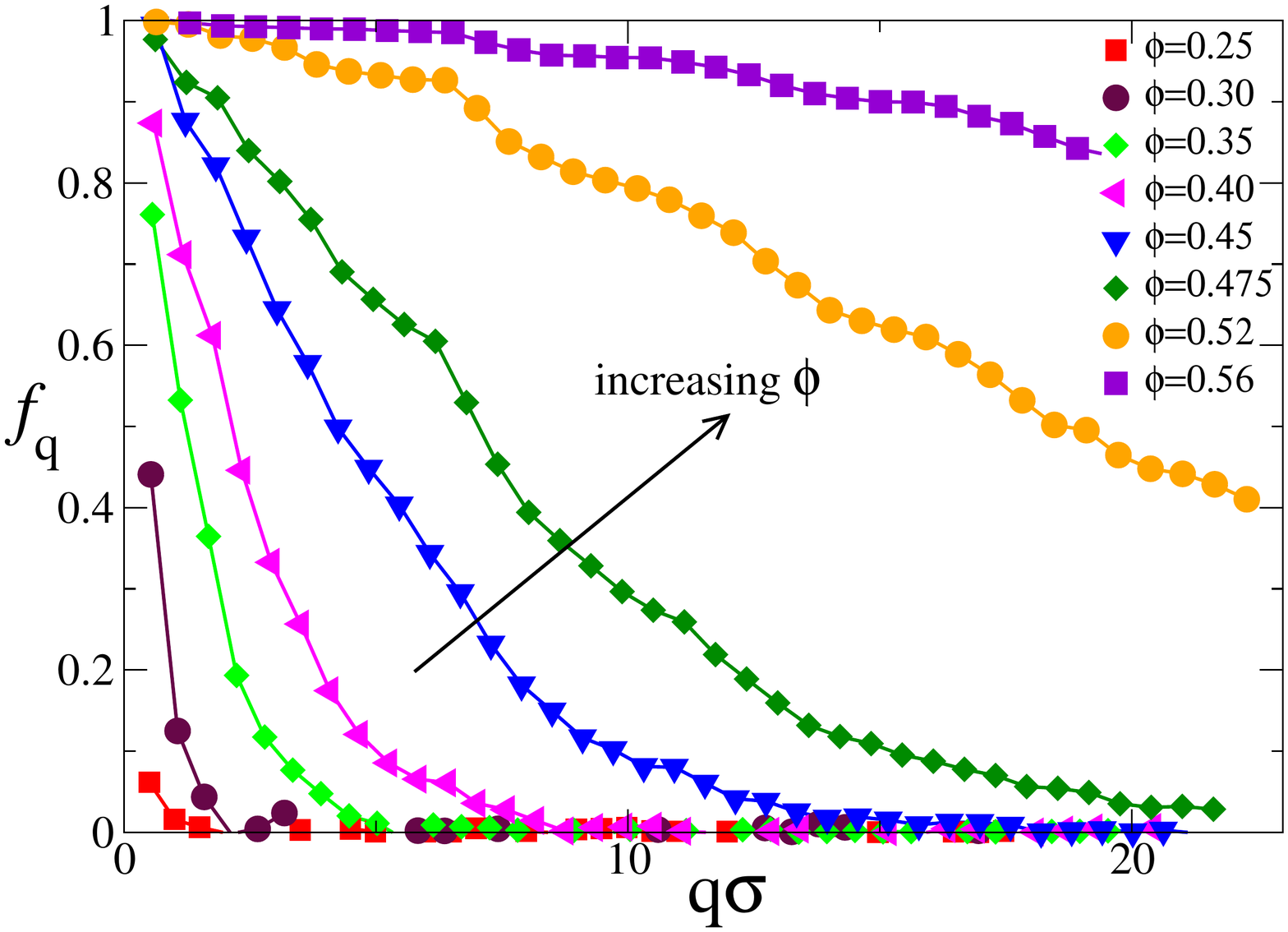}
\end{center}
\caption{$\phi$-dependence of $F_q(t)$ at the nearest-neighbour peak (left) and of the non-ergodicity parameter $f_q$
(right) for chemical gelation. Data taken from \protect\cite{Voi04a}.}
\label{fig:sqt-barrier2}
\end{figure}

To elucidate this important point, that will be frequently invoked in
the rest of the review, we provide in Fig.~\ref{fig:sqt-barrier} and
\ref{fig:sqt-barrier2} a representation of non-ergodic properties as
$\phi$ increases in the case of infinite barrier height.  In the
studied system, the percolation threshold is estimated as
$\phi_p\simeq0.23$. For $\phi < \phi_p $ (left panel in
Fig.~\ref{fig:sqt-barrier}), all studied density correlators
$F_q(t)$ for various wave-vectors, ranging from the smallest
available compatibly with the simulated box size ($q\sigma\approx
0.5$) to a large one where the decay is very fast ($q\sigma\approx
14.5$), decay to zero. However, for $\phi > \phi_p$ (right panel), a
plateau emerges. The observed plateau, and hence the non-ergodicity
parameter $f_q$, is found, at fixed $\phi$, to strongly depend on
$q$. Most importantly, with varying $\phi$ above the percolation
threshold, larger $q$-values are ergodic while small ones are not.
Starting from the smallest calculated $q$-values, which is found to
become non-ergodic just slightly above percolation (within numerical
accuracy), the system further becomes non-ergodic at larger and larger
$q$-values as $\phi$ increases. Fig.~\ref{fig:sqt-barrier2} shows the
$\phi$-dependence at a fixed wave-vector corresponding to the static
structure factor first peak $q\sigma\approx 7$ (left panel), where
a detectable non-ergodic behaviour only occurs much beyond percolation
for $\phi\gtrsim 0.35$. Also, the behaviour of $f_q$ with increasing
$\phi$ (right panel) suggests a crossover from a low-$q$ signal,
detecting the non-ergodic behaviour of just the percolating network,
to a non-ergodic behaviour at all $q$, with a signature that is
similar of that of glasses at large $\phi$. We further note that, at
percolation, $f_q$ seems to become finite in a continuous way,
starting from values close to zero (within numerical accuracy), as
opposite to the case of glasses where a discontinuous transition, also
at the essence of MCT, is found.  It is to be noted that the
$\alpha$-relaxation time at infinite barrier height diverges for each
wavevector at a different packing fraction, coinciding with the
percolation one only at the lowest studied $q$-values. Upon increasing
$q$, the divergence happens when first the $F_q(t)$ shows a finite
plateau. Thus, non-ergodicity is entirely governed by percolation in
the permanent bond case.

As soon as the bond lifetime decreases, the system at first follows
the percolation regime, as long as $\tau_B$ is longer than
$\tau_{\alpha}$, and then crosses over to a standard glassy regime in
full agreement with the lattice model findings of Del Gado {\it et
al}\cite{Del03aEL,Voi04a}.  Approaching the glass transition, all
wavevectors become simultaneously non-ergodic within numerical
resolution. An important aspect of this study is that, by rescaling
time taking into account the different bond lifetimes, all curves
superimpose onto a master curve. This indicates that $\tau_B$ only affects
the microscopic time scale, after which, when enough time has been
waited to allow bond-breaking processes, the long-time behaviour (in particular $f_q$) is
independent of the microscopic dynamics.

\section{Routes to Colloidal Gelation}
\label{sec:routes}
\subsection{ 
(Non-Equilibrium) Gelation as Arrested Phase Separation} 
\label{sec:low}
After discussing the high-density behaviour in subsection
\ref{sec:high}, we now focus on the low-density region of the phase
diagram in short-ranged attractive colloids.  As anticipated in
Fig.~\ref{fig:natmat}, a natural interpretation coming out of MCT
results\cite{Ber99a,Zaccamechanical} and supported by a suitable
comparison with experimental results\cite{Ber03a}, seemed to
corroborate the thesis that a `gel' phase observed in colloid-polymer
mixtures is due to a kinetic arrest created by the bonds, and hence it
would be just a natural extension --- {\it in equilibrium} --- of the
attractive glass to much lower densities.

\begin{figure}[h]
\begin{center}
\includegraphics[width=7.cm,angle=270.,clip]{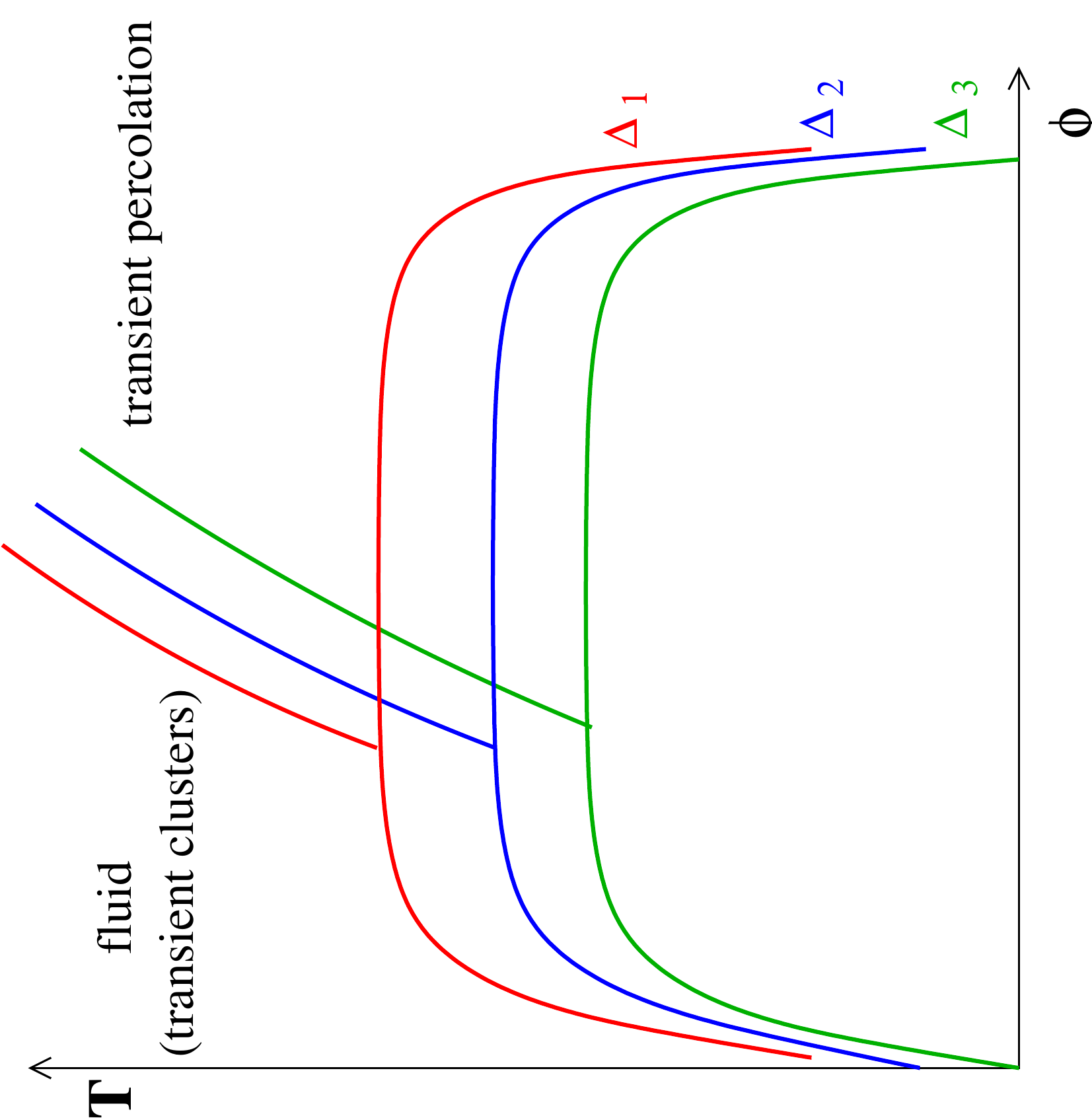}
\includegraphics[width=7.cm,angle=270.,clip]{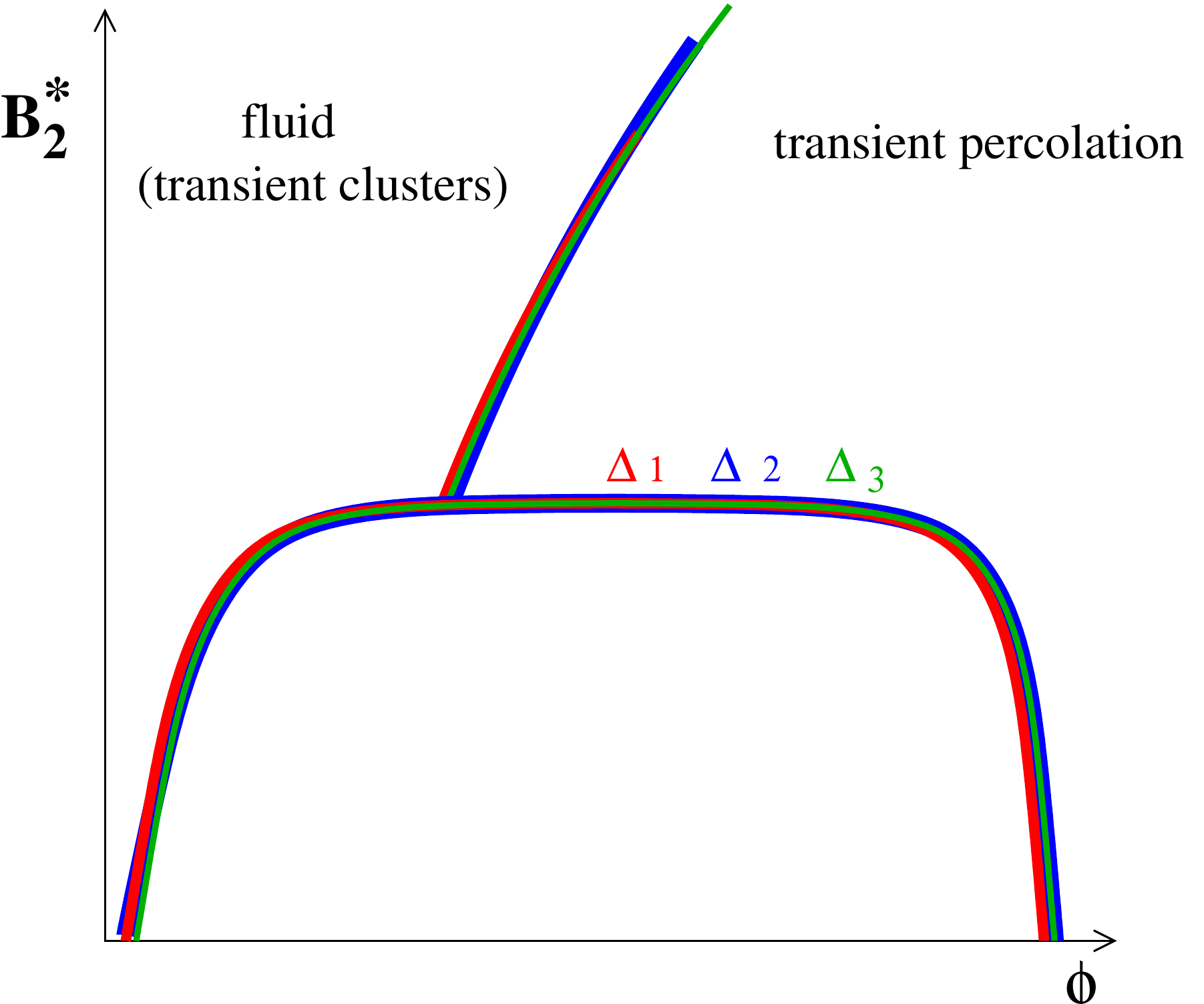}
\end{center}
\caption{Representation of the Noro-Frenkel extended law of
corresponding states for the phase diagram and (transient) percolation
line of short-ranged attractive colloids. Here $\Delta_3 \langle \Delta_2
\langle \Delta_1 \lesssim 0.10\sigma$.}
\label{fig:noro}
\end{figure}

Before discussing in detail the dynamical behaviour of short-ranged
attractive colloids, it is necessary to emphasize some important
thermodynamic features of this type of systems. Being the range of
attraction extremely short, down to a few percent of the particle
diameter, the topology of the equilibrium phase diagram is different
than that of standard atomic liquids. In particular, the gas-liquid
phase separation is metastable with respect to the gas-crystal
transition\cite{Lek92a,Ash96a,And02aNature}. Despite being metastable,
the intrinsic polydispersity of the particles helps in suppressing
crystallization and fluid properties inside the metastable region can
be studied.  A remarkable property of short-ranged attractive colloids
(with interaction range smaller than a few percent of the particle
diameter) is the invariance of thermodynamic properties with respect
to the specific potential shape and to the attractive range $\Delta$
when the normalized second virial coefficient $B_2^*\equiv
B_2/B_2^{HS}$ is used as control parameter. Here
$B_2^{HS}=2\pi\sigma^3/3$ is the second virial coefficient for hard
spheres. This invariance is known as Noro-Frenkel extended law of
corresponding states\cite{Nor00aJCP,Vli00a,Fof06PRE}. It implies that,
if we plot the phase coexistence line in the $(\phi,B_2^*)$ plane for
any short-ranged attractive potential of arbitrary shape and range
within a few percent of the particle diameter, all curves superimpose
onto each other, as sketched in Fig.~\ref{fig:noro}.  Moreover, at
fixed $B_2^*$, all thermodynamic properties such as $S(q)$ are
identical for different shapes of short-ranged attractive models with
small $\Delta$.  Also, the well-known Baxter potential (the limit of
the SW potential for infinitesimal width and infinite depth in such a
way that $B_2$ is finite)\cite{hansen06} scales in the same way.
Hence, the phase diagram of all of these systems can be represented by
the phase diagram of the Baxter model, which has been carefully
evaluated via grand-canonical Montecarlo techniques by Miller and
Frenkel\cite{Mil03a,Mil04a}. 

\begin{figure}[h]
\begin{center}
\includegraphics[width=10.cm,angle=0.,clip]{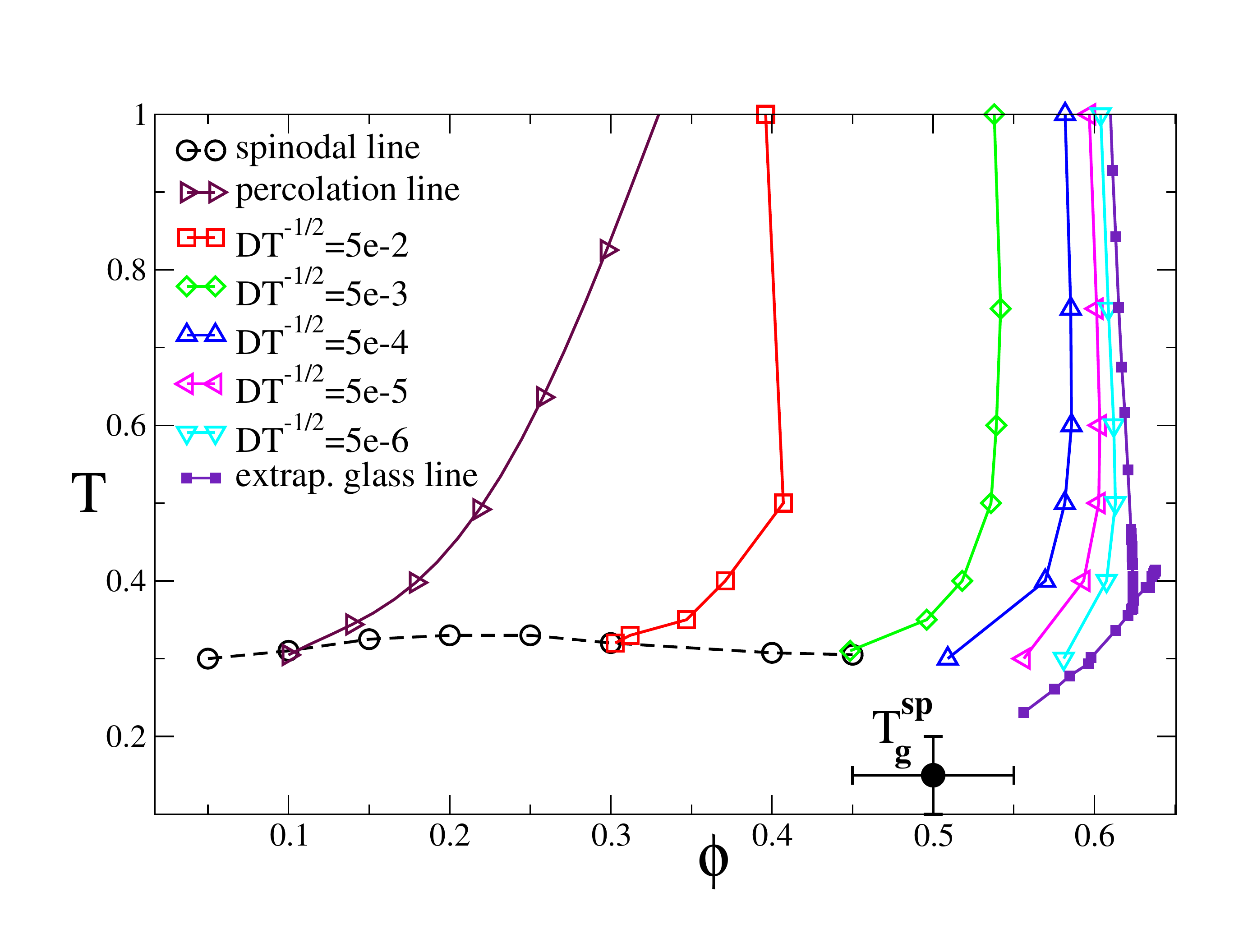}
\end{center}
\caption{Phase diagram from simulations of a $3\%$ SW binary mixture,
reporting percolation, spinodal, iso-diffusivity and extrapolated glass lines. The
latter is calculated through a mapping of the MCT glass lines onto the
points where the diffusivity is found to diverge as a
power-law\protect\cite{Sci03a}. It is found to only meet the spinodal
at $T_g^{sp}$ on the high density side, indicating the absence of an
equilibrium gel phase. Data taken from \protect\cite{zaccapri}.}
\label{fig:capri}
\end{figure}

Numerical simulations for the $3\%$-width SW model \cite{zaccapri}
focused on the dynamics also at low $\phi$. This study reported
iso-diffusivity lines, i.e. lines where the normalized diffusion
coefficient $DT^{-1/2}$ is constant, in the whole phase diagram, and
showed that no sign of dynamical arrest was present for the system
above the two-phase region at low $\phi$, as shown in the phase
diagram of Fig.~\ref{fig:capri}.  The same study showed that quenches
inside the spinodal region generate a phase separation into a gas and
an arrested dense phase. Indeed $S(q)$ is initially found to follow
the typical coarsening pattern: a growing low-$q$ peak and a peak
position moving towards lower and lower $q$-values with time. At some
point, the coarsening process stops within the observation time-window
and the structure does not evolve any further. This scenario of an
arrested phase separation was observed if the quench was performed at
a temperature below the intersection $T_{g}^{sp}$ between the spinodal
and the extrapolated glass line, also shown in Fig.~\ref{fig:capri}.
Hence the origin of arrest can be traced back to an attractive glass
transition of the denser phase. On the other hand, if the quench is
limited to $T>T_{g}^{sp}$, the system would undergo a standard phase
separation into a gas and a liquid phase\cite{zaccapri,Fof04aJPCM}.

Many previous studies of quenches inside the phase-separating region
in connection to gelation, or more generally to dynamical arrest, had
been performed, both experimentally \cite{Ver97aPhysA,Pou99aEPJB}
especially in the sticky (DLCA-like) limit, and in simulations
\cite{sciortinoyoung,JackleEPL,Onu99a}. However, it was not clear the
relation of the thermodynamic line with the glass line.  For the
$3\%$-SW case, a careful mapping \cite{Sci03a} between the simulation
data and MCT allows to have a robust estimate of the glass line
location, and hence to establish firmly for this model that arrest at
low density can only result, via a non-equilibrium route, from an
`arrested (or interrupted) phase separation' process, since the
attractive glass line crosses the spinodal line on the right hand side
of the critical point\cite{zaccapri}.

To investigate the question
whether the dependence on the width of attractive potential could
affect this picture, and allow for the existence of an equilibrium gel
state above the coexistence curve, Foffi {\it et al.}  studied the
dependence of dynamics on the attraction range for a SW binary mixture, 
down to the Baxter
limit. Ref.~\cite{Fof05a} showed that the well-width dependence of the
dynamics is also controlled by $B_2^*$.  Comparing the self-diffusion
coefficient and the bond relaxation time at the same $B_2^*$ values,
universal scaling curves were found.  Hence, within this picture, no
equilibrium gel can exist above the spinodal region.
This study suggests that the Noro-Frenkel scaling of thermodynamic
properties can be pushed forward to invoke also a scaling of dynamical
properties for short-ranged attractive colloids, up to the studied
$\phi =0.40$. However, this is a point deserving further
investigation, since (i) a rescaling with $B_2^*$ of the MCT glass
lines themselves for different well widths does not
hold\cite{Ber99a,unpub} and (ii) the study in \cite{Fof05a} was
limited to $\phi \leq 0.40$, whereas for larger packing fraction
values many-body terms could become important and alter (but only at
high $\phi$) the simple $B_2^*$ scaling.  Subsequent work by the same
authors focused on quenches inside the phase separation region for the
extremely narrow case of $0.5\%$ SW model\cite{Fof05b}.  This work
confirmed the arrested phase separation route for a quench below
$T_{g}^{sp}$, reinforced by a comparison between Molecular and
Brownian dynamics. Again, a percolating network (generated during the
phase separation process) is necessary for providing elasticity to the
final structure.  At too low $\phi$ the system forms compact clusters
which continue to coarsen.  These objects share some similarities with
the so-called `sticky gel beads' discussed within Cluster MCT
(CMCT)\cite{Kro04aPRL} and observed in experiments of salt-rich
solutions of lysozyme\cite{Sedg05}.

Cates {\it et al} \cite{Cat04aJPCM} proposed a framework for
interpreting possible quench paths inside the binodal/spinodal region
for short-ranged attractive colloids (see Fig.~2 in \cite{Cat04aJPCM}).
They suggest the possibility of two different routes to gel
formation, according to the rate of the quench. Their first route,
which coincides with what we have called above arrested phase
separation, is active when the system is quenched slowly and phase
separation has time to develop. Hence, arrest is observed if the quench is
below $T_{g}^{sp}$ and if a percolating structure of the dense phase
is generated.  If the average density is such that phase separation
does not produce a percolating structure, the system will form sticky
beads, i.e. unconnected pieces of an attractive glass. Glass beads are
expected to be internally frozen, because of the extremely low
bond-breaking probability, but also freely diffusing in suspension. In
practice, beads are prevented to form a unique aggregate by the
extremely long time-scales involved in further coarsening.
The second type of quench will be discussed below in the framework of CMCT.

We also mention recent numerical studies of a SW system undergoing a
Brownian cluster dynamics\cite{Babu06}, in which bonds are rigid,
contrarily the standard algorithms to treat a SW interaction. In this
work, a phase separation was found to arise prior to any slowing down of the
dynamics in agreement with previous SW studies. Similar results were
also obtained for a lattice model for T-shaped molecules in 2d, where
gelation was also interpreted as the continuation of the glass line
into the two-phase region\cite{Whi06a}.

The phase diagram depicted in Fig.~\ref{fig:phase2} summarizes the
results of numerical studies of a spherical short-ranged attractive
potential (hard-core plus attraction). Note that (i) no dynamic arrest
takes place above the spinodal, (ii) dynamic arrest is observed for
quenches below $T_{g}^{sp}$, inside the spinodal region.
We further note that such an arrested phase separation scheme applies
also to the case of longer-range attractive potential. Studies of a
Lennard-Jones potential \cite{Sas00PRL,SasNEW} and of a larger
($15\%$) SW model \cite{Emanuelab}, for which the Noro-Frenkel mapping
does not apply, have also reported that the glass line meets the
two-phase region on the right-hand side of the critical point.

These simulation results are at odds with attempts to interpret
colloidal gelation based on the ideal MCT predictions for short-ranged
attractive potentials.  For all potential shapes --- including SW
model, hard-core attractive Yukawa and AO model--- MCT predicts that
the attractive glass line continues down to very low $\phi$, passing
above the coexistence region, and merging into the spinodal on the
left hand-side of the critical point\cite{Ber99a,Zaccamechanical}.
These theoretical results (which do not account for the known 
difficulty, intrinsic in MCT, of predicting the exact location of $T$
and $\phi$ of the glass line) predict the possibility that a stable
`equilibrium gel' at low $\phi$ exists, without an intervening phase
separation.  When dealing with MCT predictions, it has to be taken
into account the fact that MCT always over-stabilizes the tendency to
glassify. For HS, MCT underestimates the
location of the arrest line by about 10 per cent. In the case of
attractive glasses (and consistently with the case of simple atomic
potentials), the theory overestimates the arrest temperature by 
more than a factor of two.  
Moreover, at low $\phi$, the interplay of dynamic arrest with phase
separation and critical fluctuations is not correctly described by
MCT, which is based on the assumption of a homogeneous fluid undergoing
a glass transition.  Once the theoretical MCT curves are properly
mapped (using for example a bilinear
transformation\cite{sperl,Sci03a}), the MCT attractive glass line is
seen to intersect the spinodal line on the high density branch, as shown in Fig.~\ref{fig:capri},
suggesting that a proper treatment of the theoretical curves is
consistent with the idea that arrest at low $\phi$ can not be
approached in equilibrium.

The attempt to fit MCT attractive glass line predictions with
experimental data of a gel phase was carried out by Bergenholtz {\it
et al}\cite{Ber03a} for colloid-polymer mixtures of size ratio
$\sim 0.08$. While the agreement at large $\phi$ was reasonable, it was
noted by the same authors that at low $\phi$ the theory seemed to
`insist on associating its predictions to observed equilibrium
boundaries, rather than to transient gelation'.  This seems to be a
confirmation of the fact that the gel states are only observed within
the binodal/spinodal region, although no sufficient characterization
of the samples in terms of gravity effects and residual charges was
done in those early experiments. However, the same authors still point
out the possibility that MCT could be more predictive for even shorter
attraction ranges, where the gel line was expected to become more
stable with respect to binodal line. This possibility was supported by
experimental works in the group of Zukoski
\cite{Ram02aJCP,Sha03b}. Indeed, they reported a homogeneous gel
formation (that we would associate to an equilibrium route according
to our classification of Section 2) in semi-quantitative agreement
with the corresponding MCT glass line for the very small size ratio
$0.03$. 
Static scattering experiments at low $q$ on approaching the gel line
and an investigation of the residual charge in such colloidal
suspensions could be important to firmly establish whether these
systems are indeed equilibrium gels and if they belong to the class of
uncharged short-range attractive systems.

An attempt to improve the predictability of MCT at low $\phi$ was
carried out by Kroy {\it et al}, in a generalization of the theory to
consider the effect of cluster formation in the so-called Cluster MCT
(CMCT) \cite{Kro04aPRL}. This approach was carried out to combine
aspects of standard MCT with irreversible aggregation of particles,
embedding the inhomogeneous character arising from cluster formation
into the homogeneous MCT approach.  An ad-hoc renormalization
procedure is adopted, based on the following assumptions: (i) clusters
cannot internally rearrange (although bond between clusters can be
broken and re-formed); (ii) the interplay with phase separation is
neglected (supposedly because the time-scale of coarsening should be
much longer than that of cluster aggregation). CMCT predicts an
MCT-like glass transition of clusters, that is proposed to be the
mechanism at hand when rapid quenches are performed inside the
two-phase region\cite{Kro04aPRL,Cat04aJPCM}. In this scheme, the
system experiences an initial irreversible aggregation into clusters
which would then arrest (double-ergodicity breaking), in a way that
does not really distinguish whether the glass transition of clusters
is due to jamming or to attractive bonds between them.  No
experimental evidence that CMCT actually captures the right physical
ingredients has been reported since, apart from the work of Sedgwick
and coworkers \cite{Sedg05}. In our opinion, in the purely short-range
attractive case, the assumption of neglecting the phase separation
with respect to irreversible cluster aggregation seems just too severe
to be applied, and indeed it is not confirmed by all simulation
studies of uncharged colloids.  We comment that the theory could be
potentially relevant for charged colloids, where the additional
long-range repulsions help in inhibiting the phase separation.
Indeed, the first evidence of cluster formation in colloid-polymer
mixtures evolving toward a cluster glass transition, due to cluster
jamming was reported by Segr\`e {\it et al}\cite{Seg01a}. We will
discuss this phenomenon in section (\ref{sec:repulsive}), since it was
later on established that in such system the colloid-colloid
interaction is affected by charges \cite{Din02aJPCM},
sufficiently to alter the phase diagram and gelation mechanism.

Finally we come back to experimental results addressing the interplay
between gelation and phase separation.  Several works
\cite{Jan86a,Ver94a,Ile95aPRE,Ver97aPhysA,Poo97aPhysA} reported a
clear proximity of the gel boundary with respect to the coexistence
line, as well as detection of gel-like structures inside the spinodal
region, despite no precise estimates of the relative positions of the
arrest and spinodal loci were available. Very recent experiments by
Manley {\it et al} in colloid-polymer mixtures with moderate size
ratios support the arrested phase separation scenario\cite{Man05a}.
Authors combined the experimental observations of the frozen density
fluctuations with an ad-hoc use of MCT, where the static structure
factor was taken directly from the experiments in the glass-like
regime. Although this is not an `orthodox' way to use MCT, and some
adjustments of the parameters into play were employed, they provided
evidence of an ergodicity-breaking of the system after initial
spinodal decomposition.  Therefore, this study seems to reconcile
theory, simulations and experiments supporting the picture of arrest
at low packing fraction in term of an interrupted spinodal
decomposition for (spherically-symmetric) attractive colloids.

\begin{figure}[h]
\begin{center}
\includegraphics[width=7cm,angle=0.,clip]{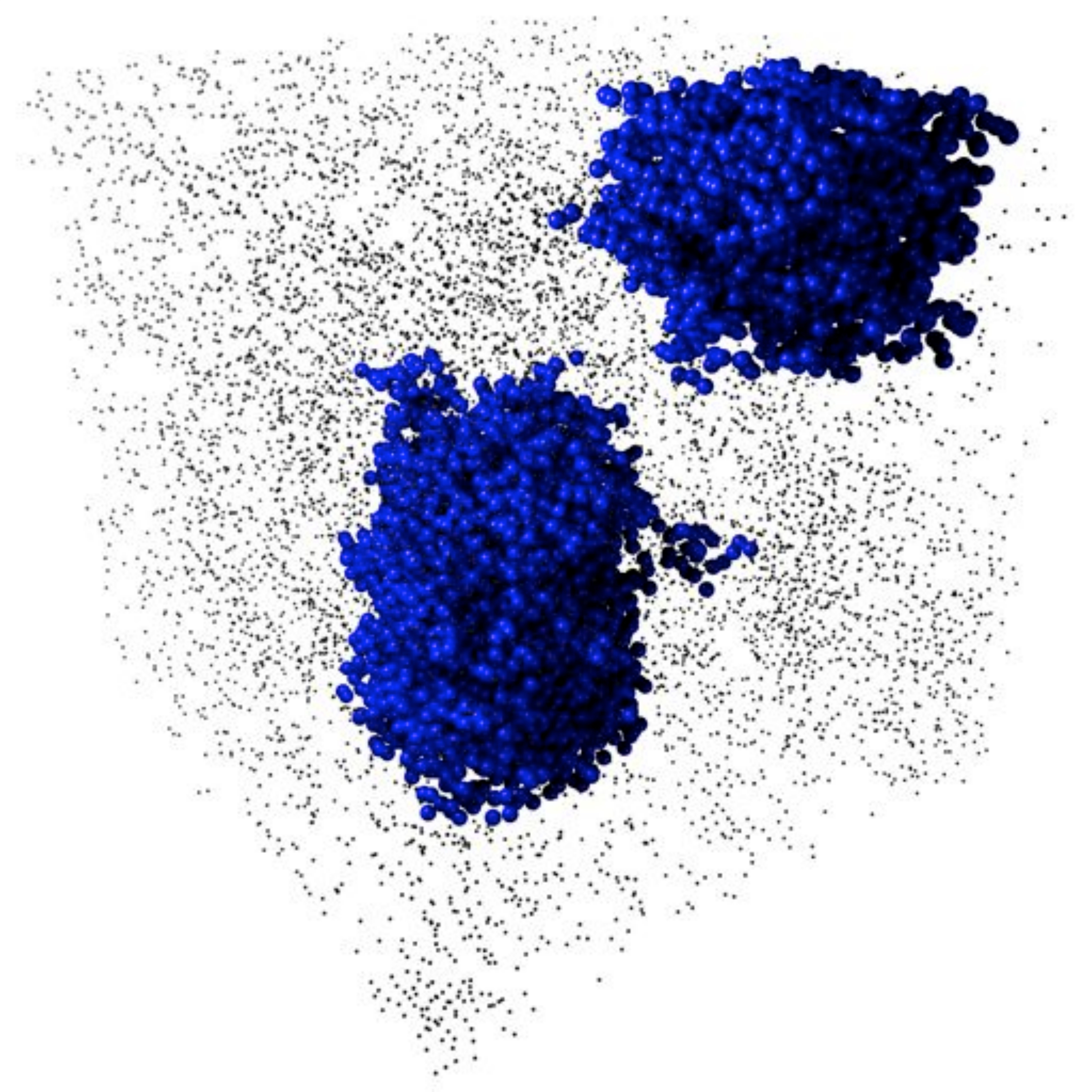}
\includegraphics[width=7cm,angle=0.,clip]{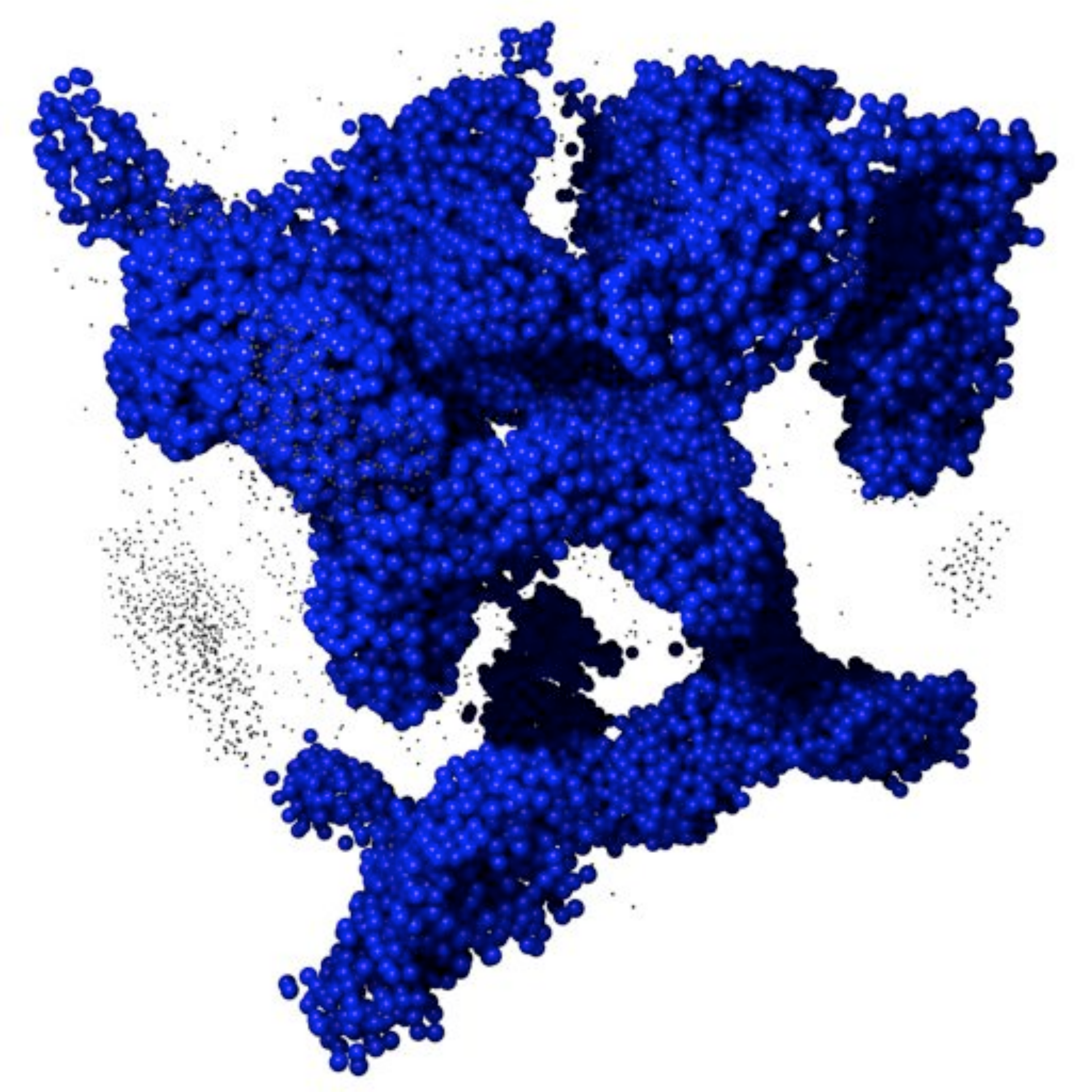}
\end{center}
\caption{Reprinted with permission from  \protect\cite{Lu06a}. Copyright 2006 by the American Physical Society. 3d-reconstruction of clusters (left) and gel image (right) made of PMMA spheres with
added polymers of size ratio $0.11$: the compact structure of the clusters and the thick strands of the gel suggest that it
results from an arrested phase separation after a not too deep
quench.  (For comparison, see Figs.~4 and 5
in \protect\cite{Fof05b} from simulations).}
\label{fig:peterlu}
\end{figure}

In a subsequent study by the same group, Lu {\it et al} \cite{Lu06a}
reported advanced confocal microscopy experiments for different
colloid-polymer size ratios and different $\phi$ in order to have a
complete gel-transition phase diagram in the 3d plane involving
attraction strength, packing fraction and size ratio (see Fig.~4 in
Ref.\cite{Lu06a}).  For $\phi=0.15$ and quite large size ratio
$\approx 0.11-0.15$,  they
observed, upon increasing added polymer concentration, a transition
from a fluid of monomers to a gel, with an intermediate state that
they call `cluster phase'.  The reconstructed 3d structures of the clusters and of the gel
are shown in Fig.~\ref{fig:peterlu}.  The local density of the gel is
rather large while it shows, on larger length-scales, a fractal
dimension of about $2.5$, consistent with random aggregation. The
almost bicontinuous thick pattern of the gel seems to indicate that it
results from an arrested phase separation, although no explanation of
the kinetics of gel formation is offered in the paper.  The so-called
clusters on the other hand are also rather compact spherical objects
of many particles (about one thousand), more indicative of an
incomplete phase separation process (perhaps due to kinetic barriers)
than of an equilibrium cluster phase. Indeed, such clusters seem to be
rather frozen, and they do not exchange particles between
them\cite{Lu06a}. Moreover, the presence of residual monomers in suspension
reinforces the idea of an initial phase separation scenario that is
somehow interrupted. For shorter polymer additives, a similar type of
clusters is also observed, although different in shape and structure:
the fractal dimension decreases down to the DLCA value of about $1.7$,
and the clusters are not spherical at all but rather small (order of
tens of particles) and chain-like.

This experiment raises a fundamental question in the study of
colloid-polymer mixtures, namely the time scale of realistic systems as
compared to the time scale accessed in simulations, especially when a
simple effective one-component theoretical treatment is chosen for
describing the colloid-colloid interactions.  As underlined in a study
regarding the reentrant glass transition at high
densities\cite{Zac04a}, the one-component effective picture is valid
only strictly in the adiabatic limit, when polymers are truly mobile,
due to the kinetic nature of the depletion
interactions\cite{Vli03c}. Increasing polymer concentration, polymers
tend to approach their own overlap concentration, at which point (i)
they will be highly non-ideal, (ii) they will not be so much
mobile. To account for such non-ideal, non-adiabatic features in
simulations is quite challenging, and some sort of multi-step
coarse-graining (similar in spirit to that of Ref.\cite{Pie06a}) could
perhaps be investigated to be able to reach very large $c_p$ close to
the gelation threshold and should be the subject of future studies.
Kinetic effects that polymers may have on the gelation processes and
the role of entropic barriers associated to the different polymer
configurations may turn out in an additional slowing down for further
aggregation that stabilizes (at least on a very long timescale)
smaller aggregates other than full phase separation and may create a gap
between experiments and theory or simulations based on simple
coarse-grained models.  These effects could ultimately lead to an
effective stabilization of bonds between particles, increasing
formally the bond lifetime by more than a simple Arrhenius factor, and
putting experiments in between MCT (dramatically over-estimating the
bond stability at glass formation) and simulations of one-component
effective models (Arrhenius bond lifetime
dependence)\cite{Zac03a,Voi04a}.

What is missing to definitely conclude that simple depletion
attraction leads to an arrested state mediated by a phase-separation
process is an independent accurate estimate of the position of the
experimental gel with respect to the phase separation.  Work is in
progress to elucidate conclusively this aspect\cite{Lu07a}, building
upon the Noro-Frenkel invariance of thermodynamics for short-range
attractive potentials of arbitrary shape\cite{Nor00aJCP}.

Finally we mention a very recent experimental observation
\cite{cardinauxpreprint} of the interplay between phase separation and
attractive glass transition in lysozyme solutions with added salt,
i.e.  in a short-ranged attractive protein solution. The addition of
salt screens the protein-protein electrostatic repulsion and enhances
the liquid-gas phase separation, so that a phase diagram of the same
kind as that reported in Fig.~\ref{fig:phase2} is found. Moreover,
Cardinaux {\it et al} provide an estimate of the attractive glass
boundary inside the spinodal region\cite{cardinauxpreprint}, through a
set of careful centrifugation experiments. They show that, while
proteins undergo a standard liquid-gas separation for $T > T_g^{sp}$,
a coexistence of a denser glass and a gas is found below this
threshold, in agreement with the results found in simulations of
short-ranged attractive colloids\cite{zaccapri,Fof05b}. This
experiment calls for further investigation inside the
spinodal region, as well as for an extension of the effective
protein-protein interaction model \cite{Card06} in the regime where
salt-addition modifies the phase diagram.

\subsection{Gels resulting from competition of attractive and repulsive 
interactions} 
\label{sec:repulsive}
The addition of a long-ranged repulsion to a short-ranged attraction
is a way to act against phase separation, since $\gamma \leq 0$ (see
section\ref{sec:gamma})\cite{Gro03a,Gro04aJPCM,Mos04a}. In this case,
particles prefer to aggregate in clusters of finite size, whose value
depends on the particular thermodynamic conditions.  Various
experimental works reported the existence of an equilibrium cluster
phase in charged (or weakly charged) colloid-polymer mixtures
\cite{Seg01a,Din02aJPCM,Strad04,Sedg04,Bartlett04,Sedg05,Bartlett05,Solomon06},
as well as in globular protein solutions at low ionic strength, such
as lysozyme \cite{Strad04,Bagl04,Strad06,Card06}. This cluster phase
arises from a delicate balance between short-range depletion (or
hydrophobic or van der Waals for proteins) attraction and long-range
screened electrostatic repulsion.  Numerous numerical and theoretical
works\cite{Sci04a,Mos04a,Imp04,Coni04,ChenPRE,Sciobartlett,Tarzia,Liu05,WuCaoPhysA,Candia,Pini06,Pini07,Elshad07}
proceeded in close connection with experiments.

\begin{figure}[h]
\begin{center}
\includegraphics[width=10cm,angle=0.,clip]{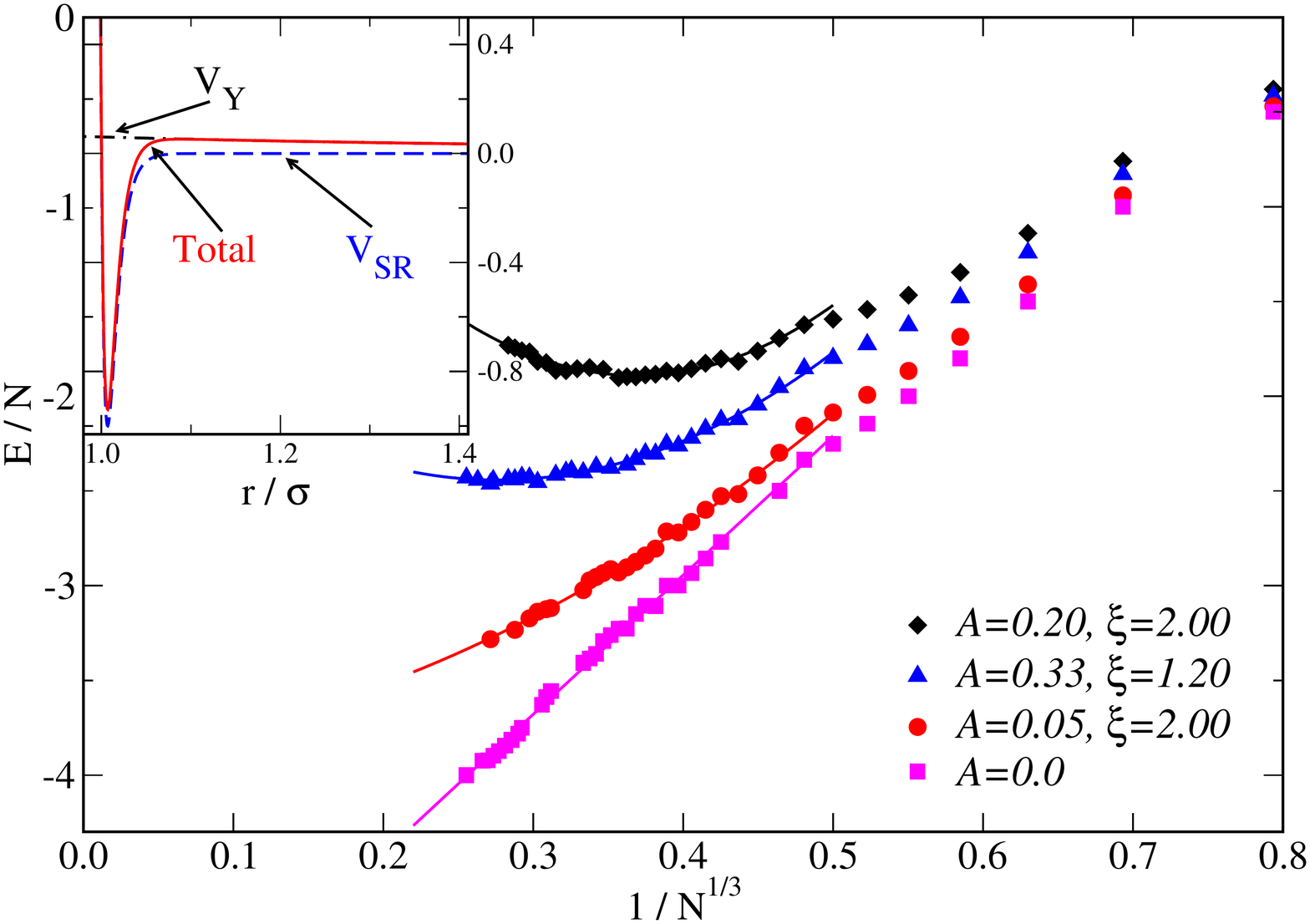}
\end{center}
\caption{Reprinted with permission from  \protect\cite{Sci04a}. Copyright 2004 by the American Physical Society. Ground state energy per particle $E/N$ for clusters of size
$N$ for particles interacting via the total potential of
Eq.\protect\ref{eq:potrep}, reported in the inset, upon variation of
the repulsion parameters $A$ and $\xi$. The $1/N^{1/3}$ behaviour is
recovered in the absence of repulsion, while a minimum appears for
larger and larger repulsions. }
\label{fig:mossa}
\end{figure}

A detailed numerical calculation of optimal cluster sizes, and related
cluster shapes, by means of ground state energy calculations was
performed by Mossa {\it et al}\cite{Mos04a}. In this study, the
short-ranged attraction was modeled for simplicity with a generalized
Lennard-Jones potential\cite{Vli99}, while the long-ranged repulsion
was chosen as a Yukawa potential, to mimic screened electrostatic
interactions. Therefore the total interaction potential is written as,
\begin{equation}
V_{total}(r)=V_{SR}(r)+V_Y(r)=4\epsilon\left[
\left(\frac{\sigma}{r}\right)^{2\alpha}-\left(\frac{\sigma}{r}\right)^{\alpha}
\right]+A \frac{e^{-r/\xi}}{r/\xi}
\label{eq:potrep}
\end{equation}
and it is shown in the inset of Fig.~\ref{fig:mossa}.
Other studies involved closely related potentials, such as
modifications of the DLVO  potential\cite{Imp04,Coni04}.
The ground state energy per particle $E/N$ of a single cluster with
varying size $N$ was calculated for the potential in
Eq.\ref{eq:potrep}, through a basin-hopping
algorithm\cite{wales97,Mos04a} and is reported in Fig.~\ref{fig:mossa}.
In the absence of repulsion, i.e. for $A=0$, the expected behaviour
for large $N$, i.e. $E/N \propto N^{-1/3}$, is recovered, for various
attraction widths ranging from Lennard-Jones to narrow
wells\cite{wales97,Mos04a}.

The addition of a long-range repulsion induces a minimum in the $E/N$
vs. $N$ behaviour, signaling the emergence of an optimal cluster size
$N^*$, whose value depends strongly on the parameters $A$ and
$\xi$. Hence, clusters with size greater than $N^*$ are energetically
disfavoured and, for $N\rightarrow\infty$, the system will prefer to
fractionate into isolated clusters of size $N^*$, rather than undergo
liquid condensation.  In particular cases, i.e. when the condition
$\gamma \simeq 0$ is realized, a clear minimum is not found, but the
energy per particle displays a flat behaviour with increasing
$N$\cite{Mos04a}, in which case the resulting cluster distribution
will be highly polydisperse.  Not only the optimal cluster size, but
also the shape of the clusters is found to be strongly
parameter-dependent. Indeed, clusters can be rather spherical or more
elongated upon variation of the repulsion parameters\cite{Mos04a}:
while spherical structures are retained only when repulsion is rather
weak and attraction is dominant, in general (quasi) one-dimensional
cluster growth is observed\cite{Bartlett04,Sciobartlett,Candia}, in
close analogy also with observations in protein
gels\cite{Ren95a,LeB99a,Pou04a,Manno04}. Such one-dimensional growth is
not typically achieved by particles growing in strings, but rather in
strands, in order to optimize locally the short-range attraction. In
this respect, the ground state of the system is expected to be
either a cluster crystal or a columnar phase at low density (depending
again on the repulsive interactions
parameters)\cite{WuCaoPhysA,Candia}, or lamellar phase at larger
densities\cite{Imp04,Candia}, in agreement with predictions for
unscreened repulsive
interactions~\cite{groussonpre,lowprl,schmalianprl,Wud92a,Sear99,dicastro},
and with those of a mean-field model for Yukawa screened
repulsion\cite{Tarzia}, as well as those of a stability
fluctuation study \cite{WuCaoPhysA}. Moreover, both the latter theoretical
studies agree with the analytical calculation of the cluster energy in
the limit $N\rightarrow\infty$ under the approximation of spherical,
homogeneous clusters\cite{Mos04a}, that the control parameter
$\gamma$, to discriminate between the presence of an equilibrium
cluster phase and that of gas-liquid phase separation, is proportional
to the product $A\xi^4$ for a repulsive Yukawa potential.

\begin{figure}[h]
\begin{center}
\includegraphics[width=7cm,angle=270.,clip]{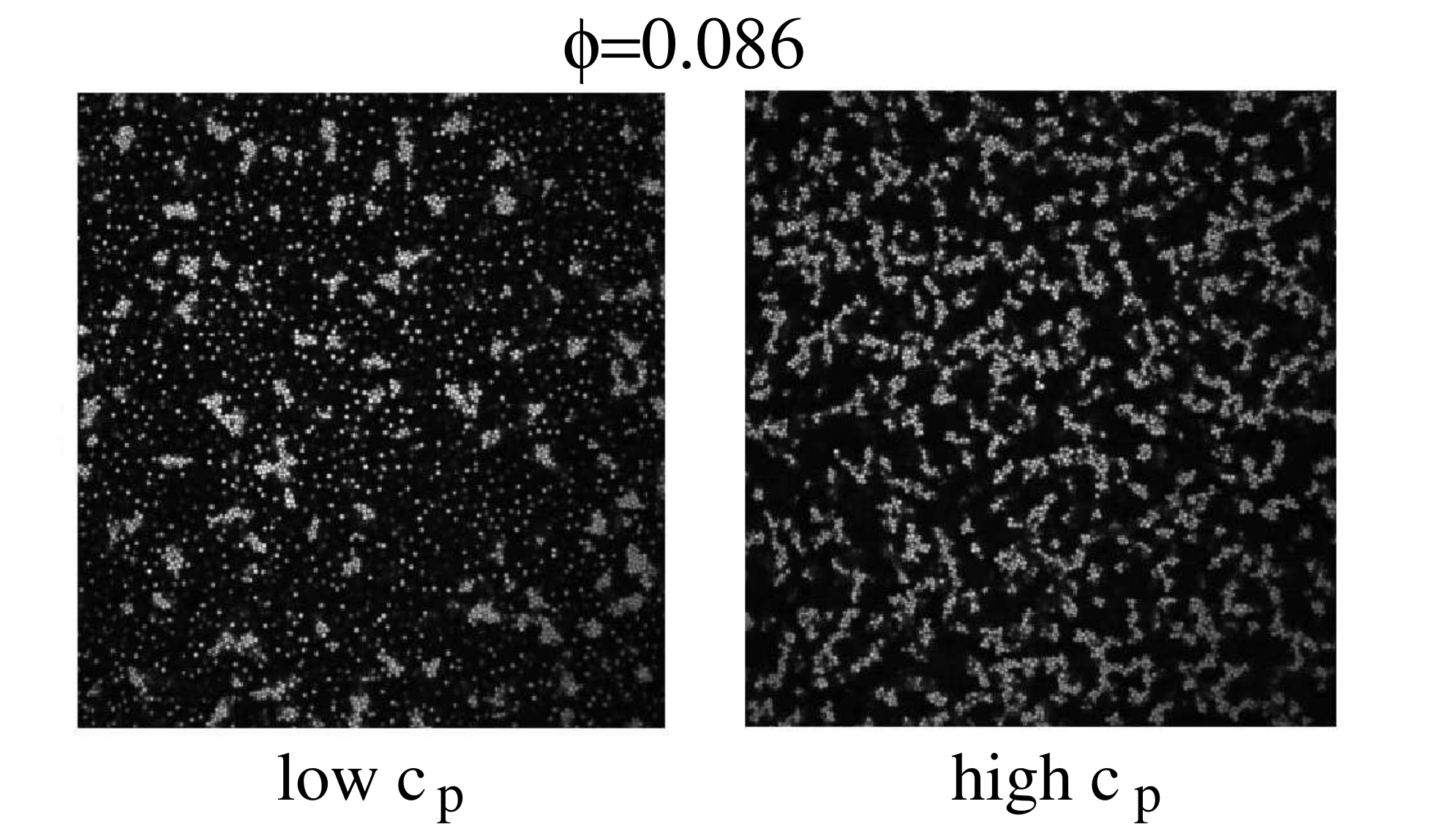}
\end{center}
\caption{Reproduced from \protect\cite{Sedg04}, with permission of IOP
Publishing.  Confocal microscopy images of charged PMMA particles in
suspension with non-adsorbing polymers. On the left panel, a fluid cluster
phase is observed, while a frozen gel network is found for increasing
attraction strength on the right panel.}
\label{fig:confocal}
\end{figure}

At finite $T$, entropic contributions will cause the cluster
distribution to be polydisperse in size, with a growing number of
clusters of the optimal size as attraction gets stronger, giving rise
to a marked peak in cluster distribution\cite{Sci04a}. However, due to
the fact that the arrangement into clusters is driven by strictly
energetic balance, they are dominating structures at low enough $T$.
Hence, clusters, not particles, can act as building blocks of a
dynamically arrested state.  
Several confocal microscopy experiments have revealed the presence of
the clusters leading, with increasing $\phi$, to an arrested state. We
report in Fig.~\ref{fig:confocal} one of those experiments taken from
Sedgwick {\it et al} \cite{Sedg04}, corresponding to charged
density-matched PMMA spheres in solutions with non-adsorbing polymers,
at two values of polymer concentration, respectively in the fluid cluster
region (left panel) and in the frozen network region (right panel).

Before discussing the various aspects related to the arrest transition
of clusters, let us focus on the signatures of a cluster phase in the
static structure of the system.  In the case of short-range attraction
plus long-range repulsion, a dominant length scale deriving from the
balance of attraction and repulsion\cite{Tarzia,WuCaoPhysA} modulates
the structure into periodic patterns\cite{Imp04,Candia}. This is a
general feature of competing interactions of any nature
\cite{Pue03aPRE,Stiak06,Bagl04}. In the static structure factor, a cluster
pre-peak at a finite length appears, clearly a distinctive feature
from the typical increase found in purely attractive systems.  A
number of experimental works have reported this feature. Firstly, the
slightly charged PMMA spheres with added polymers studied by Segr\`e
{\it et al}\cite{Seg01a}, reported the observation of a colloidal
equilibrium cluster phase at low density, although the role of charges
was not fully recognized in their original work. As the authors state
in their manuscript, a peak in $S(q)$ is observed at low $q$ even at
very low $\phi$, where the system does not form a gel but remains
ergodic. 
Indeed, it is important to note that the existence of the pre-peak
directly derives from the interaction potential, hence it is a feature
that is present both in the fluid and in the gel phase. More
specifically, it is found in the ergodic cluster phase as well as in
the presence of a percolating network.

\begin{figure}[h]
\begin{center}
\includegraphics[width=12cm,angle=0.,clip]{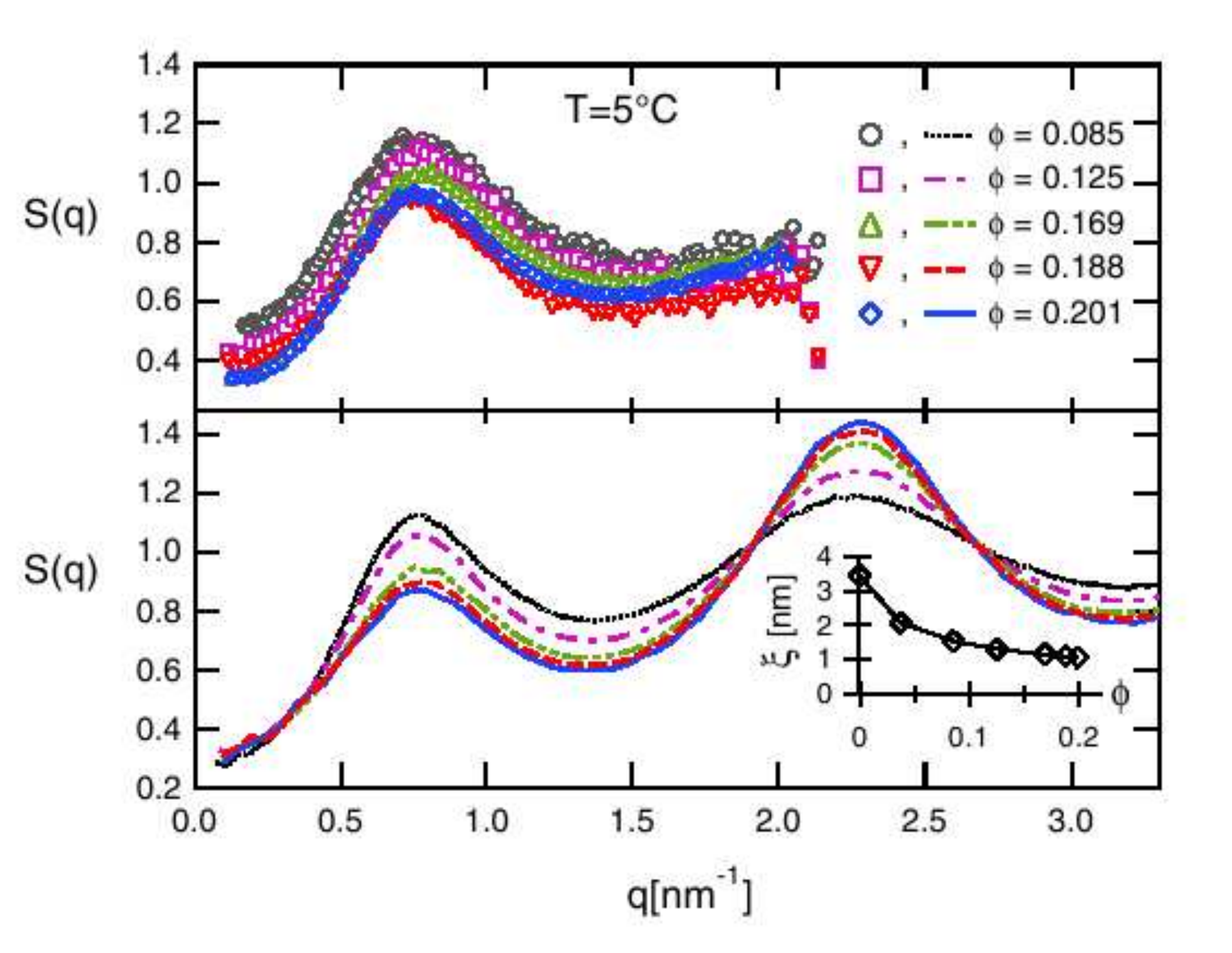}
\end{center}
\caption{Static structure factor from SAXS measurements (top) and
simulations (bottom) for salt-free solutions of lysozyme at various
studied $\phi$ and $T=5C$. A typical cluster-cluster peak, whose
position remains constant in $\phi$, is visible. Inset: the variation
of the screening length $\xi$ in $nm$ (the lysozyme molecule is
treated as a sphere of diameter $3.4 nm$). Reproduced from
\cite{Card06}. }
\label{fig:lyso}
\end{figure}

More recently, emphasis on the cluster phase and on the cluster peak
signature in $S(q)$ has been put forward by Stradner {\it et
al}\cite{Strad04} in solutions of lysozyme under low salt conditions,
as well as again in short-ranged attractive charged PMMA spheres. For
lysozyme solutions, a remarkable invariance of the cluster peak
position $q^*_c$ was reported with $\phi$.  Numerical simulations
\cite{Card06} of a total interaction potential, of the form described
by Eq.\ref{eq:potrep}, have shown that a simple effective model based
on competing interactions can indeed be successful in describing the
clustering phenomenon, even for lysozyme solutions.  A comparison of
$S(q)$ in simulations and experiments, for zero-salt lysozyme
solutions, is provided in Fig.~\ref{fig:lyso} for various studied
protein packing fractions.  Importantly, the long-range repulsive part
of the total potential is fixed by experimental conditions, with the
screening length varying with protein concentration as reported in the
inset, while the attractive depth is chosen via a single-parameter fit
to map experimental measurements. This choice is then kept fixed for
all studied $\phi$\cite{Card06}.

So far, we established that systems with competing short-range
attraction and long-range repulsion form equilibrium clusters and do
not phase separate (at least for large enough $\xi$). Hence, upon
increasing attraction strength, an arrest transition, mediated by the
clusters, not by the particles themselves, occurs.  The nature of the
cluster arrest transition will depend on the residual `effective'
cluster-cluster interactions\cite{Sci04a,Sciobartlett}.  To a first
order approximation, assuming spherical clusters homogeneously packed
and monodisperse in size, it is found that cluster-cluster
interactions have the same screening length as the particle-particle
interactions, with renormalized amplitude\cite{Sci04a}. For large
enough screening lengths, i.e. at least of the order of a particle
diameter, repulsive interactions are dominant between clusters and may
induce a repulsive Wigner glass transition at low density, preventing
the formation of a Wigner crystal of clusters thanks to the intrinsic
clusters polidispersity.  Here, repulsion is the mechanism leading to
the arrest and no percolating network is indeed detected in the
simulations, although showing non-ergodic features\cite{Sci04a}.  The
Wigner glass of clusters was observed in simulations, but also
reinforced by MCT calculations based on the renormalized
cluster-cluster interactions\cite{Sci04a}. Moreover, MCT calculations
for a double Yukawa potential (short-range attraction plus long-range
repulsion) also support this scenario\cite{ChenPRE}.

However, if the screening length is smaller, i.e. roughly of the order
of a particle radius, cluster-cluster interactions are not so relevant
and the short-range attraction for two particles at contact may act as
a glue for the gel formation. In this second case, the gel formation
can be identified with the percolation of the
clusters\cite{Sciobartlett,Coni04}.  It is to be noticed that
the difference between these two cases is very subtle, and strongly
dependent on the details of the potential parameters, but the
mechanism for arrest that results is very different: a `glass' of
repulsive clusters in one case and a branching mechanism of adiacent
particles within clusters in the other. Moreover, different paths to
achieve the low-temperature states, i.e. for example rapid quenches,
can induce a so-called `arrested micro-phase
separation'\cite{charbonneau}, where clusters, despite repulsion being
strong, remain trapped in a metastable percolating gel-like state for
a long time, without equilibrating to the underlying Wigner glass or
crystal, that would be achieved under a slow equilibration
approach\cite{Sci04a}.  We finally comment that, also in the case of
large screening lengths, a crossover from the Wigner glass behaviour
to a percolating gel transition with increasing $\phi$ should be
observed (as also suggested by Fig.~\ref{fig:phase3}).  This crossover
is currently under investigation. Interestingly, a Wigner glass of
clusters has only been predicted by simulations, but no clear experimental
observation has been provided yet, at least for simple spherical colloids.

\begin{figure}[h]
\begin{center}
\includegraphics[width=10cm,angle=270.,clip]{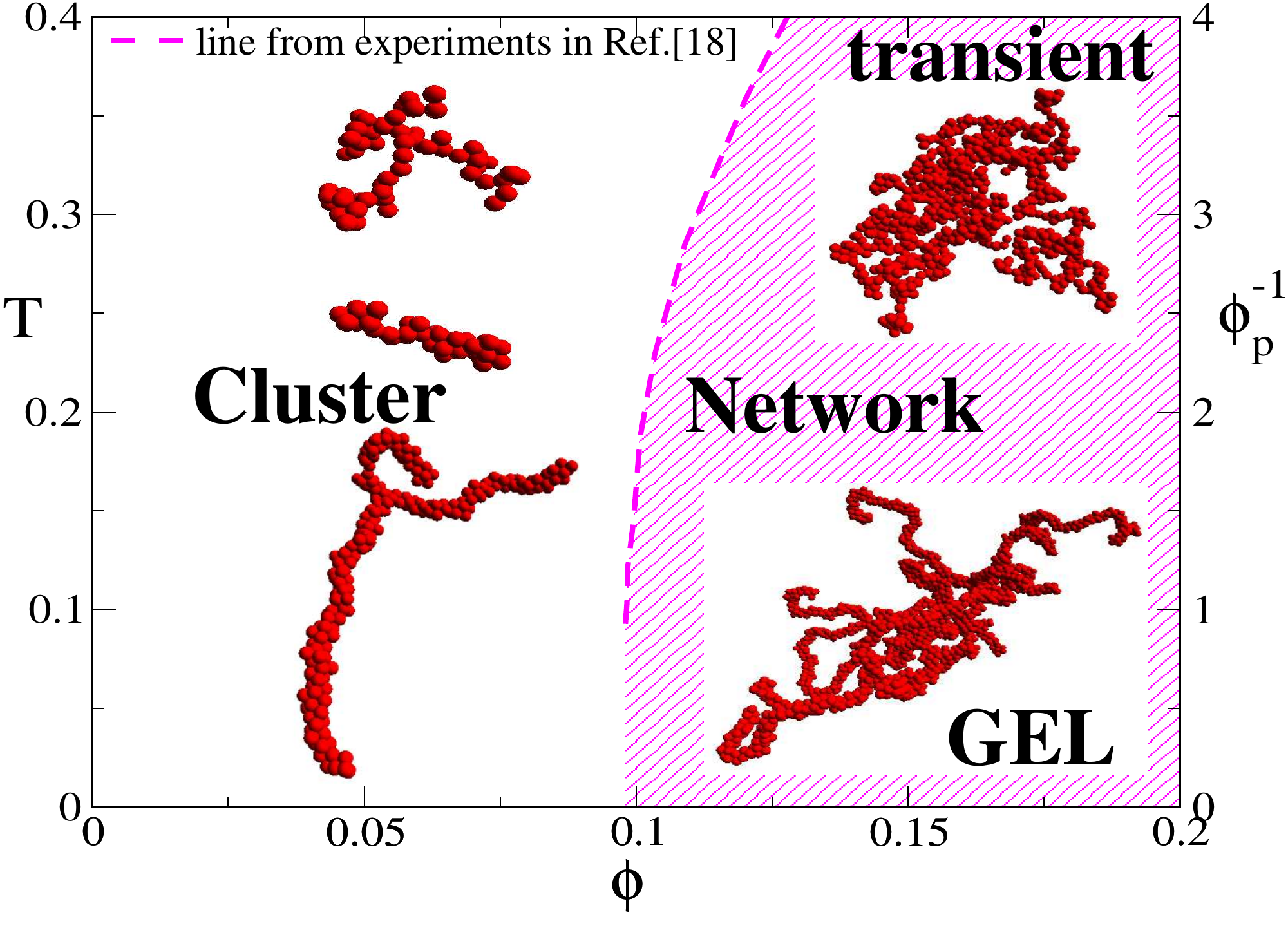}
\end{center}
\caption{Phase diagram for charged PMMA spheres with non-adsorbing
polymers from \protect\cite{Bartlett04}.  On the $x$-axis, colloid
packing fraction $\phi$ is reported, while on the $y$-axis temperature
(from simulations in \protect\cite{Sciobartlett}) and inverse added
polymer packing fraction $\phi_p^{-1}$ are shown (left and right
sides, respectively). The line, taken from the experimental work of
\protect\cite{Bartlett04}, divides the region where a cluster phase is
observed from that of network formation. Cluster and gel snapshots are
taken from simulations in \protect\cite{Sciobartlett}. A transition in
cluster shapes is shown from random transient clustering to long-lived
spirals upon increasing attraction strength. The percolating region in
simulations is in good agreement with that observed in experiments and a
gelation transition of branching spirals is found. }
\label{fig:bartlett}
\end{figure}

\begin{figure}[ht]
\begin{center}
\includegraphics[width=5.cm,angle=270.,clip]{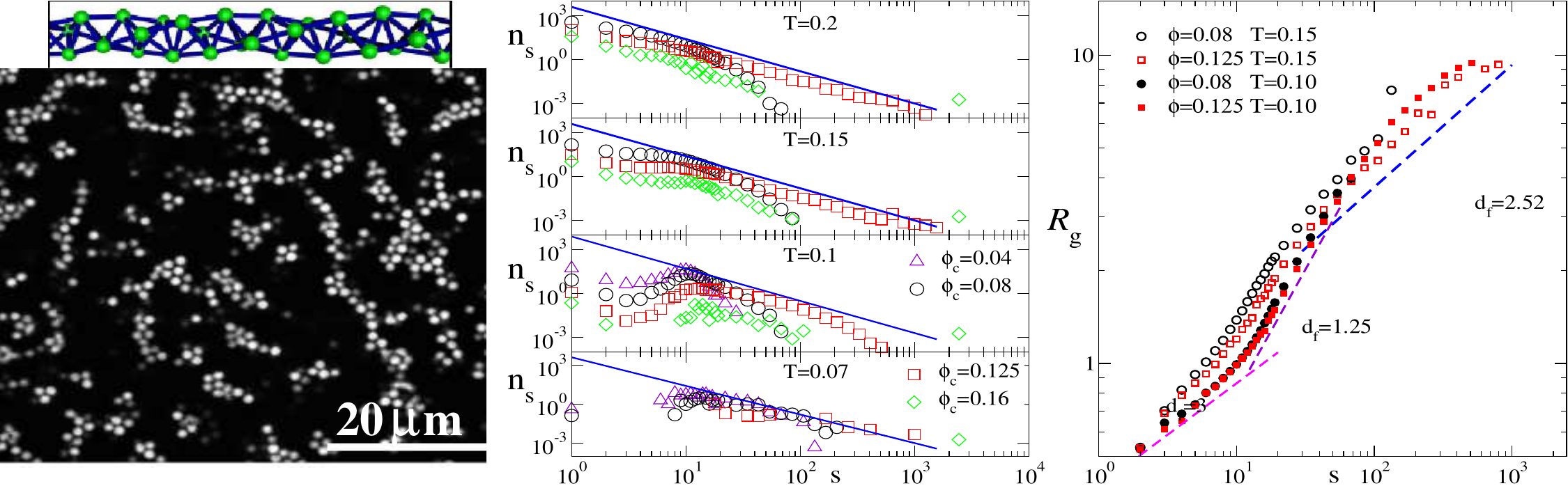}
\end{center}
\caption{Left panel: Reprinted with permission from \protect\cite{Bartlett04}. Copyright 2005 by the American Physical Society. Confocal microscopy image of the network with
spiral-like structure. The
Bernal spiral structure is also drawn. Middle panel: Cluster size
distributions $n(s)$ with varying $T$ and $\phi$ and Right panel:
Radius of gyration $R_g$ at $\phi=0.08$, i.e. below
percolation, and $\phi=0.125$, at percolation for high and low $T$:
$T=0.15$ and $T=0.1$. Middle and left panels  from simulations reproduced with permission from \protect\cite{Sciobartlett}. Copyright 2005 Am. Chem. Soc.
Data show at high $T$ the formation of
(transient) random clusters leads to (transient) random percolation,
while at low $T$ the organization of the system into the
energetically-preferred spirals provides quasi one-dimensional growth
of clusters, approaching a random percolation of spirals. }
\label{fig:bartlett1}
\end{figure}

The case of cluster branching into a gel network was reported in a
confocal microscopy experiment of charged PMMA with added
non-adsorbing polymers by Campbell {\it et al}\cite{Bartlett04}.  These authors
showed the emergence of an equilibrium cluster phase at low $\phi$
merging into a percolating network with increasing $\phi$, with the
observed phase diagram drawn in Fig.~\ref{fig:bartlett}. In
particular, they were able to identify the organization of clusters in
a well defined geometry corresponding to the so-called Bernal
spiral\cite{bernal}, i.e. a two-stranded spiral formed by face-sharing
tetrahedra.  A 2d-reconstruction of a gel slab, with an highlight of
the Bernal spiral structure is found in Fig.~\ref{fig:bartlett1}(left
panel).  Interestingly, ground-state cluster energy calculations had
been already performed for parameters strikingly close to the
experimental ones, i.e. $A\simeq 8$ and $\xi=0.5$ in \cite{Mos04a},
predicting a Bernal spiral ground state (although the specific
structure was not recognized), which with increasing size of the cluster
can only grow linearly along one-dimension.

Brownian Dynamics simulations in bulk conditions were then carried
out\cite{Sciobartlett} with an effective potential aiming to to mimic
precisely the experimental conditions of \cite{Bartlett04,Bartlett05}.
Combined results from experiments and simulations are shown in
Fig.~\ref{fig:bartlett}.  An aggregation into spiral clusters was
detected as well as a branching of spiral clusters giving rise to
non-ergodic behaviour. This is possible due to the fact that the
residual repulsive interactions between spirals are negligible, while
a branching point is a defect of the perfect spiral structure at low,
but non-zero $T$, providing rigidity to the network.  An interesting
aspect pointed out by the simulations is that percolation is observed
both at high and low temperatures, with an intermediate
non-percolating region (reentrant percolation). The cluster size
distributions $n(s)$ for the studied state points are reported in
Fig.~\ref{fig:bartlett1} (middle panel). At high $T$, a typical
transient percolation is found and no gel state is formed but only a
transient network. In this case the fractal dimension of the clusters
is consistent with random percolation. At intermediate temperatures,
the system starts to organize itself into spirals to minimize the
energy, hence a peak in the cluster distribution arises. While the
system rescales its basic units from monomers to spirals (for which
the minimal size is of order $\sim 10$ particles), the effective
packing fraction decreases, so that percolation ceases.  A further
decrease in $T$ allows for percolation of spirals rather than
particles, while monomers become absent. Interestingly, the spirals
have an almost linear fractal dimension,i.e. $d_f\sim 1.25$, while at
percolation the branching mechanism follows again the random universal
exponent, i.e. a random percolation of spirals. This scenario is well
described by the evolution of the radius of gyration $R_g\equiv 1/N
[\sum_{i=1}^N({\bf r}_i-{\bf R}_{CM})^2]^{1/2}$ reported in
Fig.~\ref{fig:bartlett1}(right panel) for two studied values of packing
fractions $\phi=0.08$ (below percolation) and $\phi=0.125$ (at
percolation) at high and low $T$. We further note that, at low $T$,
the cluster structure is independent of $\phi$  for all sizes,
suggesting the approach to the ground state structure. We can
distinguish an initial spherical growth at small $s$ compatible with a
fractal dimension $d_f=3$\cite{Mos04a}, followed by an almost-linear
growth, i.e. $d_f\sim 1.25$ in the size-interval where long spirals
are forming $10\lesssim s\lesssim 100$, and finally a recovery of the
random organization of spirals for large $s$. Beyond this point, the
high and low $T$ results are almost coincident. More details can be
found in \cite{Sciobartlett}.

It is important at this point to focus on the visualization of the
clusters. These are reported together with the phase diagram in
Fig.~\ref{fig:bartlett}. A comparison of the percolating clusters
observed at $\phi=0.125$ respectively at high and low $T$ in the
network region is offered. For $T=0.15$ a random structure, which is
transient in time, is observed, mostly organized into single-lined
chains. On the other hand, for $T=0.07$, a branching mechanism of
spiral strands has taken place to form a gel. Here, no single-line
chains are observed, but the only defects in the spiral structures are
the junction points allowing the establishment of a macroscopic
network.  In between these two behaviours, a shrinking of the largest
cluster accompanied by a reorganization into spirals is observed due
to above mentioned competition between entropic and energetic
effects\cite{Sciobartlett}. This can be visualized in the cluster side
of the diagram, where the reorganization of small clusters at
$\phi=0.08$ for $T=0.15, 0.1, 0.07$ is shown. An almost perfect spiral
is recovered at the lowest $T$.  The detailed spiral structure has
been also confirmed by the study of the rotational invariant
distributions directly compared to theoretical values and experimental
curves\cite{Bartlett04,Sciobartlett}.

\begin{figure}[h]
\begin{center}
\includegraphics[width=12cm,angle=0.,clip]{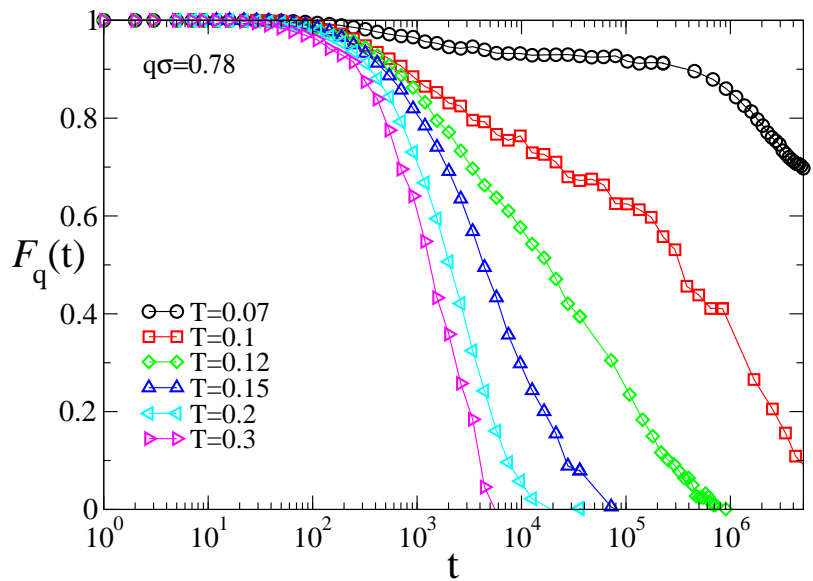}
\end{center}
\caption{Density auto-correlation functions signaling non-ergodic
behaviour of the branching gel of spirals at $\phi=0.16$ and
decreasing $T$. Note that at this $\phi$ all state points are within
the percolating region, but only those at low $T$ show an arrest
transition. Reproduced with permission from \protect\cite{Sciobartlett}. Copyright 2005 Am. Chem. Soc.}
\label{fig:bartlett3}
\end{figure}

Being the clusters ground state structures, bonds will be less and
less broken with decreasing $T$, and a gel state with marked plateau
in the MSD and density correlation functions is detected, as shown in
Fig.~\ref{fig:bartlett3}. Here, only a partial investigation of the
$q$-dependence of $F_q(t)$ was carried out at $\phi=0.16$, suggesting
however a non-ergodic behaviour for a gel structure only at the
smallest studied $q$-values, in the region of the structure factor
pre-peak, while smaller length-scales retain quasi-ergodicity, despite
a marked slowing down of the dynamics also observed in the
MSD\cite{Sciobartlett}.  The functional form of $F_q(t)$ is found to
depend strongly on $q$ and $T$. We note that, at this $\phi$, the
system always percolates for $T<0.2$, but only at low $T$ an arrest
transition is found. Hence the random percolating cluster shown in
Fig.~\ref{fig:bartlett} do not show any slow relaxation as anticipated
before.  In summary, we have reported a case of an equilibrium route
to gelation, whose phase diagram belongs to the category of
Fig.~\ref{fig:phase3}. However, in this special case, the small value
of $\xi$ is not sufficient to generate a Wigner glass.  Interestingly,
it also turns out that a total spherical potential of the kind of
Eq.\ref{eq:potrep} is capable of producing peculiar self-assembled
structures, providing in the case of the low-$T$ Bernal spirals an
effective particle coordination number strictly equal to $6$.

Similar results to those of \cite{Sciobartlett} have been reported for
a modification of the DLVO potential, for which de Candia
{\it et al} also studied the organization into ordered columnar and
lamellar phases at low $T$\cite{Candia}, pointing out that at low
$\phi$ columns are expected, being more stable than the cluster
crystal (due to the small screening length in the studied case), while
at larger $\phi$ lamellae should be formed. However, this phase is
never observed spontaneously during a simulation, due to the
intervening of disordered gel phases of the kind discussed above.

The scenario of competing interactions and formation of a cluster
phase is quite general and can be found in many examples. It was
already established in other types of systems where competing
interactions take place, such as in micelles \cite{Wud92a,Dee94a} and
nanoparticles deposited at the air-water
interface\cite{Sear99JCP,Sear99}.  Recently it was also discussed in
other charged, neutral or magnetic colloidal systems.  For example, in
laponite suspensions different arrested states are observed upon
varying concentration or ionic
strength\cite{Tan04aPRE,NicoLapo,Ruz06Lang}. It has been hypothesized
that different mechanisms are responsible for arrest: a transition
mediated by the clusters at low concentrations, followed by an
arrested state mediated by particles at larger concentrations, which
is still debated in
nature\cite{Ruz04APRL,Ruz04aJPCM,NicoLapo,Ruz06Lang}.  More recently,
a similar behaviour was observed for other clay suspensions, such as
Cloisite\cite{Schurt}, where a low-density solid was interpreted in
terms of a Wigner glass of clusters at low ionic strength.  Also, for
charged liposomes in solutions with charged polyelectrolytes, large
equilibrium clusters are observed, these being stabilized by charge
complexation\cite{Bordi03,Bordi04,Bordi05,Bordi05b}, while, in
mixtures of star polymers and linear chains\cite{Stiak05,Stiak06},
entropic effects give origin to a long-ranged repulsion and depletion
effects induced by the linear chains induce a short-ranged attraction
among the stars which form small clusters. Also, clustering has been
observed in water solutions of silver iodide\cite{Kegel03}, molybdenum
oxide\cite{Kegel04}, as well as paramagnetic colloids in
2d\cite{Hoff06}.  In protein solutions, clusters have been reported
not only for lysozyme, but as well as for cytochrome C
proteins\cite{Bagl04}, hemoglobins\cite{VekBIOJ} and
ferritins\cite{Boutet}, thereby suggesting an underlying similarity of
the competing interaction mechanism, which is applicable to both
colloid and protein solutions.  A case where cluster formation has a
different origin, arising from purely repulsive interactions, is that
of ultrasoft systems\cite{Mladek06}.

\subsection{Patchy models}
\label{sec:patchy}
The formation of an equilibrium gel, and the exploration of the
dynamics close to the ideal gel state, can also be achieved without the
need to invoke additional forces such as long-range repulsion. Indeed,
purely attractive interactions can be tuned to explore the case of
small $\gamma$. To this aim, it is sufficient to ensure a low
coordination number for aggregation, so that there is no driving force
for the system to form a bulk liquid, while network formation is
enhanced. In this framework, hence, $e_{bulk}\simeq e_{surface}$
and for $T\rightarrow 0$ the system will tend to form a disordered
fully-connected network. The concepts of limited-valency and of patchy
particles (or as explained later of patchy-like particles) have been
recently emerging as a new class of materials to build, among various
issues, ideal colloidal gels.

\begin{figure}
\begin{center}
\includegraphics[width=12cm,angle=0.,clip]{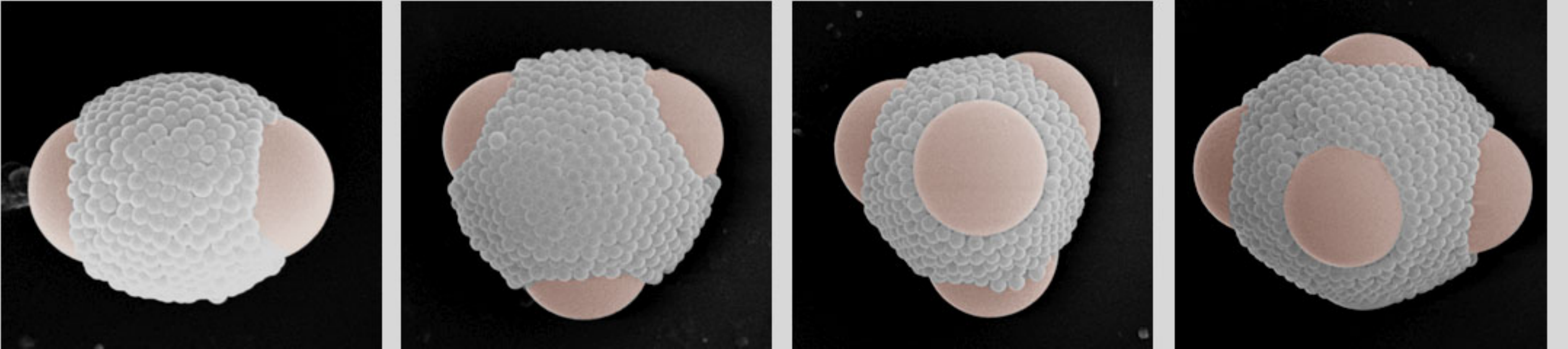}
\includegraphics[width=12cm,angle=0.,clip]{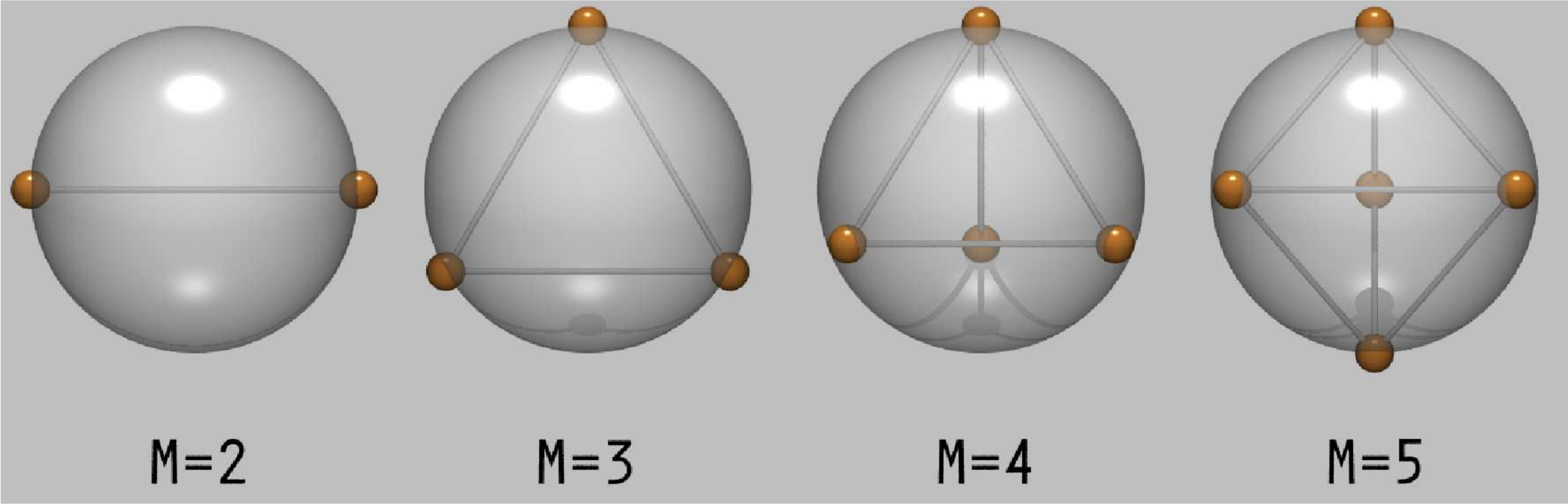}
\end{center}
\caption{Top: Adapted with permission from \protect\cite{Cho_05}. Copyright 2005: American Chemical Society.  Experimental particles realized from bidisperse colloids
in water droplets. Courtesy of G.-R. Yi. Bottom: Reprinted with permission from \protect\cite{Bianchi_06}. Copyright 2006 by the American Physical Society.  Primitive models of
patchy particles used in the theoretical study of
Ref.\protect\cite{Bianchi_06}. }
\label{fig:patchy}
\end{figure}

Experimental realization of such systems is growing at fast pace,
through sophisticated engineering of `colloidal molecules'
\cite{Manoh_03,Yi_04,Klein_05,Cho_05,Liddell_03,Zerro_05}, as well as
use of relative interactions to design colloids with valency
\cite{mohovald}. An example of experimental patchy
colloids is shown in Fig.~\ref{fig:patchy}-top, reproduced from
\cite{Cho_05}.  The aim of these studies is, of course, not limited to
gelation. Indeed, such particles offer the possibility to be used as
building blocks of specifically designed self-assembled structures
\cite{bottomup,Glotz_04,Zhang_03,Zhang_04,Glotz_Solomon,Zhang_05,Doye},
with in mind the ambitious goal to realize a colloidal diamond
crystal\cite{Manoh_03,vanblaad,science-diamond}, which may offer the
possibility of a large photonic band gap for many industrial and
technical purposes \cite{fotonic,Lu_}. Moreover, the implications of
patchy colloids towards proteins\cite{Doye2}, which are patchy by
nature, could be significant.  Literature on this extremely new and
emerging topic is growing fast, due to the many possibilities offered
by the realization of new colloidal molecules\cite{vanblaad}.
Numerical studies are being used to design specific self-assembly
\cite{Glotz_04,Zhang_03,Zhang_04,Glotz_Solomon,Zhang_05,Doye}, as well
as to determine optimized circularly (spherically-)symmetric
interactions in 2d (3d) for producing targeted self-assembly with
low-coordination numbers: by inverse methods, square and honeycomb
lattices\cite{TorquatoPRL,TorquatoPRE} have been assembled in 2d, and
a cubic lattice in 3d\cite{TorquatoPRE2}.  In this paragraph, we will
only focus on the knowledge about phase diagram and gelation of patchy
colloids that is being recently established.

Models have started to appear in the literature, taking into account
not only a spherical attraction, but also angular constraints for bond
formation\cite{Kern03}. Similar ideas have been exploited in the study
of protein phase diagrams\cite{Lom99a,Sear_99}. However, these earlier
works have not addressed the important question of how to
systematically affect the phase diagram of attractive colloids in
order to prevent phase separation and allow ideal gel formation. To
this end, we recently revisited \cite{Zac05aPRL,Zac06JCP} a family of
limited-valency models introduced by Speedy and
Debenedetti\cite{Spe94,Spe96}, where particles interact via a simple
square well potential, but only with a pre-defined maximum number of
attractive nearest neighbours, $N_{\rm max}$, while hard-core
interactions are present for additional neighbours. This model can be
viewed as a toy model for particles with randomly-located sticky
spots, due to the absence of any angular constraint. Moreover, the
sticky spots are not fixed, but can roll onto the particle surface,
also relatively to each other. The disadvantage of such model is that
the Hamiltonian contains a many-body term, taking into account how many
bonded neighbours are present for each particle at any given
time. Notwithstanding this, the model is the simplest generalization
of attractive spherical models, and its study can be built on the vast
knowledge of phase diagram and dynamics for a simple SW potential.

\begin{figure}
\begin{center}
\includegraphics[width=12cm,angle=0.,clip]{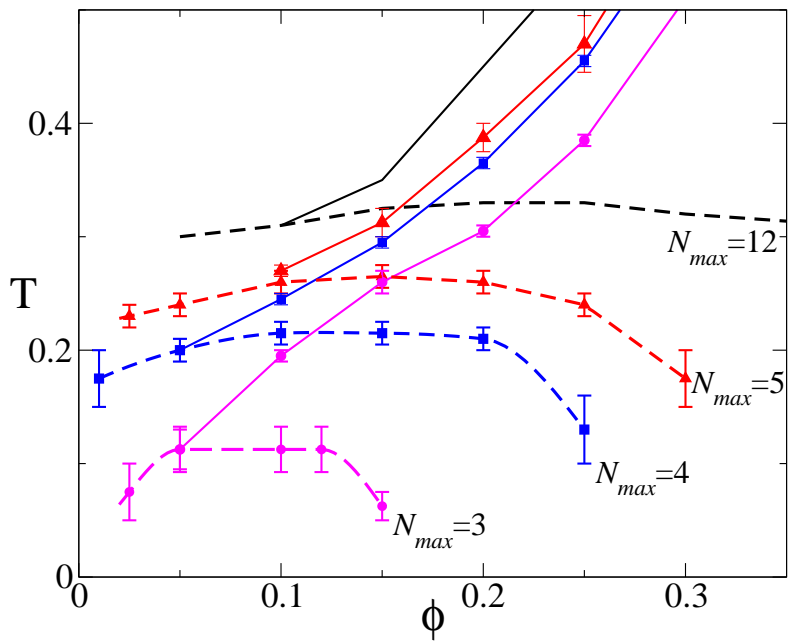}
\end{center}
\caption{Reprinted with permission from \protect\cite{Zac05aPRL}. Copyright 2005 by the American Physical Society. Phase diagram variation with $N_{\rm max}$: the phase-separating region is pushed to
lower temperatures, but most importantly to lower and lower
densities. Similar fate is found for the transient percolation lines.}
\label{fig:nmax}
\end{figure}

The addition of the $N_{\rm max}$ constraint effectively
reduces the tendency of the system to phase separate, when $N_{\rm
max}$ is less than 6 neighbours. In Fig.~\ref{fig:nmax}(left), the liquid-gas spinodal and percolation lines are drawn for
$N_{\rm max}=3,4,5$ and for the standard SW, where by geometric
constraints $N_{\rm max}=12$.
Not only the temperature or energy scale where phase separation takes
place decreases with decreasing $N_{\rm max}$, as also previously
observed with increasing angular constraints \cite{Kern03}, but most
importantly a significant shift in the critical packing fraction is
observed. The shift of the coexistence region is accompanied by that
of the transient percolation line. Hence, a wide region emerges at low
densities, where the system can be equilibrated down to very low
$T$ without an intervening phase separation. In this
region, for example at fixed $\phi$ upon lowering $T$ (e.g. $\phi=0.20$ for $N_{\rm max}=3$ or $\phi=0.30$
for $N_{\rm max}=4$), the bond lifetime grows by
orders of magnitude with respect to the unconstrained SW case\cite{Zac05aPRL},
allowing for the persistence in time of the percolating network, since
the system is already well within the percolating region. Therefore,
gelation can be approached in equilibrium. 
\begin{figure}
\begin{center}
\includegraphics[width=7cm,angle=0.,clip]{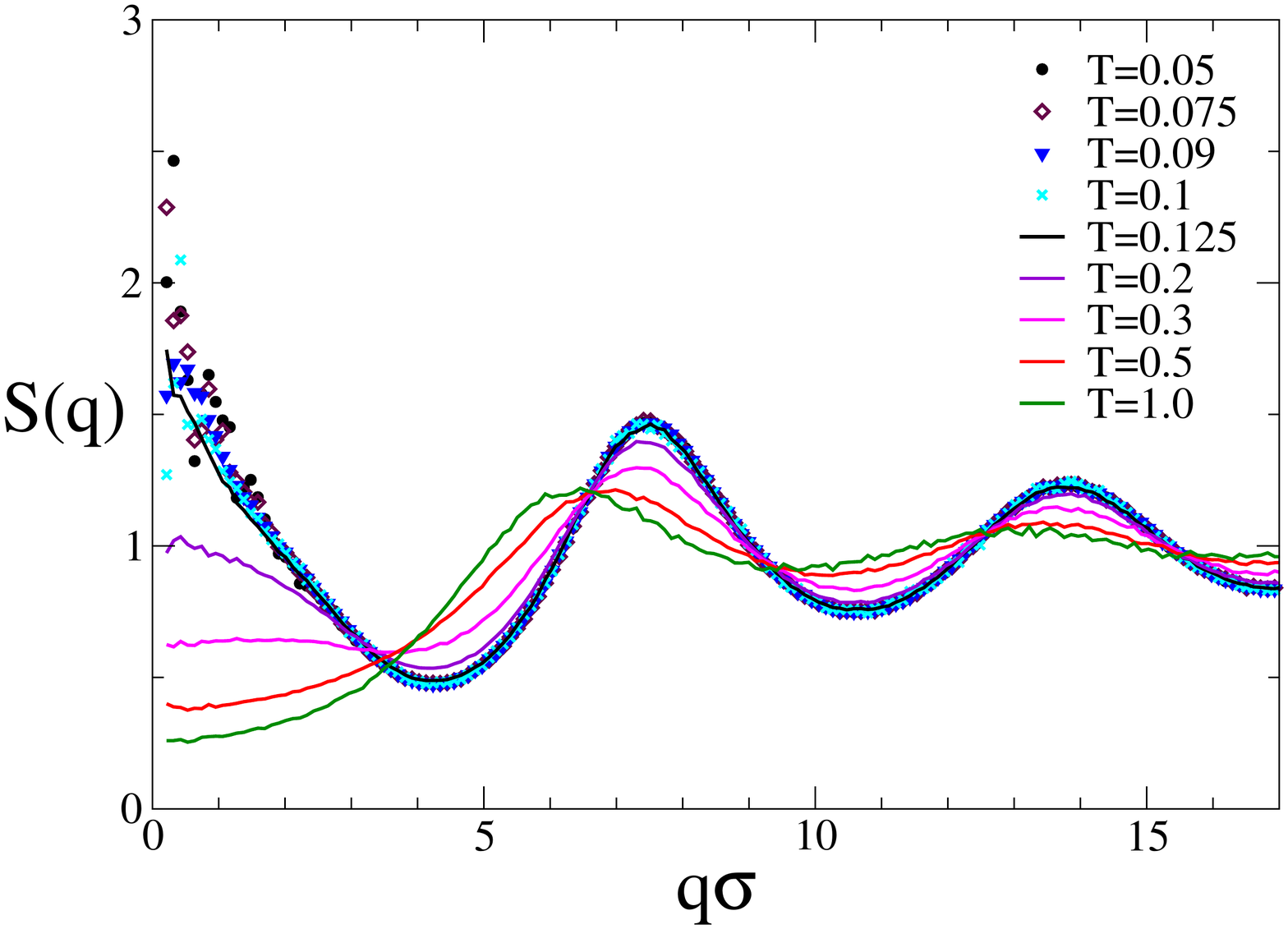}
\includegraphics[width=7cm,angle=0.,clip]{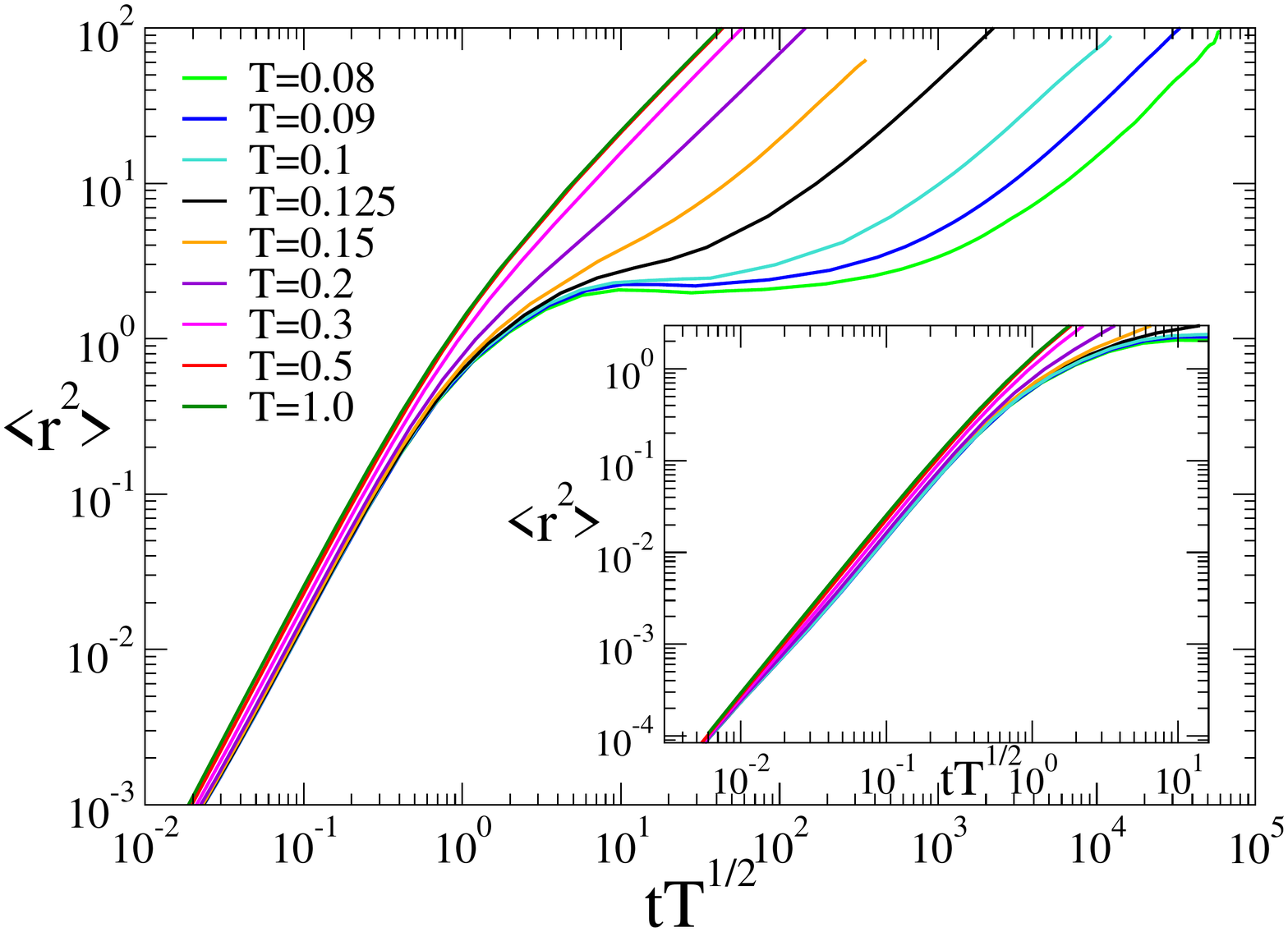}
\end{center}
\caption{$S(q)$ and MSD for $N_{\rm max}=3$ and $\phi=0.20$ at
different $T$. Below $T\lesssim 0.1$ the structure does not evolve
further and the system displays an equilibrium gel transition. The MSD
data are plotted as a function of $tT^{1/2}$ to take into account the
difference in thermal velocities. Data taken from \protect\cite{Zac06JCP,Zac05aPRL}. }
\label{fig:nmax2}
\end{figure}

Let us focus on the properties of this low-$\phi$ equilibrium gel
states. We show in Fig.~\ref{fig:nmax2} the evolution of $S(q)$ (left)
and of the MSD (right) for various studied temperatures at $\phi=0.20$
and $N_{\rm max}=3$. 
The
static structure factor displays progressive structuring of peaks at
$q\sim 2\pi/\sigma$ and multiples thereof as $T$ decreases. Most
importantly, an increase of the low-$q$ signal is detected which
saturates to a finite value. Indeed, around $T\sim 0.1$, the system
has already formed about $99\%$ of the possible bonds, so that no
further restructuring is allowed, and indeed $S(q)$ remains constant
for lower $T$. The increase at low $q$ is an echo of the nearby phase
separation, and indicates that the system is highly compressible due
to the large voids separating the network branches. Hence, large
inhomogeneities are indicative of the equilibrium gel structure.

The MSD, shown in Fig.~\ref{fig:nmax2}(right), displays with
decreasing $T$ a clear plateau, i.e. a significant slowing down of the
dynamics. Such a plateau arises at a very large length scale, of the
order of $\sigma$, suggestive of the large localization length that is
due to the large voids between the network branches. Indeed, the
localization length was found to depend sensitively on $\phi$, but not
on $T$. It is highlighted in the inset that, not only the long-time
plateau displays a marked $T$-dependence, but also a
short-intermediate time behaviour shows difference for the different
studied $T$. This is due to caging within the attractive well, and
indeed appears for  $<r^2> \sim \Delta^2$, compatibly with the bond distance.

Also, the density correlators display marked plateaus\cite{Zac05aPRL}
at small $q$-values, from which it was possible to establish that (i)
the arrest transition shows a marked $q$-dependence in contrast to
what is found in standard glasses and (ii) the non-ergodicity
parameter $f_q$ appears to grow continuously from zero (within
numerical accuracy) and displays a finite signal only at very small
$q$, much smaller than the typical nearest-neighbour
distance. Importantly, at all studied $T$ where it was possible to
equilibrate i.e. $T>0.05$, the system after a long time recovers
ergodicity. For lower $T$, the bond lifetime becomes longer than the
observation time.  Hence, it applies here exactly what we defined as
an equilibrium gel, a disordered state, approached through successive
equilibrium states, with a `long' relaxation time. Ideally, if the
observation time was infinite, the ideal gel at $T=0$ would be
accessed. However, we notice that at fixed $\phi$, while $T$ is so
low that the system has reached the (almost)-fully connected state
(i.e. $T\lesssim 0.1$), the properties of the gel, i.e. localization
length, $S(q)$, $f_q$, do not show further dependence on $T$, hence the
ideal gel is just a continuation of the equilibrium gel to longer and
longer bond lifetime and slower and slower relaxation time.

\begin{figure}
\begin{center}
\includegraphics[width=12cm,angle=0.,clip]{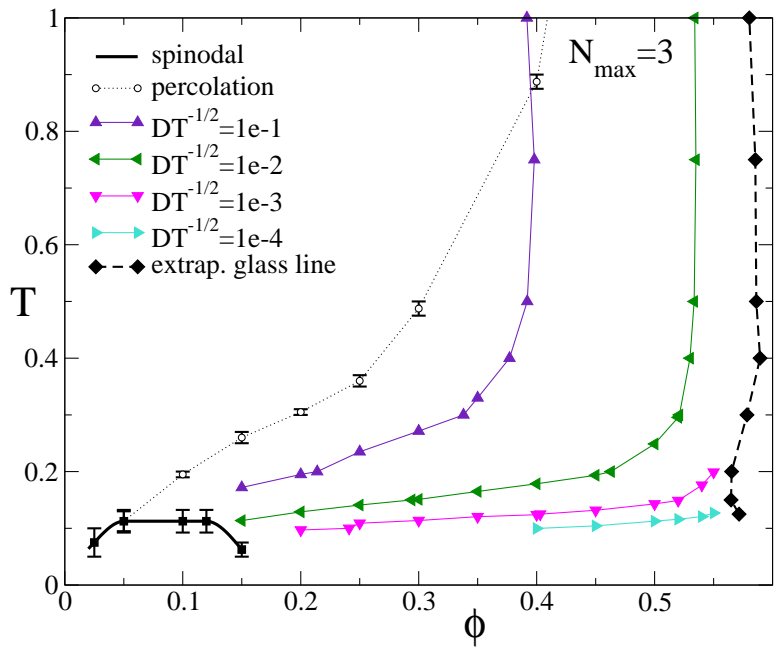}
\end{center}
\caption{Adapted with permission from  \protect\cite{Zac06JCP}.
Copyright 2006, American Institute of Physics.
 Extended phase diagram and iso-diffusivity
lines for $N_{\rm max}=3$.}
\label{fig:nmax3}
\end{figure}
The study of the dynamical behaviour in the $(\phi,T)$ plane, combined with
the phase diagram, is reported in Fig.~\ref{fig:nmax3} for $N_{\rm
max}=3$, revealing that there are two distinct arrest transition
mechanisms\cite{Zac06JCP}.  Indeed, the iso-diffusivity curves,
covering a slowing down of four orders of
magnitudes, show a clear vertical shape at high $\phi$ turning into a
rather flat horizontal shape at low $T$.  For large $\phi$, arrest is
governed by the hard-core only. Indeed, the diffusion coefficient
follows, along isotherms, a power law as $(\phi-\phi_c)^\gamma$, where
the extrapolated glass transition values $\phi_c(T)$ do not show
signification variation with $T$ and remain always close to the
hard-sphere characteristic glass value ($\approx 0.58$).  For smaller
$\phi$ and low $T$, the iso-diffusivity lines cross from vertical to
horizontal, meeting the spinodal region with an almost flat
slope. This horizontal region is where the system forms an almost
perfect network, with $\gtrsim 99\%$ of the particles having saturated
the $N_{\rm max}$ available bonds, but with a high degree of disorder,
signaled by a finite configurational
entropy\cite{Moreno2005,MorenoJCP}. Dynamics becomes increasingly
slow and an equilibrium gel is formed. In this region, $D$ follows a
purely Arrhenius dependence on $T$, suggesting that the activation
barrier to break bonds is the key quantity controlling the dynamics.

\begin{figure}
\begin{center}
\includegraphics[width=12cm,angle=0.,clip]{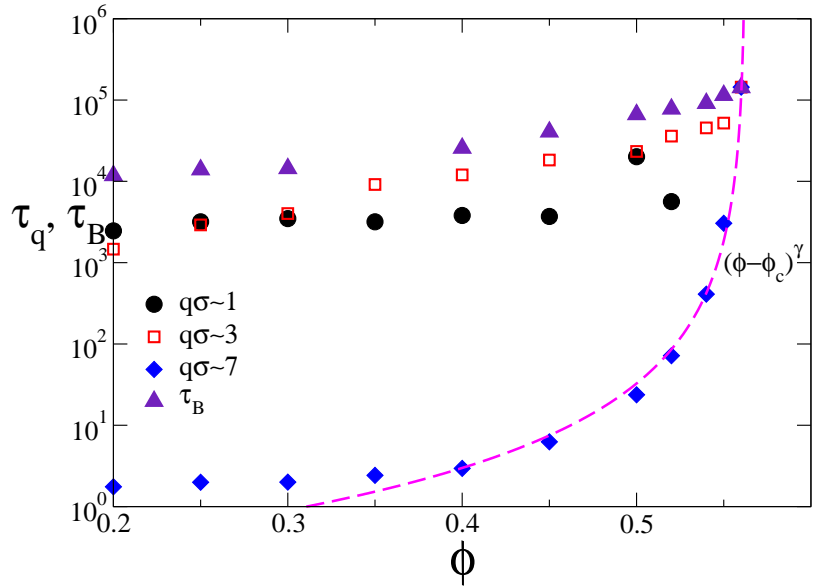}
\end{center}
\caption{Adapted with permission from  \protect\cite{Zac06JCP}.
Copyright 2006, American Institute of Physics. 
Bond lifetime and relaxation times at different wavevectors as
a function of $\phi$ for fixed low $T=0.1$ and $N_{\rm max}=3$. While
low-$q$ dynamics is slaved to the bond lifetime, indicating the gel
nature of the arrest, the large $q$ dynamics is completely decoupled
and only becomes slow at the glass transition with a power-law
behaviour. Here the exponent $\gamma$ is $\approx 2.5$. }
\label{fig:tau}
\end{figure}
\begin{figure}
\begin{center}
\includegraphics[width=7cm,angle=0.,clip]{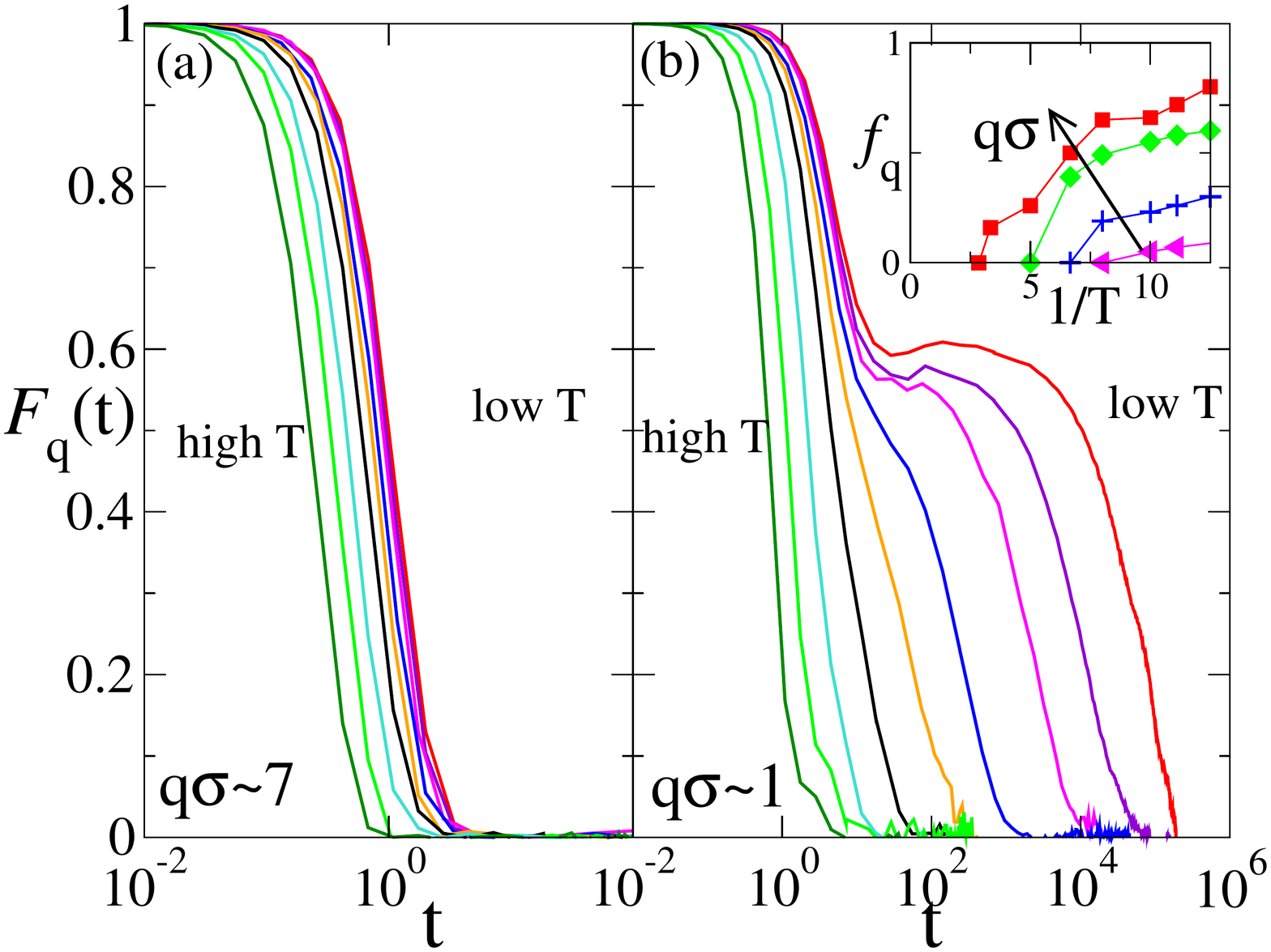}
\includegraphics[width=7cm,angle=0.,clip]{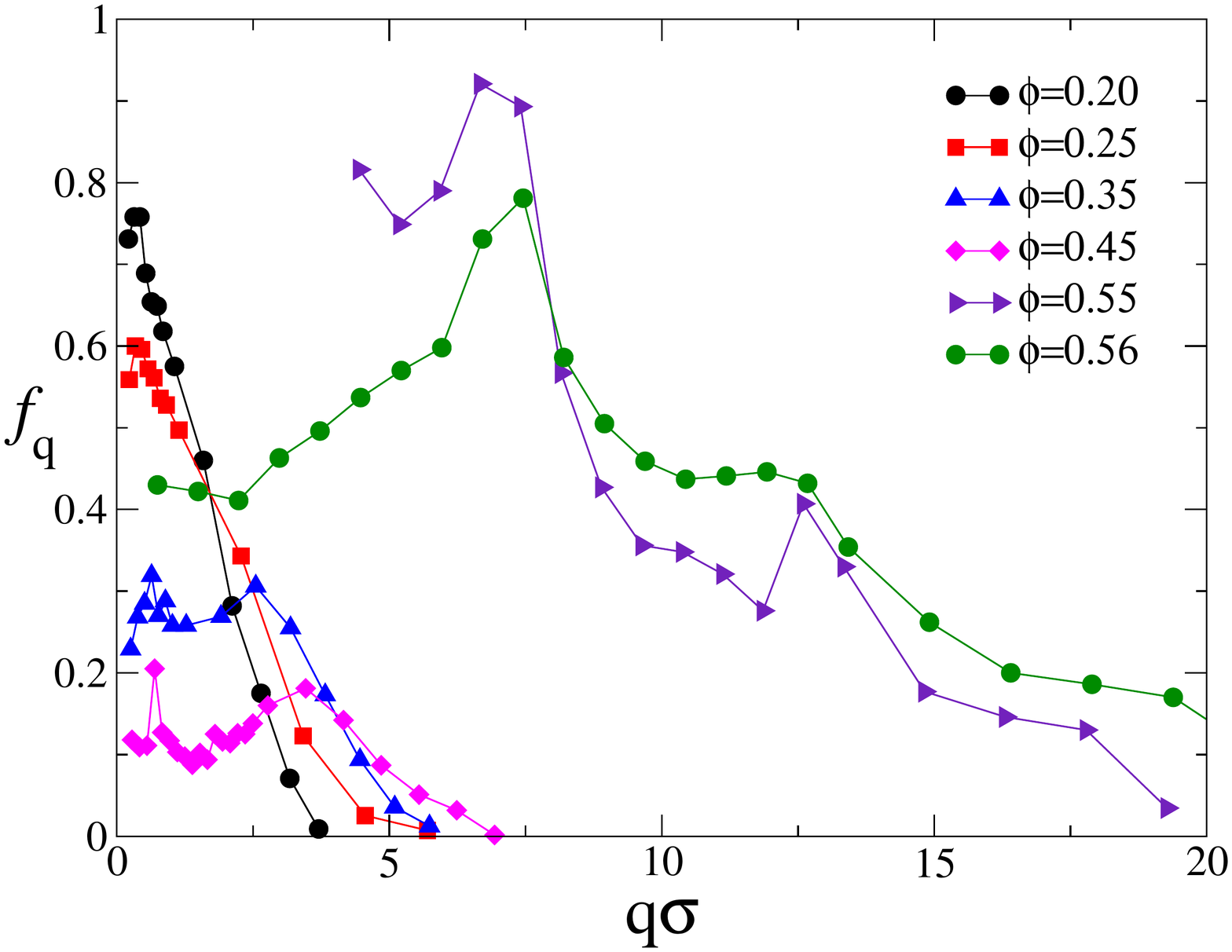}
\end{center}
\caption{Left: $F_q(t)$ behaviour with $T$ at $\phi=0.20$ and $N_{\rm
max}=3$. No slow relaxation is detected at the nearest-neighbour
length scale (left). Inset: plateau value $f_q$ as a function of $1/T$ for $q\sigma=0.2,1,2,3$ from top to bottom. Data from \protect\cite{Zac05aPRL}; Right: Adapted with permission from  \protect\cite{Zac06JCP}.
Copyright 2006, American Institute of Physics.
 Non-ergodicity parameter as a function of $\phi$
at $T=0.1$ and $N_{\rm max}=3$.}
\label{fig:nmax-non-ergo}
\end{figure}

To elucidate better this argument, we show in Fig.~\ref{fig:tau} the
$\phi$-dependence of the bond lifetime $\tau_B$ and of the relaxation
timescale $\tau_q$ at three different $q$-values for fixed $T=0.1$.
It is clear that, at all $\phi$, the bond lifetime is the slowest
timescale of the system, governing entirely the dynamics at this low
$T$. Moreover, the low $q$ dynamics is ruled by $\tau_B$, being
$\tau_q$ proportional to $\tau_B$ in this region, but the large $q$
relaxation is completely decoupled at low $\phi$.  Indeed, it follows
a typical (glassy) power-law increase and joins $\tau_B$ at large $\phi$.
Note that large $q$ here indicates the typical nearest-neighbour
distance, i.e. the relevant length-scale for a glass transition.
Indeed, looking at the full decay of density auto-correlation
functions, we observe that, in the gel phase, for $q\sigma\sim 7$
no slowing down is detected at all, while a clear plateau is observed
for smaller $q$ as reported in Fig.~\ref{fig:nmax-non-ergo}-left. 
These results raise the  fundamental problem that only a certain $q$-window is appropriate
for looking at gel dynamics. Therefore, with respect to glasses,
experimental studies have to focus strictly in the low $q$-limit. The
gel-to-glass transition is evident from the behaviour of the
non-ergodicity parameter $f_q$ with increasing $\phi$, shown in
Fig.~\ref{fig:nmax-non-ergo}-right.

Hence, we clearly identify a crossover from gel to glass dynamics,
that appears to be quite sudden in a small window of $\phi$. It would
be interesting to focus future studies on the investigation of this
crossover, if it is smooth or discontinuous, given also the fact that
evidence of a competition between the two transitions is found at
intermediate $\phi$, signaled by anomalous dynamics similar to that
encountered in the SW potential between attractive and repulsive
glasses, i.e. sub-diffusive MSD and logarithmic decay for the density
correlation functions.  Finally we remark that, in the $N_{\rm max}$
model, no evidence of an attractive glass is found, due probably to
the low-coordination structure which forbids the formation of
attractive cages at low $T$ (here perhaps the study of even lower $T$
would be appropriate, if feasible), as well as the fact that MCT would
predict results similar to those for a standard SW, hence an
attractive glass instead of a gel line\cite{Zac06MCT}.

Despite the simplicity of the $N_{\rm max}$ model, it allows to
establish distinct gel features with respect to standard glasses: (i)
a very large localization length, much larger than that typical of
glasses; (ii) non-ergodicity properties strongly dependent on the
length-scale of observation; (iii) a static structure factor
displaying a growing, but finite signal at low $q$.  Interestingly, in
agreement with previous studies of the bond lifetime influence of the
dynamics discussed in Section \ref{sec:bond}, the density
autocorrelation functions start to display non-ergodic features,
i.e. the emergence of a clear plateau, at first at very low $q$, then
growing in $q$ with further decreasing $T$. Thus, the gel transition
temperature strongly depends on the wave-length of observation, as shown in the inset of Fig. ~\ref{fig:nmax-non-ergo}, in
contrast to what commonly found in glasses, where all length-scales
become non-ergodic simultaneously at a single, well-defined
$T_g$. Hence, if one looks at the typical nearest neighbour distance,
the dynamics appears completely ergodic, in analogy to what observed
in chemical gelation. Only with further increase of density or
decrease of temperature, successively all length-scales become non-ergodic,
crossing over to a glassy behaviour, in agreement with earlier
models\cite{Del03aEL,Voi04a} and with studies in polymer gel-to-glass
transition\cite{corezzietal}.

\begin{figure}[ht]
\begin{center}
\includegraphics[width=10cm,angle=0.,clip]{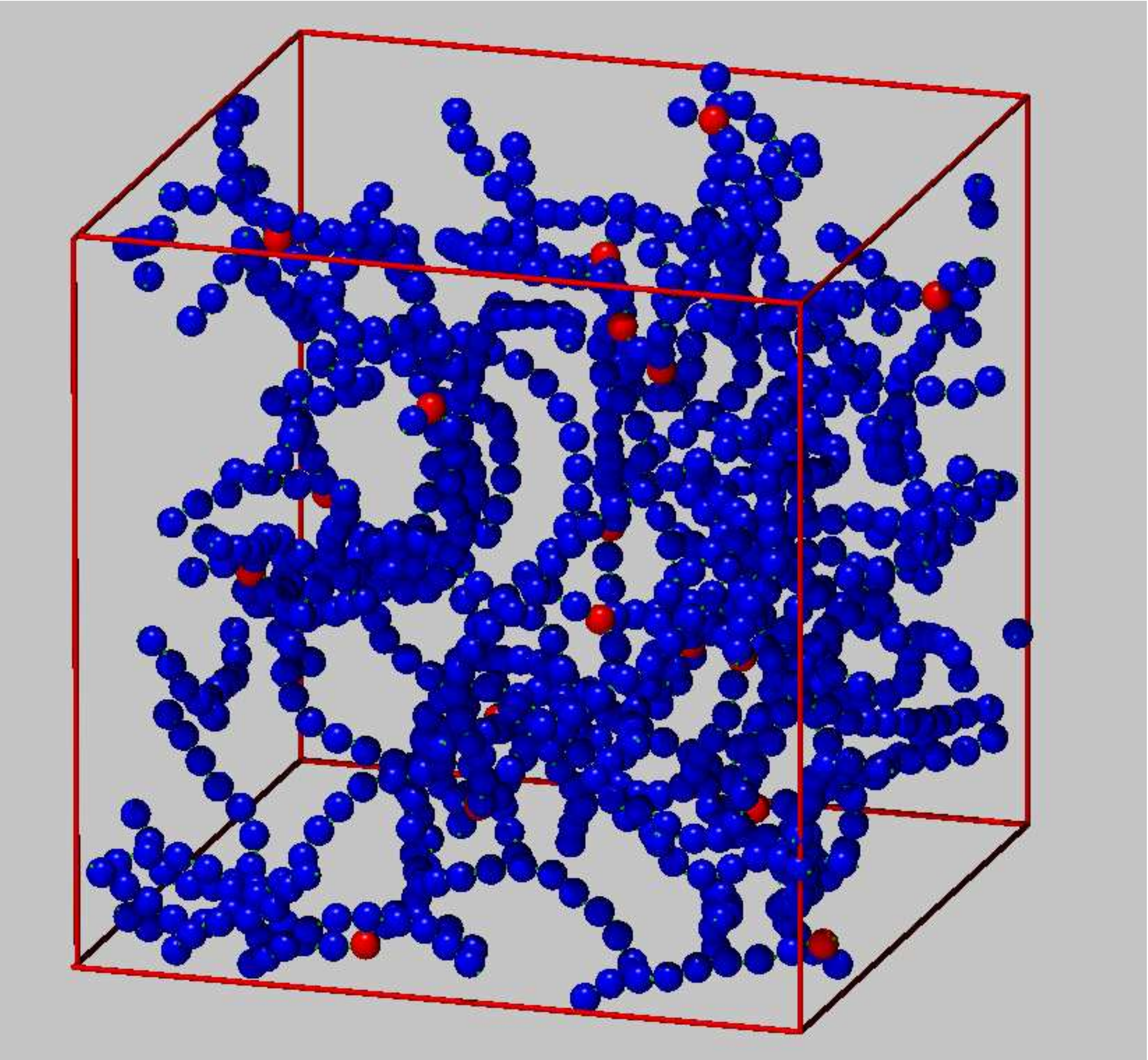}
\end{center}
\caption{A snapshot from simulations of a gel network made of a mixture of $2$ and $3$-coordinated particles of Fig.~\protect\ref{fig:patchy} with $<M>=2.025$ and $\phi=0.033$, at very low $T$. Red
particles are those with $3$ neighbours, i.e. the branching points
giving rigidity to the gel, while blue particles have $2$ neighbours
and provide persistence to the length of network chains. Courtesy of F. Sciortino.}
\label{fig:gel-patchy}
\end{figure}

From the $N_{\rm max}$ model, a step towards more realistic models
with fixed, geometrically organized sticky spots, to mimic
experimentally available particles\cite{Manoh_03} can be done using
ideas already established in the physics of associating liquids
\cite{Kol_87,Nez_89,Nez_90,Sear_96,Monson_98,Ford_04}, such as water,
or silica. These models, that we name `patchy' models, are based on
hard sphere particles, decorated with a small number $M$ of identical
short-ranged square-well attraction sites per particle (sticky spots)
at fixed positions\cite{Bianchi_06}. These particles are shown in
Fig.~\ref{fig:patchy}(bottom). Only when two attractive sites are
within the attractive well distance, a bond occurs. Multiple bond
formation between more than two sites is avoided by a sufficiently
small choice of the attractive range $\delta$, namely $\delta <
0.5(\sqrt{5-2\sqrt{3}}-1) \approx 0.119$.  Such models are amenable to
a thermodynamic perturbation theory treatment, developed by
Wertheim\cite{Werth1},
as well as numerical
simulations\cite{Bianchi_06}.  We find that, also in this case, a
reduction of the number of sticky spots per particle shifts
systematically the critical point and the phase coexistence region
towards lower and lower $\phi$ and $T$. Moreover, the use of binary mixtures
of particles with $2$ and $3$ sticky spots of varying compositions
allows to explore non-integer $\langle M
\rangle$. When $\langle M \rangle\rightarrow 2$, also the critical
point tends continuously to zero, allowing for the possibility to
create `empty' liquids, and accordingly equilibrium gels at very low
temperatures. The dynamics of such systems is currently under
investigation. However, we report a snapshot from a gel obtained from simulations of such a mixture with low $\langle M\rangle$ in Fig.~\ref{fig:gel-patchy}.

We further note the intriguing result that, while a reduction of the
attractive range for spherical potentials has the effect of shifting
the critical packing fraction $\phi_c$ to larger values ($\phi_c\simeq
0.14 $ for the van der Waals limit, while $\phi_c\simeq 0.27$ for the
Baxter model), the reduction of number of sticky spots goes in the
opposite direction, down to $\phi_c=0$, which is the limit for particles
with 2-sticky spots only. The latter can only form strings and hence no liquid phase. It
would be interesting in the near future to complete the phase diagram
knowledge with studies of the crystal phases.

It is interesting to point out here, referring the reader to the
literature growing at fast pace on the topic, the deep analogy between
these models and those `primitive models' for associating
liquids\cite{Kol_87,Nez_89,Nez_90,Sear_96,Monson_98,Ford_04}. Indeed,
recent studies for the dynamics of primitive model for
water\cite{DeMichele_05} and silica\cite{DeMichele_06}, both models
with $4$ bonds per particles, reinforce the robustness of the shape of
the $N_{\rm max}$ phase diagram, shown in Fig.~\ref{fig:nmax3}(bottom), 
suggesting that patchy models for ideal gels
and network glass-formers belong to a separate category of liquids
with a phase diagram where the glass transition does not end into the
coexisting region, as it happens for spherical attractive models, but
into a `new' region of network formation with different dynamical
properties. This corresponds to a new topology of phase
diagrams, that we schematized in Fig.~\ref{fig:phase4}.

A model for colloidal gels, with directional interactions and more
sophisticated than the $N_{\rm max}$ model, was recently introduced by
Del Gado and Kob\cite{DelG05,DelG07}. In this model, the particle is
decorated again by a number of attractive sticky spots (12 sites at
fixed geometry). However, a penalty cost in energy is introduced to
avoid that all attractive spots are simultaneously saturated, through
an angular constraint between bonds. Choosing ad-hoc the involved
parameters (penalty angle, penalty energy), it is possible to
equilibrate the system at very low $\phi$, as low as $5\%$, and to
study the formation of a gel in equilibrium.  The study of the
coordination number of the particles reveals that, at low $T$, most
particles forming the network are involved in only 2 bonds, forming
long chains which provide persistence to the network, while a few
3-bonded particles provide rigidity to the network bridging different
chains. Hence, this model can be considered, in principle, as an
effective mixture of 2 and 3-coordinated particles with very low
$<M>$, very similar to that shown in Fig. ~\ref{fig:gel-patchy}, so
that the absence of phase separation down to very low $\phi$ is in
agreement with the results of Bianchi {\it et al}\cite{Bianchi_06}.
The DelGado-Kob model is naturally richer in ingredients than the
$N_{\rm max}$ toy model, and allows for a careful study of the length
scale dependence of the dynamics.  It indeed confirms the main
predictions for the characteristic gel properties (i-iii). First of
all, a very large localization length, that at such low packing
fractions, can be $10$ times larger than the particle size,
corresponding to the mesh size of the network. Moreover, the static
structure factor displays a finite value at $q\rightarrow 0$, with a
shape similar to that reported in Fig.~\ref{fig:nmax2} for the $N_{\rm
max}$ model, suggesting a significant compressibility of the network
and large inhomogeneities in the structure. However, we note here that
a large contribute to the $S(q)$ increase at low $q$ comes from the
chains contribution\cite{Sci07a}.  Last but not least, a marked
length-scale dependence in the density correlation function is
detected\cite{DelG07}, again suggesting a distinction between large
and small $q$ values, the latter being those significant to
characterize in details the dynamics of the gel network. However, this
model was studied so far at one single packing fraction value, not
allowing a clear characterization of the phase diagram and its
relation to dynamics. This should be the subject of future studies.

Similar behaviour to patchy colloids and the formation of an
equilibrium gel has also been recently reported in dipolar colloids by
Blaak {\it et al}\cite{Blaak}, where the dipolar interactions favor
the formation of chain-like structures (effectively 2-coordinated
particles in the language of patchy colloids) and the formation of a
network is made possible by the use of slightly elongated dipolar
dumbbells. The dumbbell geometry allows for branching of the chains at
low $T$, providing structure to the network which indeed is found to
display a slow relaxation of the dynamics, in close analogy to studies
discussed above where patchy interactions were imposed not via
electro-static interactions\cite{Zac05aPRL,DelG05}.

Finally, there remains a crucial distinction in all these models
between the point where transient networking, i.e. simple percolation,
takes place and where gelation, intended as a substantial increase in
the relaxation times, occurs. This, as we have seen in details in the
introductory part of this review, is a natural consequence of the
decoupling between bond lifetime and network formation at high
$T$. However, once microscopic models for equilibrium gelation are
available, it is interesting also to think backwards and find a way to
reunify ad-hoc percolation and gelation, but maintaining the essential
character of physical gelation, that is reversible bond formation.
This can be achieved by introducing the specificity of the bond, in
order to mimic biological interactions \cite{hiddessen1,hiddessen2},
as well to functionalize colloidal particles with DNA
strands\cite{dna,russel,Luk04a,Luk06a}.  In the latter case, controlling the
length of the chains allows for multiple bond formation at once, so
that, despite bonds being reversible, effectively the bond energy
increases. Dynamics of such systems has been recently studied,
allowing for detection of  gelation in equilibrium very close to the
percolation region\cite{Starr_06,Largo_07}, and with properties close
to that of ideal gels (i-iii). Interestingly, studies of effective
potentials taking into account the specificity of the bonds, manifesting
a natural breakdown of pair-wise additivity similar in spirit to the
$N_{\rm max}$ constraint, have been appearing in the
literature\cite{LargoPRL}. This indicates that, for patchy colloids,
models have to include some level of complications, either in containing
many-body interactions as in the $N_{\rm max}$ and in the Delgado-Kob
models, or geometrical constraints breaking down the spherical symmetry
as in primitive models, or a modification of particle shape as in
dipolar dumbbells. 

\section{Discriminating Different Gels: Static and Dynamic Features; 
A closer look to Experiments}
\label{sec:experiments}

We have presented here different routes to gelation. As explained in
details, we can clearly distinguish the physical mechanisms at hand
resulting in different types of gels. However, in experimental
systems, it is not always clear what interactions play a dominant role
in determining a particular arrest transition, hence it would be
desirable to be able to classify gels accordingly to
the different routes proposed here.

For each of the three routes discussed above, we reported a
typical image of the gel. In two cases, for the arrested
phase-separation (Fig.~\ref{fig:peterlu}) and the competing
interactions(Figs.~\ref{fig:confocal},~\ref{fig:bartlett1}),
experimental examples have been offered. In both cases, simulation
snapshots are very similar to the experimental systems, as for example
in Foffi {\it et al}\cite{Fof05b} for the non-equilibrium gels and in
Sciortino {\it et al} \cite{Sciobartlett} for competing interactions systems (Fig.~\ref{fig:bartlett}), where the role of the Bernal spiral as building block of the gel structure was confirmed. For patchy colloidal gels, experiments are still at
the highly non-trivial level of particle production, so that, in this
case, only numerical simulations can offer snapshots of the structure, such
as in Fig.~\ref{fig:gel-patchy} (see also\cite{DelG05,Blaak,Starr_06}).
Already by simply looking at these gels, enormous differences can be appreciated by eye. 
The aim of this paragraph is to provide some further reference framework for classifying
gels, based on the observation of static and dynamic features.We have the precise scope to offer specific examples, when
available from the literature, to classify the different gels
according to our definitions. By no means we are reporting an
exhaustive list of experimental works on gels.

We start by discussing structural features of the gels. Structural
inhomogeneity, characterized by a non-trivial low$-q$ signal in the
scattering intensity, is an ingredient that is often observed in gels,
and that allows to distinguish gels from glasses, the latter being
generally structurally homogeneous at all length scales. Tanaka {\it et
al}\cite{Tan04aPRE}, in a recent attempt to propose a classification
scheme for gels, with respect to attractive and repulsive glasses,
individuated in the low-angle scattering signal a distinctive gel
feature with respect to glasses.  However, do the structural
inhomogeneities also allow to distinguish between different types of
gels?
In chemical gels, the spatial inhomogeneity is instantaneously induced
by the formation of random irreversible bonds\cite{Adam96}. However,
also if reorganization of the bonds is possible, in colloidal (or
thermoreversible) gels, a similar scenario can be
found\cite{Ikk99,Sol01aPRE}. Indeed, even if particles are left to
equilibrate, they may end up forming rather long chains with a few
crosslinks giving rigidity to the network. Thus, the system can be
locally dense, but with several empty regions, whose typical size is
dictated by thermodynamic parameters, such as density, attraction
strength, etc. Typically, in these conditions, 
a significant signal at low $q$ in the static structure factor is observed.

\begin{figure}[h]
\begin{center}
\includegraphics[width=7.5cm,angle=270.,clip]{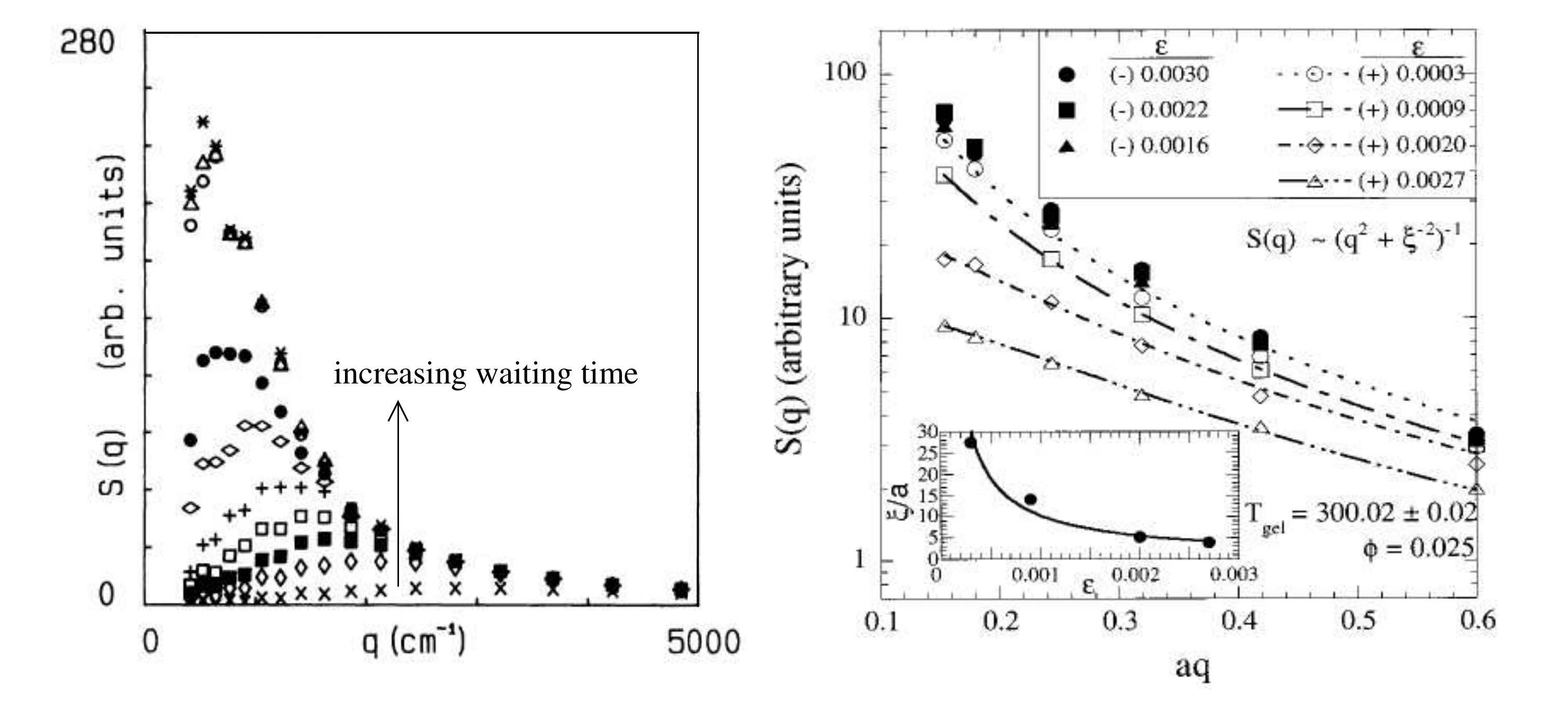}
\end{center}
\caption{Reprinted with permission from \protect\cite{Car92a} and  \protect\cite{Sol01aPRE}. Copyright 1992 and 2001 by the American Physical Society.  Two different experimental reports on the low-$q$ behaviour of the $S(q)$ for: (Left
panel) a DLCA gel as function of waiting time, varying from $722$ to $60753 s$ from bottom to top. The last three curves ($t > 18000 s$)  show a saturation effect; (Right panel) a thermoreversible gel of
adhesive spheres approaching the gel transition with
$\epsilon=(T-T_{gel})/T_{gel}$.  Here $a=40nm$ is the particle radius.
Curves are fits to the scaling $S(q)\sim (q^2+\xi^{-2})^{-1}$ in the
fluid phase (open symbols), with $\xi$ a (phenomenological) correlation length. In the gel phase (full symbols), $S(q)$ does not change
further.}
\label{fig:sq-exp}
\end{figure}

However, it is important to be able to distinguish between an
equilibrium finite value of the compressibility, i.e. a finite value of
$S(q\rightarrow 0)$, from an incipient phase separation, where
eventually $S(q=0)\rightarrow \infty$, or from an interrupted one, where the
initial coarsening (i.e. a growing peak at finite small $q$ in $S(q)$) is stopped at some point. To carefully assess this
issue, in principle, the time behaviour of the low-$q$ region of
$S(q)$ should be monitored during aggregation, as it was done in the
remarkable experiment by Carpineti and Giglio \cite{Car92a} for the case of 
colloidal DLCA.  The behaviour of $S(q)$ was shown to obey growth rules 
similar to  spinodal decomposition of fractal aggregates. The growth of $S(q)$ 
stops at  large waiting times, as shown in Fig~\ref{fig:sq-exp} (left panel). 
This behaviour was interpreted as an imprinting of phase separation on irreversible DLCA
\cite{Sci95a}.  
Cipelletti {\it et al} also reported the time evolution of
the low-$q$ behaviour of $S(q)$ for polystyrene DLCA-like
gels\cite{Cip00a}, where aggregation  proceeds with
time until gelation is reached. At the gel point,  $S(q)$  varies only on a
time-scale of days, apparently towards a more locally compact structure\cite{Cip00a}.

To include the reversibility of the bonds in the picture, we report in
Fig.~\ref{fig:sq-exp} (right panel) the temperature dependence of
$S(q)$ for thermo-reversible adhesive spheres, studied by Varadan and
Solomon \cite{Sol01aPRE}, in the vicinity of the gel temperature.  In
this system, a marked increase of the low-$q$ signal is observed,
becoming more and more pronounced with the proximity of the transition
and that can be well fitted by an Ornstein-Zernike form (lines in
figure), signaling the approach to criticality with an increasing
correlation length. However, when the gel point is reached, the
structure freezes, so that phase separation is
interrupted by gelation.
More experimental studies  of colloidal gels at finite attraction strengths
(where bonds are reversible) should be carried out 
 to properly monitor, in time, the arrested phase separation
scenario. In this way, an often invoked unifying picture between DLCA
and colloidal gels could be ultimately proven.  A
detailed $S(q)$ vs. time experiment would also be desirable for uncharged PMMA
spheres (or similar well-characterized systems), especially to differentiate the two scenarios above and below
$T_g^{sp}$, where the glass line meets the spinodal line (see
Fig.~\ref{fig:phase2}): in the former case phase separation should proceed without arrest, with a monotonic increase
of $S(q\rightarrow 0)$ over time, while in the latter phase separation is interrupted by gelation, similarly to what observed in the experiments mentioned above.  In this way, also the question of how the glass line is affected by crossing the spinodal region could be tackled.

The examples of low-$q$ behaviour of $S(q)$ reported above refer to a
non-equilibrium approach to gelation, both for irreversible and
reversible bond formation.  On the other hand, if gelation is
approached in equilibrium, through one of the routes anticipated in
section \ref{sec:intro}, no hint of spinodal decomposition scaling in
$S(q)$ should be present. Still, however, some remarkable low-$q$
features of $S(q)$ are expected approaching the gel phase.
When competing interactions are at hand, as for example a long-range
repulsion in addition to a short-range attraction, a stable pre-peak
emerges at a finite wave-vector $q^*_c$ much lower than that typical of
nearest-neighbour peak. This was discussed in section
\ref{sec:repulsive} and a typical $S(q)$ was shown in the case of
lysozyme (Fig.~\ref{fig:lyso}). Similar behaviour was also reported
for charged PMMA spheres by Segr\`e {\it et al}\cite{Seg01a}.
Concerning simulations and theory, this is a well-established
phenomenon as we discussed previously. What remains to be established
is a theoretical approach capable to describe in details the evolution
of the cluster peak positions with changing screening length and
energy strength, as well as how a variation of the potential
parameters tunes a crossover from micro to macrophase separation\cite{Pini00}. This
aspect could be particularly interesting for those systems where the
screening length changes rapidly
with particle concentration so that different scenarios may be
observed at different densities, such as in lysozyme\cite{Card06}.
Finally, in the case of gels formed by patchy particles, no
experimental results are yet available. However simulations of different
models\cite{Zac06JCP,DelG07} agree in showing  a finite increase in $S(q)$ 
 at $q\rightarrow 0$,  signaling  a finite
compressibility increase associated to  the formation of an open
network.  This is thus different from the presence of a peak at finite $q$  as well as from a critical-like scenario.

Studies of the dynamical density
fluctuations on approaching the gel line are crucial to characterize the arrest transition.
A pioneering work putting forward the analogy between gel and glass
transition was carried out by Ren and Sorensen\cite{Ren93a}, who
studied a thermo-reversible gel-forming system (gelatin) by dynamic
light scattering, identifying two relaxations that would be the
analogous of the $\alpha$ and $\beta$ relaxation in
glasses\cite{goetze}. The $\alpha$ relaxation equivalence was
identified with a stretched exponential decay approaching the gel
point, while the $\beta$ relaxation was associated to the so-called
`gel mode', i.e. a power-law decay at the gel transition which is
typical of chemical gels\cite{Martin88a,Bro97a}. Such a power law
decay was also found in irreversible colloidal gels\cite{Ell03a}. Most
importantly, Ren and Sorensen discussed the importance of wave-length
dependence in relaxation of gels with respect to glasses, and
highlighted the importance of the low $q$ region to detect
non-ergodicity in gels.
They anticipated the idea that while glasses are localized around
nearest-neighbour lengths, in gels this length should be much larger
than a particle diameter, calling for experiments at different
wave-lengths to study the different dynamics of gels and glasses.
As discussed in Sections \ref{sec:bond} and \ref{sec:routes}, this
concept is crucial in discussing gelation and calls for the need to
use also  small-angle and ultra-small-angle scattering
ranges to properly address colloidal gelation.

\begin{figure}[h]
\begin{center}
\includegraphics[width=10cm,angle=0.,clip]{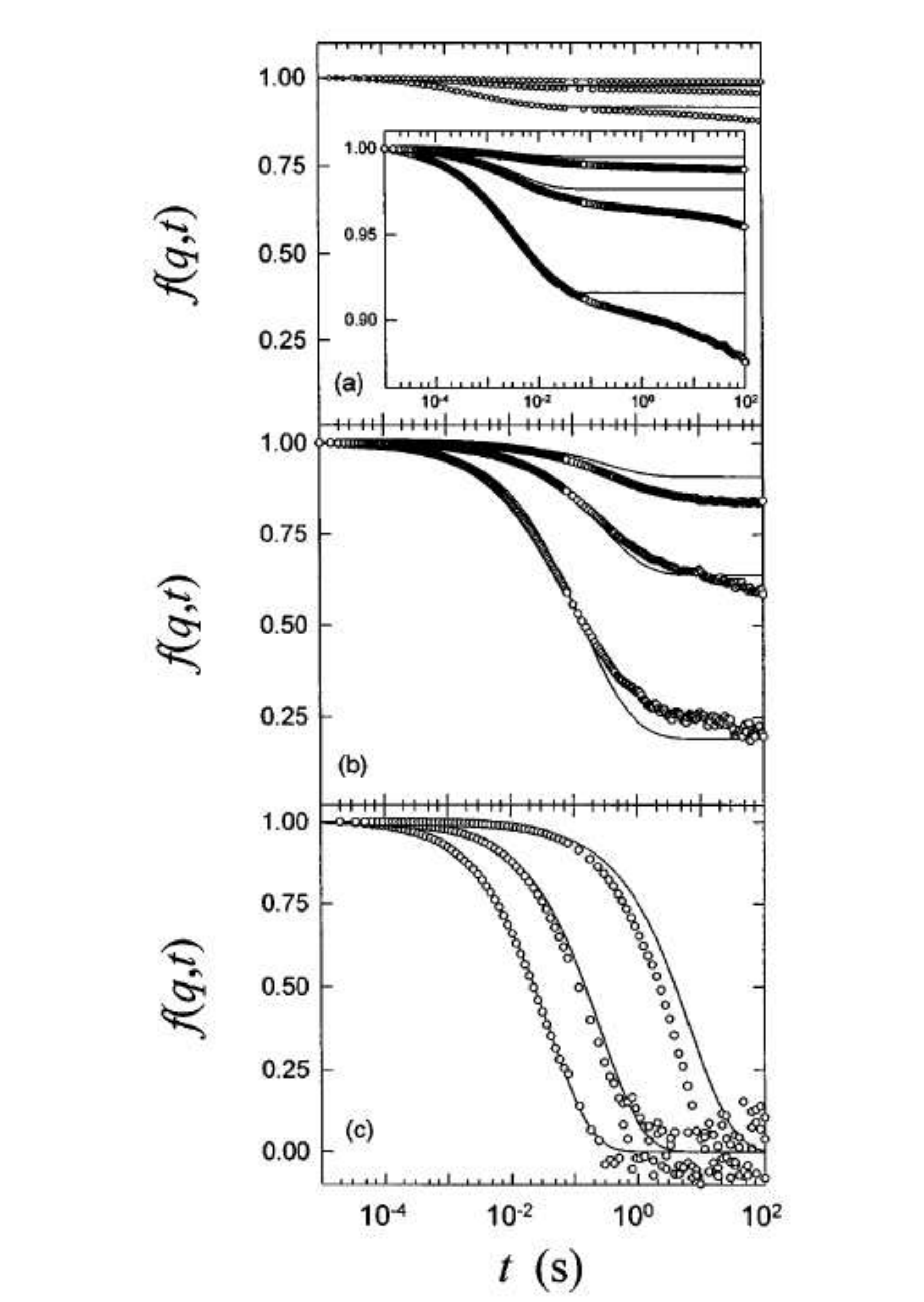}
\end{center}
\caption{Reprinted with permission from \protect\cite{Kra98a}. Copyright 1998 by the American Physical Society. Density correlation functions for a fractal gel at $\phi=5
\times 10^{-3}$ (a), $1.5 \times 10^{-3}$ (b) and $1.7 \times 10^{-4}$
(c) for various wave-vectors ranging from $\sim 0.08 q\sigma$ to $\sim 0.42
q\sigma$.}
\label{fig:krall}
\end{figure}

For fractal colloidal gels close to the DLCA limit (i.e. at very low
$\phi$ and $d_f\approx 1.9$), dynamics of density fluctuations was
studied by Krall and Weitz\cite{Kra98a} and a significant
$q$-dependence was reported in the non-ergodic behaviour.  In
Fig.~\ref{fig:krall}, density auto-correlation functions $F_q(t)$ are
shown for different packing fractions (from $5 \times 10^{-3}$ to $1.7
\times 10^{-4}$) and scattering vectors values (from $0.08 q\sigma$ to
$0.42 q\sigma$) lying in the low-$q$ peak region, well below the
static structure factor peak. In this way, only the cluster-cluster
correlations are probed.  Data show a remarkable $q$-dependence with
packing fraction.  At the smallest $\phi$ the system remains ergodic,
while with increasing $\phi$ a non-ergodic behaviour is observed,
which is much more evident for smaller than for larger $q$. The
observation time is indeed $100s$ and hence, the larger $\phi$-value
state points (a-b) can be considered as gels. The relaxation law of
the scattering curves is well described by a stretched exponential
with exponent $\approx 0.66$, which appears to be valid for all
studied $q$ and $\phi$ values.  Interestingly, given the strong
$q$-variation of relaxation plateau and relaxation time, it should be
possible, in principle, at certain length-scales to observe a
(quasi-)ergodic behaviour at large $q$ and a non-ergodic one at small
$q$.  The fact that the gel transition is strongly wave-vector
dependent is discussed by Krall and Weitz in the
conclusions\cite{Kra98a} and coincides with what is found in
simulations of colloidal gels, both in the case of competing
interactions (Fig.~\ref{fig:bartlett3}) as well as for patchy colloids
(Fig.~\ref{fig:nmax3}). Moreover, it agrees with the $q$-dependent
scenario put forward in simulations of chemical gels\cite{Voi04a}
(Figs.~\ref{fig:sqt-barrier} and \ref{fig:sqt-barrier2}).

Thus, an interesting aspect of the gel dynamics is the wave-length
dependence of the relaxation. However, it is much more common to find
in the literature scattering data at fixed angle, reporting a change
in behaviour as a function of the proximity to the transition,
e.g. varying attraction strength or density. This is a valid approach
for example in glasses, where localization acts at all relevant
length-scales, but it can provide only partial information about
gelation.  In general, various relaxation decays are observed: from
stretched or log-like to compressed or single exponentials. It appears
still difficult to properly associate these features with
inter-particle potentials and length-scale of observation.  The
DLCA-like gel studied by Krall and Weitz was well described by
stretched exponentials at low $q$, while it was found to turn to a
compressed exponential with compressing exponent $\approx 1.5$ at
large $q$ in the aging regime by Cipelletti and
coworkers\cite{Cip00a}.  On the other hand, the thermoreversible
adhesive spheres of Varadan and Solomon\cite{Sol01aPRE} are reported
to approach gelation at low $q$ also in a stretched exponential
manner, while in the gel regime a two-step decay is observed where the
departure from the plateau is exponential. The same authors offer an
interpretation, aimed to explain the different results for the two
cases, which is based on the interactions involved. They suggest that
the different strengths of attraction involved in the gel formation
process (irreversible in one case, giving rise to a network with small
$d_f$, reversible in the other with more compact structure) might be
the key for explaining the different features.  Interestingly, in both
cases, even in the gel phase, ergodicity is restored at very long
times, probably due to the chosen scattering vector that is too large
to detect the length scale of the network.  It would be highly
desirable to be able to associate to specific interactions a certain
behaviour of $F_q(t)$ and in this respect simulation studies may be
valuable (\cite{Voi04a,DelG07}. We note that the shape of the
correlation functions calculated from simulations may be affected by
the choice of the microscopic dynamics (e.g. Newtonian vs. Brownian)
and by neglecting hydrodynamics interactions.  A more detailed
analysis of $F_q(t)$ both on approaching the gel transition and in its
wave-vector dependence should be carried out for ideal gel models: a
DLCA gel, an arrested phase separation one, an equilibrium one from
competing interactions and one from patchy particles. A comparison of
these studies could help to characterize the relaxation in terms of
the different gelation mechanisms.

Finally, we briefly report on visco-elastic gel properties. This is the
subject for which plenty of experiments are available, as a
visco-elasticity study is the most natural way to characterize a
gel, while numerical simulations are rare due to the very demanding
computational effort.  For chemical gels \cite{Adam96}, the viscosity
coefficient on approaching the gelation/percolation transition, as
well the elastic modulus within the gel phase, are found to obey a
power law behaviour, whose exponents can sensitively vary among
different samples. For the viscosity, $\eta\sim (p-p_c)^k$, with $k$
varying in the range $0.6-1.5$, while for the elastic modulus $G\sim
(p-p_c)^t$, with $1.9 \leq t \leq 3$. Several models have been used to
rationalize these differences, pointing to possible different
universality classes due to different bonding units or bond
rigidity.
For colloidal gels, a power law behaviour as a function of increasing
packing fraction is found.  In the case of the fractal gel studied by
Krall and Weitz\cite{Kra98a}, the elastic modulus exponent was
determined to be $t \sim 3.9$. For thermoreversible gels, Grant and
Russel\cite{Grant93} found an exponent close to $3$, while Rueb and
Zukoski\cite{Rueb97,Rueb98} found a slightly different functional
form, but still compatible with a power-law increase with packing
fraction with different exponents, for a similar thermoreversible
suspension.  For PMMA particles with added polymers, it was found that
the power-law exponent varies with the range of depletion attraction,
increasing from $\sim 2.1$ to $\sim 3.3$ as the attraction range is
decreased \cite{Pra03a}.  The onset of elasticity in such colloidal
gels has been interpreted in terms of rigidity
percolation\cite{Tra01aNature,Pra03a} and the change of exponents
attributed to an increase of the gel resistance to bond-stretching for
large ranges, well described by the exponent $t \approx 2.1$, to
bond-bending for shorter ranges, where $t \approx 3.3$. Hence the
difference in the exponents should be associated to different
stress-bearing properties in the gel network\cite{Tra04a}. We also
mention a study of a DLCA-like gel, at moderate densities $\phi\simeq
0.2$, where a power-law increase of the moduli was observed as a
function of time during aggregation\cite{Ell03a}.  From viscosity
measurements, a similar scenario can be inferred from micellar
solutions\cite{Lafleche,Mal05aCM}, since viscosity is found at first
to undergo a sudden increase close to the percolation packing
fraction, later followed by a true divergence close to the glassy
behaviour at large $\phi$.  Also, this suggests that in attractive
systems at first there is the establishment of a network with elastic
properties, only later, at much higher density, crossing over to a
standard glass behaviour.

\section{Conclusions and Perspectives}
\label{sec:conclusions}
In this review we proposed a classification of colloidal gels.  We
highlighted  the difference between transient percolation and gelation
for physical gels and we discussed three different routes to
gelation.  We also tried to address how it should be possible to discriminate among them.

As for colloidal glasses, also in the case of colloidal gelation,
the interplay between simulation, theory and experiments is starting to
provide a coherent picture of arrest at low packing fraction.
In the case of spherically-symmetric potentials in which excluded volume repulsion 
 is complemented only by attraction,  there is now growing consensus that arrested phase separation is the mechanism driving gelation. For competing interactions, combined and closely related experimental, theoretical  and numerical
studies are also providing a coherent view of the arrest processes involved. 
Finally for patchy colloids it is foreseeable, 
that the same will happen in the near future, when
functionalised or patchy particles will become readily available for experimental studies.

A final comment is owed to the relationship between gels and glasses.
Gels and glasses have often been viewed in an unifying framework due to
unambiguous similarities in the arrest transition, i.e. the presence of
long-time plateaus in the relaxation observables which can be also
well-fitted by power-laws\cite{Kum01a,Seg01a}. For this reason,
MCT-like approaches have been often applied to describe also the gel
transition\cite{Ber99a,Kro04aPRL}. However, we pointed out in this
review the limits of such applications, both in relation to the
arrested-phase separation scenario and towards the development of a
Cluster-MCT. While both gels and  attractive glasses
undergo  arrest dictated by bond formation, many differences are
present. In our opinion, the main difference is that  arrest occurs within the 
presence of a spanning network
in the gel case (thus with a large localization length dictated by the
mesh size of the gel)  and through local cages arising from the
short-range bonds in the attractive glass case (where the localization length is provided by
the bond distance). Consequently a completely different scenario arises
in the non-ergodic behaviour of gels, shifting the interesting
$q$-range of arrest to values that are 
smaller than that of glasses (more than two orders of magnitude in
the case of attractive glasses).  
A notable case is, in this context,
the Wigner glass, which  can be made either of particles
or of clusters.  In this review, we proposed  to classify it as a glass, despite
the low density,  due to the absence of connectivity and 
of attractive interactions in stabilizing the arrested disordered
structure. However, we might speculate  that the Wigner glass is an  intermediate state
between gels and glasses as it should share some properties with gels
(structural inhomogeneity, large localization
length) and some with glasses (low shear modulus, repulsive
caging). Therefore, a detailed study of the Wigner dynamical arrest
transition, also focusing on the $q$-dependence of the density
correlation functions, will be important to clearly establish its
classification.

Throughout this review, we pointed out several times that 
gelation is strongly associated to a connectivity transition. 
Hence, if the lifetime of the bonds is sufficiently long, 
ergodicity should strictly be broken starting from
$q=0$. Within the gel region, larger and larger $q$ will display then
non-ergodic behaviour, but each $q$ at a different distance from the
$q=0$-gel transition.  It will be important to characterize this
aspect, firstly for chemical gels\cite{Mattson}, and then possibly for
physical gels of various kinds.  A closer look should be taken to
the low-$q$ behaviour both of $S(q)$ as well as of $F_q(t)$, to
definitely identify the distinctive features of gels, and the peculiar
characteristics arised from the different arrest mechanisms. To
develop an ideal theory of gelation, previous studies on reversible
polymer gelation should be taken into account\cite{Rub99a}, combined
with the central idea of the $q=0$-transition at long-lived
percolation. Clearly, the lack of a predictive framework constitutes a
problem in the present-day study of colloidal gelation.

On the other hand, the close similarity between gels and glasses, both
being phenomena of dynamical arrest, suggests to look for the
characteristic arrest signatures also in gels. In particular, in close
analogy with investigations in colloidal glasses\cite{Wee00aScience},
recent, intense activity is aiming to characterize dynamical
heterogeneities of colloidal gels.  Through diffusing wave
spectroscopy and time resolved correlation
techniques\cite{Cip03PCCP,Cip05a}, confocal microscopy
experiments\cite{Solomon06,kilfoil}, and
simulations\cite{Pue04b,coniglio07,hurtado}, the complexity of gels
emerges also in terms of different populations of slow and fast
particles, which become more and more evident approaching the gel
transition, and remarks a close analogy of both gel and glass
transition in terms of dynamic cooperativity. Such studies are extremely
useful for the establishment of a unifying theoretical framework of a
(generic) dynamical arrest transition, although, again, differences
are expected among the different mechanisms. For example, we
expect that, for an arrested phase-separated gel, dynamical
heterogeneities should look very much different from those arising in
equilibrium gels.

This review shows that the study of colloidal gels is still
challenging, despite the progress that has been made in the last
decade. Colloidal gelation is a rich field of scientific
investigation, which offers the possibility to develop new theories
and models, as well to design new classes of materials (as in the case
of patchy particles). The progresses in the study of colloidal
gelation will hopefully also provide a deeper understanding of several
protein-related aggregation processes\cite{Sear_99,Sear06,DoyePoon}.

\section{Acknowledgments}
First of all it is a pleasure to thank Francesco Sciortino and Piero
Tartaglia for many discussions on colloids, glasses and gels, as well
as for illuminating comments on this manuscript. Also, I deeply thank
Hartmut L\"owen for suggesting that I write a review on this topic,
and Silvia Corezzi and Gi-Ra Yi for providing me figures. 
Warmest thanks to all colleagues who granted me permissions to
reproduce their figures as well as to the close collaborators who
helped me, with their work, to understand relevant aspects of
colloidal gelation: I. Saika Voivod, G. Foffi, S. Mossa, E. Bianchi,
J. Largo, A. J. Moreno, E. La Nave, C. De Michele, S. Buldyrev,
F. Ciulla, B. Ruzicka.  I am also grateful to C. N. Likos, P. J. Lu,
and F. Cardinaux for many stimulating discussions.
I acknowledge support from MIUR-Prin and Marie Curie
Network on Dynamical Arrest of
Soft Matter and Colloids MRTN-CT-2003-504712.
\section{References}
\bibliographystyle{./iopart-num} 
\bibliography{./articoli,./altra,./advances,./biblio_patchy.bib,./star-star}

\end{document}